\documentclass[aps,preprintnumbers,nofootinbib,eqsecnum]{revtex4}
\usepackage{graphicx}
\usepackage{epsf}
\usepackage{amsmath}
\usepackage{epstopdf}
\usepackage{bm}
\usepackage{color}
\usepackage{tabularx}
\usepackage{enumitem}
\usepackage{float}
\usepackage{array,booktabs}
\usepackage{footnote}
\usepackage{threeparttable}
\usepackage{graphicx}
\usepackage{hyperref}
\usepackage{amssymb,epsf}
\usepackage{latexsym}
\usepackage{epstopdf}
\usepackage{epsfig}
\usepackage[normalem]{ulem}
\usepackage{caption}
\usepackage{subfig}
\usepackage{amssymb}

\newcommand{\tcb}{\textcolor{blue}}

\begin{document}

 \title{Neutron Stars in Mimetic Gravity}

 \author{
 Hajar Noshad$^{1,2,3}$\footnote{email address: noshad.hajar2@gmail.com},
 Seyed Hossein Hendi$^{1,2,4}$\footnote{email address: hendi@shirazu.ac.ir}
 Behzad Eslam Panah$^{4,5,6}$\footnote{email address: eslampanah@umz.ac.ir}}

 \affiliation{
 $^{1}$Department of Physics, School of Science, Shiraz University, Shiraz 71454, Iran\\
 $^{2}$Biruni Observatory, School of Science, Shiraz University, Shiraz 71454, Iran\\
 $^{3}$Department of Physics, College of Science, Yasouj University, Yasouj, 75918-74934, Iran
 $^{4}$Canadian Quantum Research Center 204-3002 32 Ave Vernon, BC V1T 2L7 Canada\\
 $^{4}$Department of Theoretical Physics, Faculty of Science, University of Mazandaran, P.O. Box 47415-416, Babolsar, Iran\\
 $^{5}$ICRANet-Babolsart, University of Mazandaran, P. O. Box 47415-416, Babolsar, Iran\\
 $^{6}$ICRANet, Piazza della Repubblica 10, I-65122 Pescara, Italy}

\begin{abstract}
In this paper, a modified version of the hydrostatic
equilibrium equation based on the mimetic gravity in the presence
of perfect fluid is revisited. By using the different known
equation of states, the structural properties of neutron stars are
investigated in general relativity and mimetic gravity. Comparing
the obtained results, we show that, unlike general relativity, we
can find the appropriate equation of states that support
observational data in the context of mimetic gravity. We also find
that the results of relativistic mean-field-based models of the
equation of states are in better agreement with observational data
than non-relativistic models.\\ \\
\end{abstract}

\maketitle

\section{Introduction}

General Relativity (GR) and its black hole solutions are interesting topics
in gravitating systems. Despite having acceptable information from the
region beyond the event horizon of black holes, the interior solution of
horizonless massive objects is one of the open questions in physics. Indeed,
the interior properties of stars and their time evolution are of interest
during recent decades. The Tolman-Oppenheimer-Volkov (TOV) equation \cite{TOV1,TOV2} is one of the primitive attempts to describe the
interior properties of spherical static perfect fluid objects.
However, for solving the field equations and find the physical
properties of a typical star, we have to consider an equation of
state (EoS) explaining the relationship between two physical
quantities; pressure and density.

We should emphasize that the interior solutions are based on three
components: i) metric ansatz, ii) gravitational theory with an
appropriate energy-momentum tensor such as GR with a perfect fluid
which leads to the equation of hydrostatic equilibrium known as
TOV, iii) EoS such as polytropic relation between pressure and
density. Since the theoretical results of the TOV equation arisen
from GR are not consistent with the observational evidence,
different attempts have been considered to modified one (some) of
the mentioned components. All these attempts are motivated by
obtaining a better agreement between theory and observation, or
testing alternative theories of gravitation. In this regard,
different models of gravity are considered to investigate the
compact objects and their modified TOV equations. For example
neutron stars in an energy dependent spacetime are studied in Ref.
\cite{TOVrainbow}. Modified TOV equation in
vector-tensor-Horndeski theory and dilaton gravity are reported in
Refs. \cite{TOVHorndeski} and \cite{TOVdilaton}, respectively.
Moreover, the neutron star structure is investigated in
the context of $F(R)$, $F(G)$, $F(T)$, $F(R,T)$ and $F(R,L)$
theories of gravity
\cite{TOVfR1,TOVfR2,TOVfR3,TOVfG,Astashenok2015,Capozziello2016,Astashenok2017,Astashenok2020,Lobato2020,Carvalho2020,labato2021,Astashenok2021,Pretel2021,Lin2022}.
Furthermore, neutron stars in massive gravity and magnetized
neutron stars in gravity's rainbow are analyzed in
\cite{massiveTOV} and \cite{Magnetized}, respectively. The
structure of neutron stars in massive scalar-Gauss-Bonnet gravity
is studied in Refs. \cite{Saffer2021,Staykov2022,Xu2022}. Besides,
neutron stars in the context of Eddington-inspired Born-Infeld
theory of gravity are extracted in Refs. \cite{Prasetyo2021}.
Considering the scalar-tensor theories of gravity, some
interesting properties of neutron stars are studied in Refs.
\cite{Harada1998,Barausse2013,Doneva2013,Shibata2014,Mende2016,Silva2019}.
In addition, neutron star structure in Hořava and
Hořava-Lifshitz theories of gravity are evaluated in Refs.
\cite{Barausse2019} and \cite{Kim2021}, respectively. See Refs.
\cite
{ModTOV1,ModTOV1b,ModTOV2,ModTOV2b,ModTOV3,ModTOV4,ModTOV4b,ModTOV6,ModTOV7,ModTOV7b,ModTOV7c,ModTOV7d,ModTOV8,ModTOV9,ModTOV10,ModTOV10b,ModTOV10c,ModTOV11,ModTOV12,ModTOV13,ModTOV14,ModTOV15,ModTOV16,ModTOV17,ModTOV18,ModTOV19,ModTOV20,ModTOV21}
for additional information about neutron stars in modified gravity
theories.

In this paper, we regard a modified TOV equation comes from the mimetic
gravity in the presence of perfect fluid. As a matter of fact, mimetic
gravity is a Weyl-symmetric extension of GR in which a dust-like perfect
fluid can mimic cold dark matter from the cosmological point of view \cite%
{chamseddine2013mimetic}. So, one of the main motivations of considering
mimetic gravity is providing an interesting geometrical description for
challenging topics such as the late-time acceleration \cite%
{odintsov2016accelerating}, inflation \cite{zhong2018inflation} and dark
matter \cite{chamseddine2013mimetic}. Indeed, considering the mimetic scalar
field in gravitating systems, one finds a dynamical longitudinal degree of
freedom in addition to two traditional transverse degrees of freedom of GR.
It is shown that such an additional degree of freedom could play the role of
mimetic dark matter. The main idea of mimetic gravity backs to the
interesting work of Chamseddine and Mukhanov \cite{chamseddine2013mimetic},
that proposed isolation of the conformal degree of freedom of gravity by
introducing a parameterized physical metric $g_{\mu \nu }$ in terms of an
auxiliary metric $\tilde{g}_{\mu \nu }$ and a (mimetic) scalar field $\phi $%
, as follows
\begin{equation}
g_{\mu \nu }=-\tilde{g}_{\mu \nu }\;\tilde{g}^{\gamma \delta }\;\partial
_{\gamma }\phi \partial _{\delta }\phi ,
\end{equation}%
where confirms that the physical metric is invariant under conformal
transformations of the auxiliary metric. Besides, taking a consistency
condition into account, it is easy to show that the mimetic scalar field
should satisfy the following constraint
\begin{equation}
g^{\gamma \delta }\partial _{\gamma }\phi \partial _{\delta }\phi =-1.
\end{equation}

In this work, we follow the conventional notation of mimetic
gravity in which this constraint can be appeared at the level of
action formalism by including a Lagrange multiplier. Following
Ref. \cite{chamseddine2013mimetic}, it is worth mentioning that
the conformal degree of freedom becomes dynamical even in the
absence of matter field, and therefore, the mimetic gravity may
admit non-trivial solutions.

Motivated by what was mentioned above, the mimetic gravity has arisen a lot
of attention in the past few years. From the cosmological point of view, it
is a powerful theory to explain the flat rotation curves of spiral galaxies
\cite{vagnozzi2017recovering}, gravitational wave \cite%
{casalino2018mimicking,casalino2018alive} and the cosmological
singularity \cite{chamseddine2017resolving}. This theory can
resolve the singularity in the center of black holes
\cite{chamseddine2017nonsingular} and introduce new black hole
solutions from the gravitational viewpoint. Thus, it will be
interesting to investigate its effect on astrophysical objects
such as neutron stars and study their structures.

Without exaggeration, neutron stars are proper laboratories to
examine fundamental physics at high energy regime; from strong
gravitational field to considerable nuclear densities. Actually,
neutron stars help us to investigate the high density low
temperature regime of matter field that is complementary to
terrestrial laboratories. The density of neutron stars is
distributed from few $\mathrm{g/cm^3}$ to more than $10^{15}
\mathrm{g/cm^3}$ at their surface and the center, respectively.
The different layers of a neutron star can be categorized into the
atmosphere, the outer and inner crusts, and also outer and inner
cores. The outer crust can be described by comparing it with
experimental data of atomic nuclei. Nevertheless, uncertainty in
the EoS of neutron stars for regions with densities above nuclear
matter density is still a substantial challenge since analyzing
the astrophysical observations depends properly on the neutron
star structure. So, there are different proposed EoSs that can
describe the inner neutron stars with different structures such as
the nucleon, hybrid, and the strange quark. The main theoretical
techniques for determining EoSs are based on variational method,
relativistic mean-field (RMF) models, effective interactions,
perturbation expansion, effective energy-density functionals, and
so on \cite{24}.

In this work, we consider the nuclear and hadronic stars which are
in the range of the gravitational wave event GW170817
\cite{abbott2018gw170817} and they have the maximum mass
$M_{max}>2M_{\odot}$. For pure nuclear matter, we consider
different EoSs: SLy4 \cite{douchin2001unified} of the Lyon group
that is based on energy density functional, WFF1
\cite{wiringa1988equation} that is obtained by the variational
method. We also include BSK21 \cite{fantina2012unified} and MPA1
\cite{muther1987nuclear} constructed from the generalized Skyrme
nuclear interaction arising from the Argonne $V18$ potential plus
a microscopic nucleonic three-body force and
Brueckner-Hartree-Fock theory, respectively. Besides, we
choose the density-dependent relativistic mean-field (RMF) HS
(DD2) EoS \cite{MHempel2010,Fischer2014} and the recently proposed
FSU2H \cite{Tolos2017,Negreiros2018,Providencia2019}. For the
hadronic stars, we use various hadronic EoSs: $BHB\Lambda\Phi$
\cite{SBanik2014} and SFHoY
\cite{MFortin2018,MHempel2010,AWSteiner}.

In this paper, we consider the different observational
evidence of mass-radius relation of neutron stars, such as
$GW170817$, PSR $J0740+6620$, PSR $J2215+5135$, NICER data on PSR
$J0030+0451$ and GW190814. In fact, taking different observational
data into account, we want to demonstrate how the mimetic gravity
can be adapted to the appropriate EoS to obtain theoretical
results consistent with observational evidence.

The outline of this paper is as follows. In the next section, we
consider the spherically symmetric metric and obtain the required
equations for solving, numerically, the generalized TOV in mimetic
gravity. In section III, we introduce a set of nuclear matter EoSs
and three potentials to drive $M-R$ diagrams and discuss the
properties of neutron stars. Finally, we finish our paper with
conclusions.

\section{Hydrostatic equilibrium equation in mimetic gravity \label{FE}}

The starting point is the action of mimetic gravity in four dimension in the
presence of a matter field as
\begin{equation}
S=\frac{c^{4}}{16\pi G}\int d^{4}x\sqrt{-g}\left[ \mathcal{R}+\lambda \left(
g^{\mu \nu }\partial _{\mu }\phi \partial _{\nu }\phi -\epsilon \right)
-V(\phi )\right] +\mathcal{I}_{m},  \label{actin}
\end{equation}%
where $\mathcal{R}$ and $\phi $ are, respectively, the Ricci
scalar and the mimetic scalar field, $\lambda $ is the Lagrange
multiplier while the constant $\epsilon =\pm 1$ can be fixed by
spacelike or timelike nature of $\partial _{\mu }\phi $. Besides,
$V(\phi )$ is the potential related to the mimetic scalar field
and $\mathcal{I}_{m}$ denotes the action of matter field which we
consider it as a perfect fluid. It is straightforward to vary the
action (\ref{actin}) with respect to the metric tensor $g_{\mu \nu
}$ and the mimetic scalar field $\phi $ to obtain the equations of
motion as
\begin{equation}
G_{\mu \nu }=\lambda \partial _{\mu }\phi \partial _{\nu }\phi -\dfrac{1}{2}%
V(\phi )g_{\mu \nu }+\kappa T_{\mu \nu },  \label{FE1}
\end{equation}%
\begin{equation}
\frac{1}{\sqrt{-g}}\partial ^{\mu }\left( \sqrt{-g}\lambda \partial _{\mu
}\phi \right) =-\frac{1}{2}\frac{dV(\phi )}{d\phi },  \label{FE2}
\end{equation}%
where $\kappa =\frac{8\pi G}{c^{4}}$ and $G$ is the four dimensional
gravitational constant. In addition, $G_{\mu \nu }$ is the Einstein tensor and $%
T_{\mu \nu }=\frac{-2}{\sqrt{-g}}\frac{\delta \mathcal{I}_{m}}{\delta g^{\mu
\nu }}$ is the energy-momentum tensor with the following explicit form
\begin{equation}
T^{\mu \nu }=(\rho +\frac{P}{c^{2}})u^{\mu }u^{\nu }-Pg^{\mu \nu },
\label{Tab}
\end{equation}%
where $\rho $ and $P$ are, respectively, the density and pressure of the
perfect fluid measuring by a local observer, and $u^{\nu }$ is the fluid
four-velocity ($u_{\nu }u^{\nu }=c^2$). Moreover, one can vary the above
action with respect to the Lagrange multiplier to obtain
\begin{equation}
g_{\mu \nu }\partial ^{\mu }\phi \partial ^{\nu }\phi =\epsilon .
\label{epsilonEq}
\end{equation}%

Taking the trace of Eq. (\ref{FE1}) and inserting Eq. (\ref{epsilonEq}), one
finds%
\begin{equation}
\lambda =\frac{(G_{\alpha }^{\alpha }-\kappa \,T_{\alpha }^{\alpha }+2V)}{%
\epsilon }.  \label{lambda}
\end{equation}%

Now, we can insert $\lambda $ (Eq. (\ref{lambda})) into the field equations
to achieve%
\begin{equation}
G_{\mu \nu }-\epsilon (G_{\alpha }^{\alpha }-\kappa \,T_{\alpha }^{\alpha
}+2V)\partial _{\mu }\phi \partial _{\nu }\phi +\frac{1}{2}g_{\mu \nu
}V(\phi )-\kappa \,T_{\mu \nu }=0,  \label{FE11}
\end{equation}%
\begin{equation}
\frac{\epsilon }{\sqrt{-g}}\partial ^{\mu }\left( \sqrt{-g}\left[ G_{\alpha
}^{\alpha }-\kappa \,T_{\alpha }^{\alpha }+2V\right] \partial _{\mu }\phi
\right) +\frac{1}{2}\frac{dV(\phi )}{d\phi }=0.  \label{FE22}
\end{equation}%

In order to find the equation of hydrostatic equilibrium for compact stars,
we have to regard a suitable line element. Here, we assume a spherical
symmetric spacetime with the following ansatz
\begin{equation}
ds^{2}=c^{2}f(r)dt^{2}-\dfrac{dr^{2}}{g(r)}-r^{2}\left( d\theta ^{2}+\sin
^{2}\theta d\varphi ^{2}\right) ,  \label{metric}
\end{equation}%
where we should analyze the properties of the radial metric functions $f(r)$
and $g(r)$. Using the metric introduced in Eq. (\ref{metric}), we can obtain
the components of energy-momentum tensor (\ref{Tab}) as%
\begin{equation}
T_{0}^{0}=c^{2}\rho \quad \&\quad T_{1}^{1}=T_{2}^{2}=T_{3}^{3}=-P.
\label{Tab2}
\end{equation}%

Taking into account Eq. (\ref{epsilonEq}) with the metric ansatz
(\ref{metric}), the mimetic scalar field can be given by the
spatial metric function with the following form
\begin{equation}
\phi (r)=\int \left( \frac{-\epsilon }{g(r)}\right) ^{\frac{1}{2}}dr.
\label{Phi}
\end{equation}%

Inserting the metric ansatz (\ref{metric}) with Eqs. (\ref{Tab2}) and (\ref%
{Phi}) into the gravitating field equation, we can simplify the components
of Eq. (\ref{FE11})
\begin{eqnarray}
eq_{1} &=&-\kappa c^{2}\rho r^{2}+\dfrac{1}{2}V(\phi )r^{2}-rg^{\prime
}-g+1=0,  \label{eq1} \\
eq_{2} &=&\left[ r^{2}g^{2}\left( 2ff^{\prime \prime }-f^{\prime 2}\right)
+r^{2}gff^{\prime }\left( g^{\prime }+\frac{4g}{r}\right) +2f^{2}g\left(
\kappa r^{2}\left\{ c^{2}\rho -3P-\frac{2V}{\kappa }\right\}
+[2r(g-1)]^{\prime }\right) \right] \phi ^{\prime \,2}+  \notag \\
&&2\epsilon rfgf^{\prime }+\epsilon f^{2}\left( -2\kappa
r^{2}P-r^{2}V+2g-2\right) =0,  \label{eq2} \\
eq_{3} &=&rfgf^{\prime \prime }-\frac{rg}{2}f^{\prime 2}+\frac{ff^{\prime }}{%
2}\left( rg^{\prime }+2g\right) -f^{2}\left( 2\kappa rP+rV-g^{\prime
}\right) =0,  \label{eq3}
\end{eqnarray}%
where $f$, $g$, $\rho $ and $P$ are functions of $r$ and also the prime and
double prime are, respectively, the first and second derivatives with
respect to $r$. Taking Eq. (\ref{Phi}) into account, we can remove $\phi
^{\prime }$ in Eq. (\ref{eq2}) as follows
\begin{equation}
eq_{2}=2r^{2}fgf^{\prime \prime }-r^{2}gf^{\prime 2}+rff^{\prime }\left(
rg^{\prime }+2g\right) +f^{2}\left( 2\kappa c^{2}r^{2}\rho -4\kappa
r^{2}P-3r^{2}V+4rg^{\prime }+2g-2\right) .  \label{eq22}
\end{equation}%

Considering Eq. (\ref{eq1}) as a differential equation, we can obtain $g(r)$
as a function of from $\rho $ and $V(\phi )$
\begin{equation}
g(r)=1-\frac{1}{r}\kappa c^{2}\int \rho \,r^{2}dr+\frac{1}{2\,r}\int
r^{2}V(\phi )dr.  \label{g(r)}
\end{equation}%

Now, we derive, algebraically, $g^{\prime }(r)$ from Eq. (\ref{eq1}) and
regard $g(r)$ from Eq. (\ref{g(r)}), and insert them into Eq. (\ref{eq22}).
After some manipulations, we can find that Eq. (\ref{eq22}) can be rewritten
as
\begin{eqnarray}
eq_{2} &=&\left( \frac{r^{2}f^{\prime \,2}-rff^{\prime }}{2}%
-r^{2}\,ff^{\prime \prime }+f^{2}\right) \int r^{2}V(\phi )dr+r^{3}\left(
f^{\prime \,2}-2ff^{\prime \prime }\right) \left( 1-\frac{\kappa Mc^{2}}{%
4\pi r}\right) +  \notag \\
&&rf\,f^{\prime }\left( c^{2}\kappa r^{3}\rho +\frac{\kappa Mc^{2}}{4\pi }-%
\frac{1}{2}r^{3}V(\phi )-2r\right) +r^{3}f^{2}\left( V(\phi )+2\kappa \,%
\left[ \rho c^{2}+2P-\frac{Mc^{2}}{4\pi r^{3}}\right] \right) =0,
\label{eq222}
\end{eqnarray}%
where $M=\int {4\pi r^{2}\rho \,dr}$. Here, we can define a new variable $w=%
\frac{df}{dr}$ to change Eq. (\ref{eq222}) into the following first-order
differential equation
\begin{eqnarray}
eq_{2} &=&\left( \frac{r^{2}w^{2}-rfw}{2}-r^{2}\,fw^{\prime }+f^{2}\right)
\int r^{2}V(\phi )dr+r^{3}\left( w^{\,2}-2fw^{\prime }\right) \left( 1-\frac{%
\kappa Mc^{2}}{4\pi r}\right) +  \notag \\
&&rf\,w\left( c^{2}\kappa r^{3}\rho +\frac{\kappa Mc^{2}}{4\pi }-\frac{1}{2}%
r^{3}V(\phi )-2r\right) +r^{3}f^{2}\left( V(\phi )+2\kappa \,\left[ \rho
c^{2}+2P-\frac{Mc^{2}}{4\pi r^{3}}\right] \right) =0.  \label{last2}
\end{eqnarray}

Now, we are in a position to talk about the equation of
hydrostatic equilibrium by use of the conservation of
energy-momentum tensor, $\nabla _{\mu }T^{\mu \nu }=0$, in which
for $\nu =r$, we can write
\begin{equation}
\frac{dP}{dr}=-\left( P+\rho c^{2}\right) \frac{f^{\prime }}{2\,f}.
\label{dpdr}
\end{equation}

Now, we have Eqs. (\ref{Phi}), (\ref{eq1}), (\ref{last2}),
(\ref{dpdr}) and we should solve them numerically for a given EoS
($P(\rho)$).

\section{EoS, neutron star structure and observational data}

There is a theoretical challenge to applying a proper EoS
for the description of the internal structure of neutron stars.
Here, we consider different known models of EoS in the context of
GR and mimetic gravity to obtain an appropriate conclusion. So
far, a large number of the EoS with various motivations have been
introduced, and in this paper, we examine eight
relativistic/non-relativistic EoSs that seem more realistic and
have already received more attention.

The first non-relativistic EoS is SLy4 has been derived
from energy density functional methods with an effective
nucleon-nucleon interaction \cite{douchin2001unified}. The second
model of non-relativistic EoSs refers to BSK21 that is constructed
from the generalized Skyrme nuclear interaction arising from the
Argonne $V18$ potential plus a microscopic nucleonic three-body
force \cite{fantina2012unified}. The third non-relativistic model
is defined according to a variational method known as WFF1
\cite{wiringa1988equation}.

Now, we point out the relativistic category of EoSs that
we considered in this paper. The first model is devoted to MPA1
arising from the relativistic Dirac-Brueckner-Hartree-Fock
approach with the $\pi$-$\rho$ mesons exchange
\cite{muther1987nuclear}. We also choose the HS (DD2) EoS
\cite{MHempel2010,Fischer2014} as the second model of this
category that is based on the density-dependent relativistic
mean-field (RMF) interactions of Typel et al.
\cite{Typeletal2010}. The third model is known as FSU2H EoS. The
nucleonic FSU2 model of \cite{Chen2014} is modified in the context
of relativistic mean-field theory to describe both the nucleon and
hyperon interactions, and so FSU2H EoS is created
\cite{Tolos2017,Negreiros2018,Providencia2019}.

We also use two hadronic EoSs, $BHB\Lambda\Phi$
\cite{SBanik2014} and SFHoY
\cite{MFortin2018,MHempel2010,AWSteiner}, as other realistic
models. The $BHB\Lambda\Phi$ EoS additionally includes $\Lambda$
hyperons and hyperon-hyperon interactions mediated by $\Phi$
mesons. The $\Lambda$ hyperon makes the EoS stiffer
resulting in $2.1M_{\odot}$ maximum mass neutron star
corresponding to a radius $11.58$
$km$. Also, SFHoY is based on the statistical model with an
excluded volume and interactions of Hempel and
Schaffner-Bielich(HS) \cite{MHempel2010} with RMF interactions
SFHo \cite{AWSteiner}. Although these EoSs are obtained with RMF
interactions, we may exclude them from the full relativistic
category since they are formulated in densities below saturation
density and low temperatures.

Before proceeding, it is notable that we examine the
mentioned models of EoSs with different accepted functions of
scalar potentials $V(\phi)$.

It is worth mentioning that in this paper, for nucleonic EoSs, we use
from the parameterized piecewise-polytrope representation applied in \cite%
{read2009constraints,raaijmakers2018pitfall,kumar2019inferring}, which we
now describe in brief.

A piecewise polytropic EoS with four segment pieces satisfy the following
polytropic relation,
\begin{equation}
P(\rho)=\left \{
\begin{array}{ll}
K_0 \rho^{\Gamma_0} \quad \rho \leq \rho_0 &  \\
K_1 \rho^{\Gamma_1} \quad \rho_0 \leq \rho \leq \rho_1 &  \\
K_2 \rho^{\Gamma_2} \quad \rho_1 \leq\rho \leq \rho_2 &  \\
K_3 \rho^{\Gamma_3} \quad \rho > \rho_2 &
\end{array}
\right.
\end{equation}
where $P$ is the fluid pressure, and $K_i$ and $\Gamma_i$ denote,
respectively, the polytropic constant and the adiabatic index. The first
polytrope piece shows the crust EoS that is determined by $K_0= 35939 \times
10^{13} [\mathrm{cgs}]$ and $\Gamma_0=13572$ \cite{read2009constraints}.
Furthermore, two dividing high densities between core pieces are chosen as $%
\rho_1=10^{14.7} g/cm^{3}$ and $\rho_2=10^{15} g/cm^{3}$. In this method,
for a chosen of four free parameters $P_1$, $\Gamma_1$, $\Gamma_2$, $%
\Gamma_3 $ (where $P_1=P(\rho_1) $ and $\Gamma_1$, $\Gamma_2$ and $\Gamma_3$
are the adiabatic indices) and by considering the continuity of the
pressure, we can fix the polytropic constants $K_1$, $K_2$, and $K_3$ in the
following form
\begin{equation}
P(\rho_i)=K_i \rho^{\Gamma_i}=K_{i+1}\rho^{\Gamma_{i+1}}.  \label{Prho}
\end{equation}

Considering the mentioned points and following Ref. \cite%
{read2009constraints}, the parameterized EoS is completely determined (For
the details see Ref. \cite{read2009constraints}).

\begin{figure}[tbp]
\centering {\includegraphics[width=0.5\textwidth]{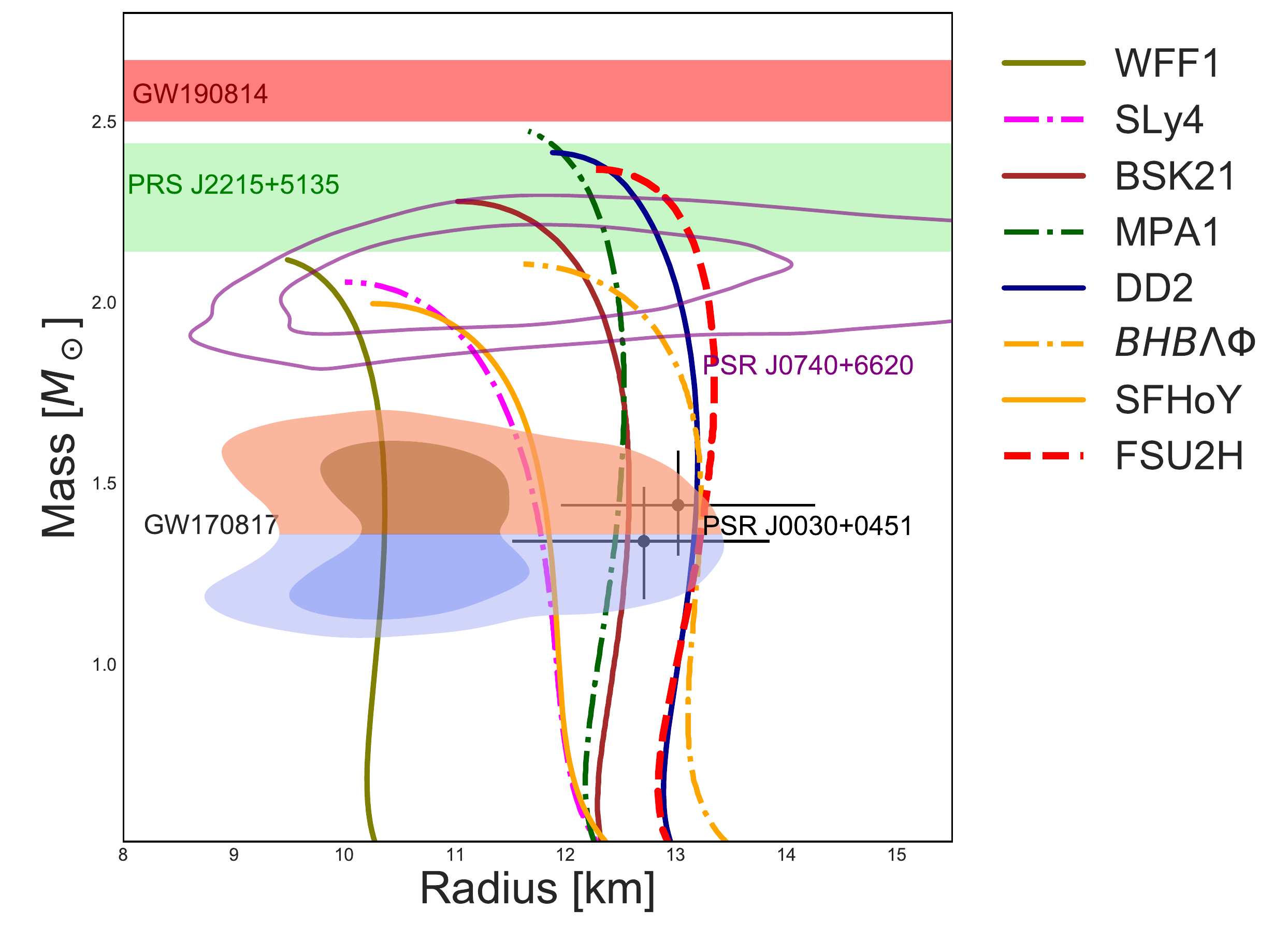}}
\caption{The mass-radius diagram of neutron stars in GR for the
different EoSs. The blue and orange regions are the mass-radius
constraints from the GW170817 event. The purple region represents
the pulsar J0740+6620 and the light green region amounts to the
pulsar J2215+5135 and also the black dots with error bars, are the
NICER estimations of PSR J0030+0451. The mass of the compact
object observed by the GW190414 event is shown as a red band.}
\label{sly4_fig0}
\end{figure}

Moreover, taking the GR limit into account, one can find that all
of the chosen EoSs are in the range of the gravitational wave
event GW170817 \cite{abbott2018gw170817} and they produce a
neutron star with maximum mass $M_{max}> 2m_{\odot}$. However,
these EoSs did not support some of the observational evidence (see
Fig. \ref{sly4_fig0} and table \ref{table_GR} for more details).

\begin{figure}[!ht]
    \centering
    \subfloat[${V_1(\phi)=\frac{\alpha_1}{\phi^2}}$]{\includegraphics[width=0.4%
        \textwidth]{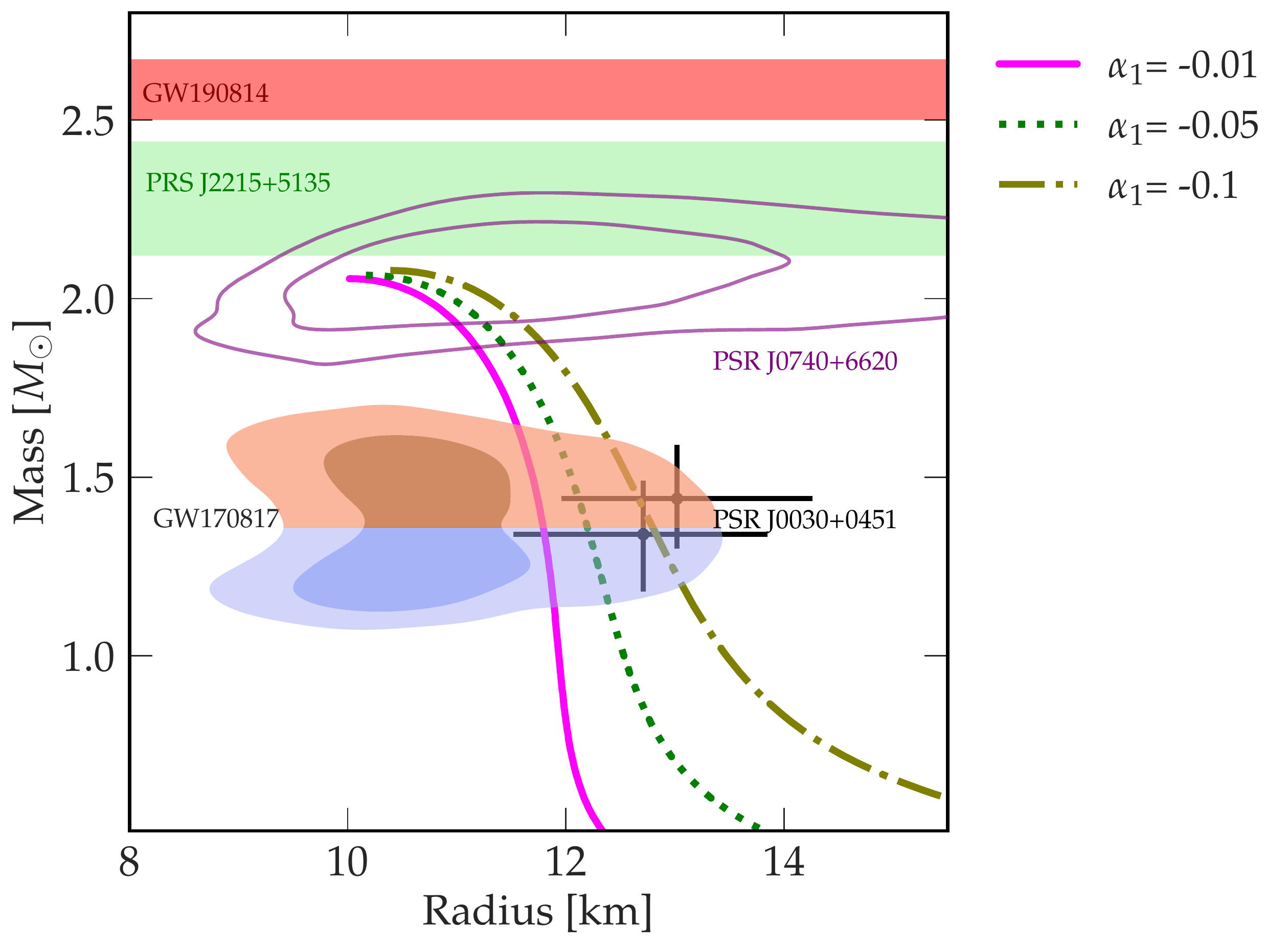}} \quad
    \subfloat[${V_1(\phi)=\frac{\alpha_1}{\phi^2}}$]{
        \includegraphics[width=0.4\textwidth]{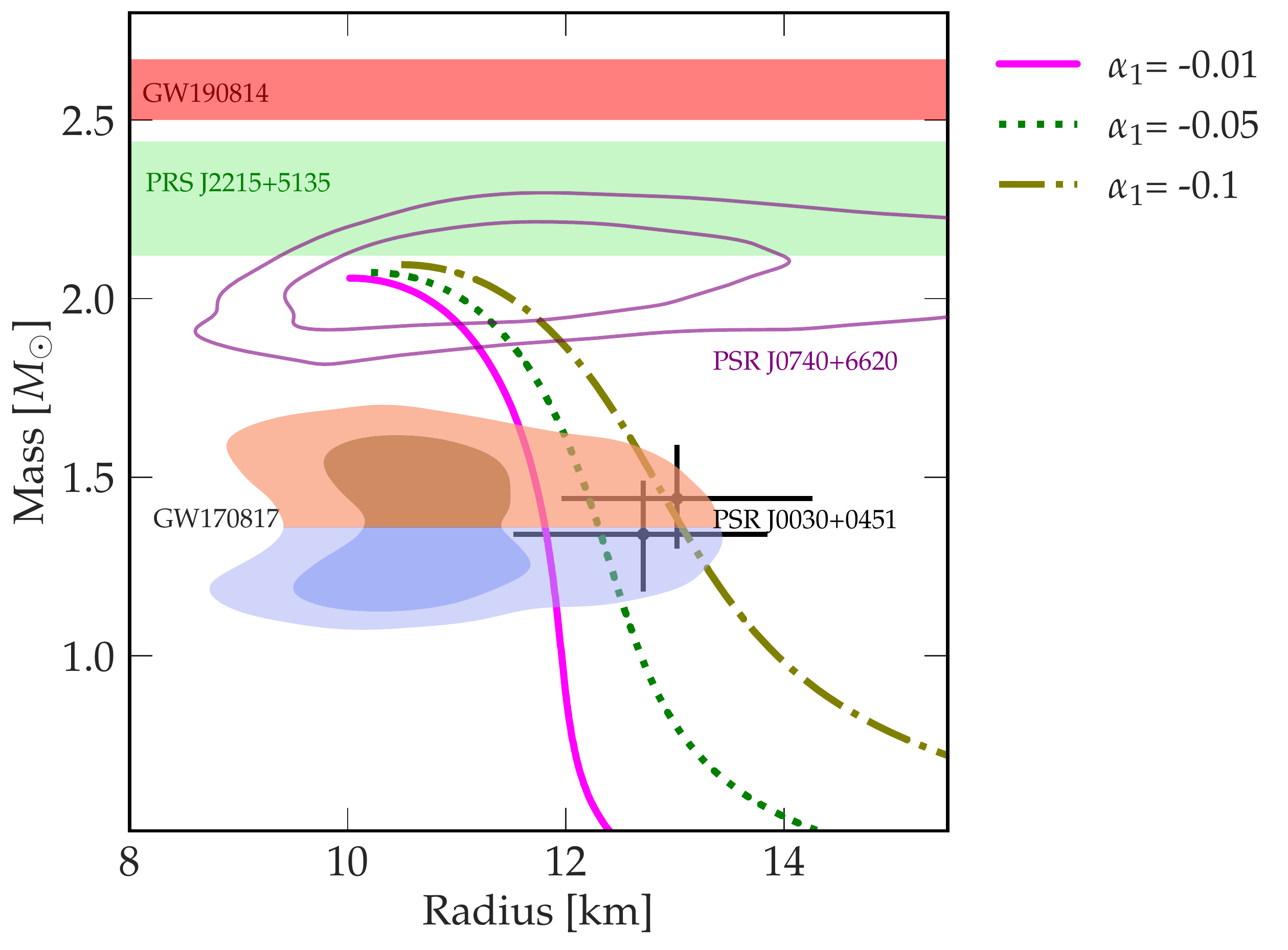}} \newline
    \subfloat[$V_2(\phi)=\frac{\alpha_2}{e^{K\phi^2}}$]{
        \includegraphics[width=0.4\textwidth]{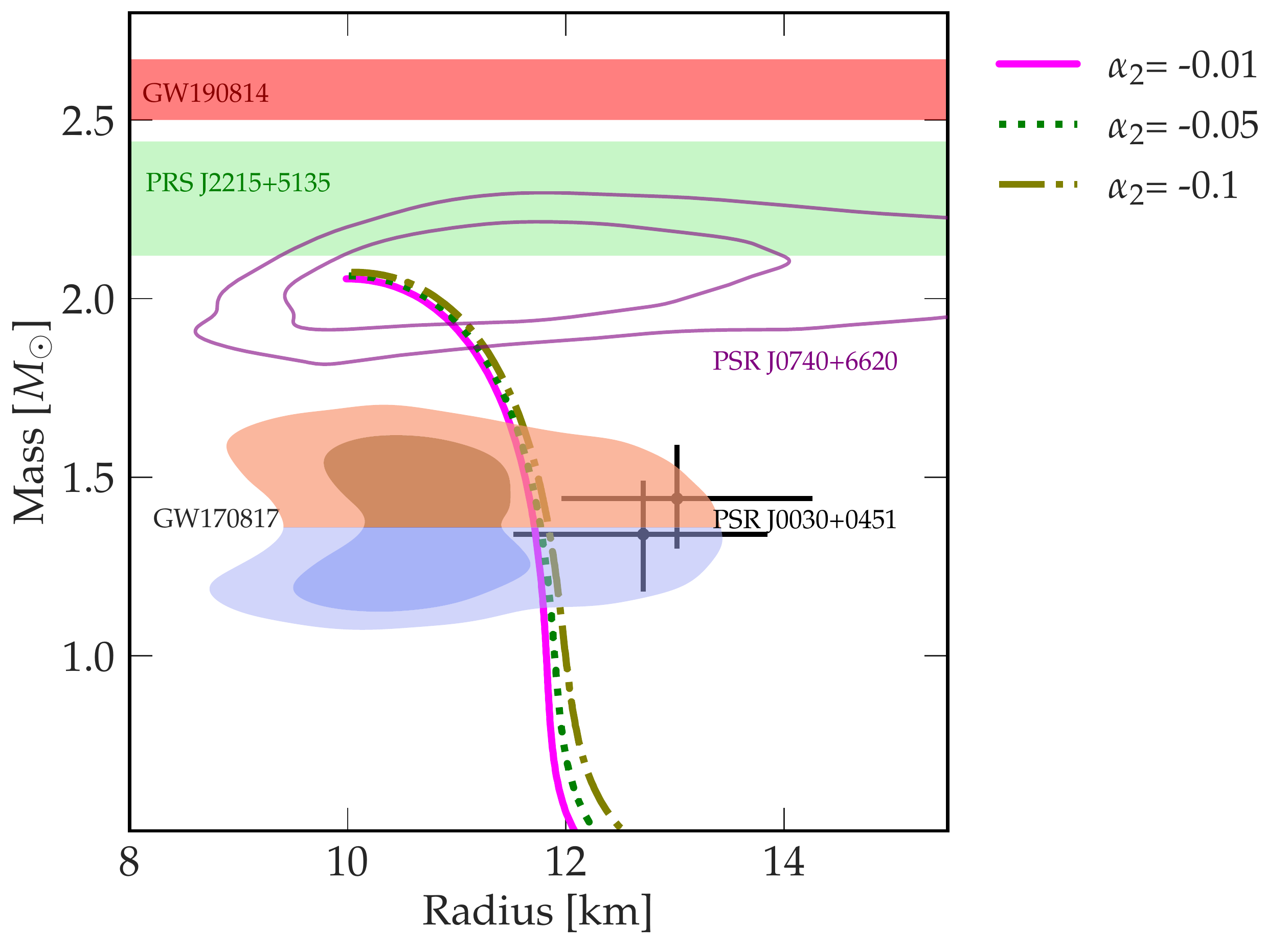} } \quad
    \subfloat[$V_2(\phi)=\frac{\alpha_2}{e^{K\phi^2}}$]{\includegraphics[width=0.4\textwidth]{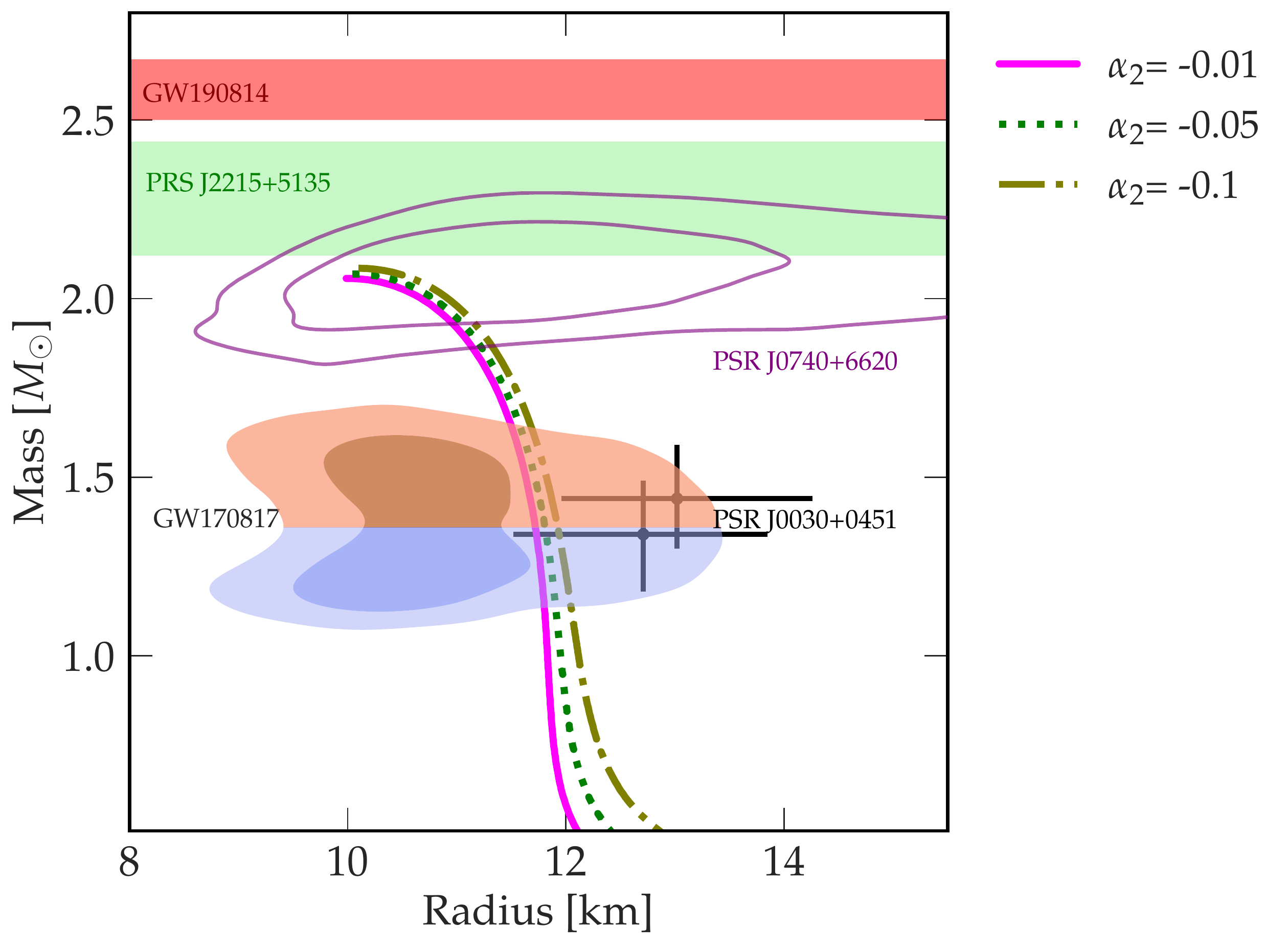} }
    \newline
    \subfloat[${V_3(\phi)=\frac{\alpha_3\phi^2}{1+e^{K\phi^2}}}$]{\includegraphics[width=0.4\textwidth]{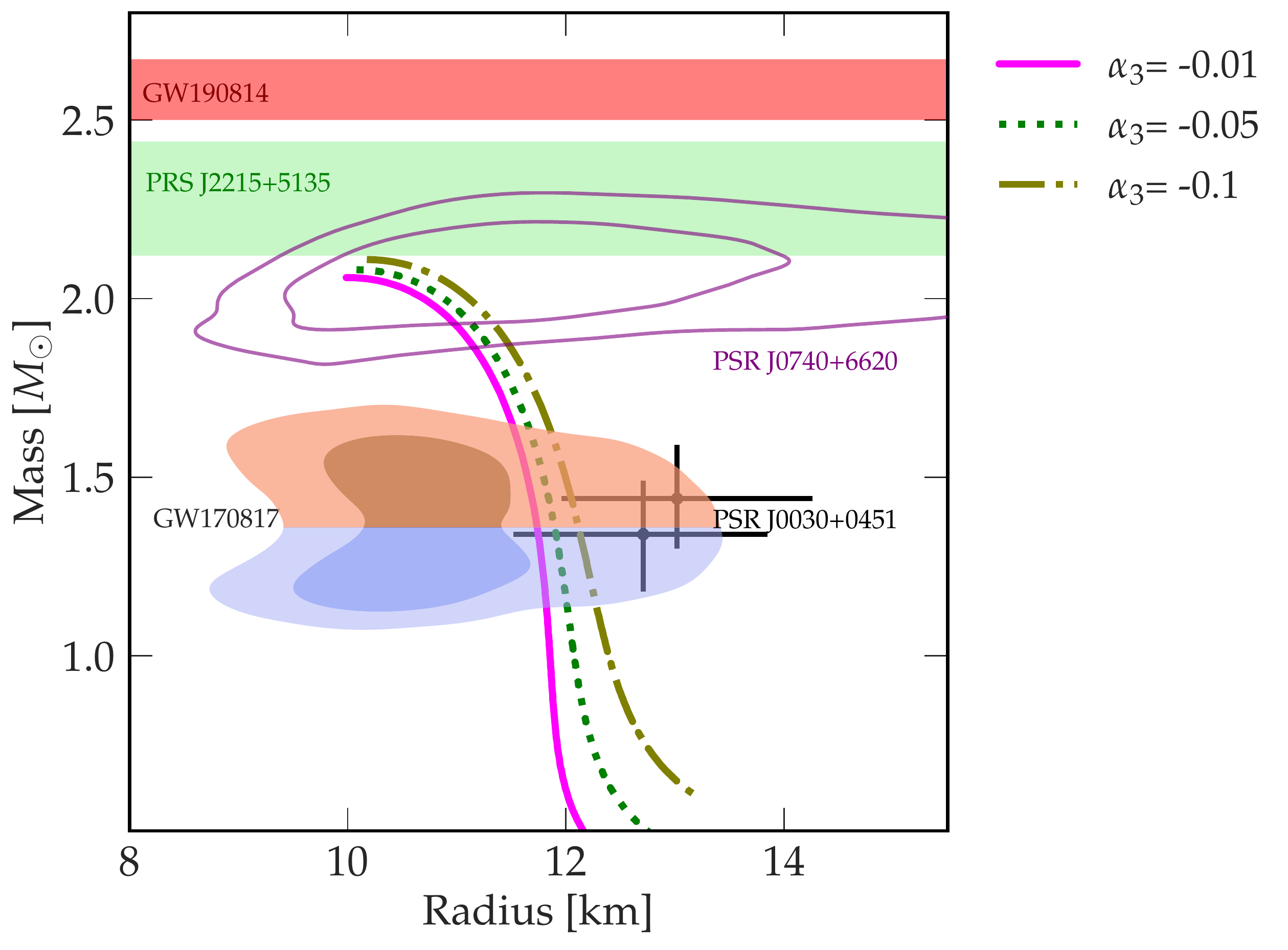}
    } \quad
    \subfloat[${V_3(\phi)=\frac{\alpha_3\phi^2}{1+e^{K\phi^2}}}$]{\includegraphics[width=0.4\textwidth]{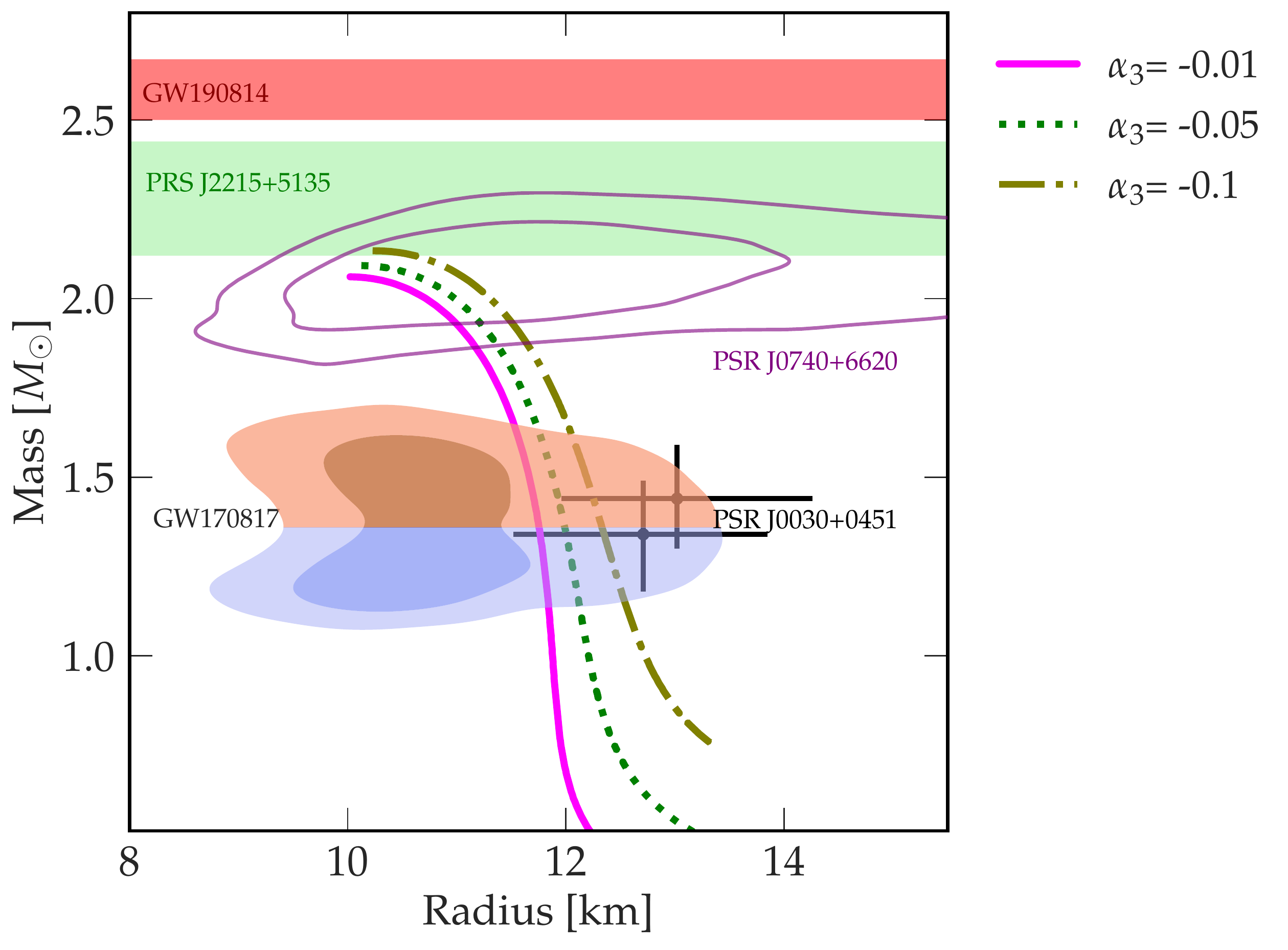}
    } \caption{The mass-radius
        relation of neutron stars in mimetic gravity for SLy4 EoS. $\protect\phi(0)=1$: left panels and $\protect%
        \phi(0)=0.8$: right panels and $K=0.5$.}
    \label{sly4_fig}
\end{figure}

\begin{figure}[!ht]
    \centering
    \subfloat[${V_1(\phi)=\frac{\alpha_1}{\phi^2}}$]{\includegraphics[width=0.4%
        \textwidth]{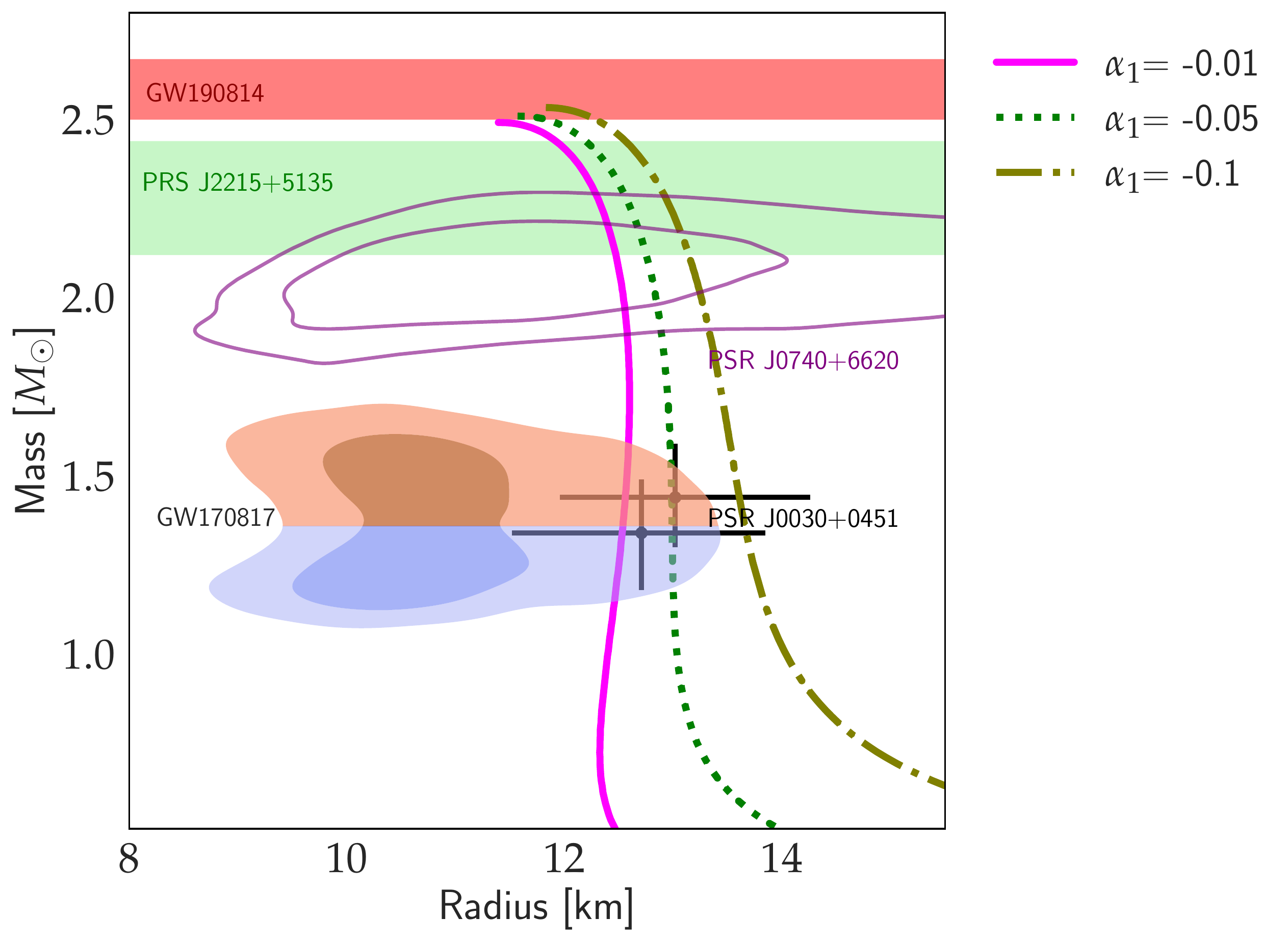}} \quad
    \subfloat[${V_1(\phi)=\frac{\alpha_1}{\phi^2}}$]{
        \includegraphics[width=0.4\textwidth]{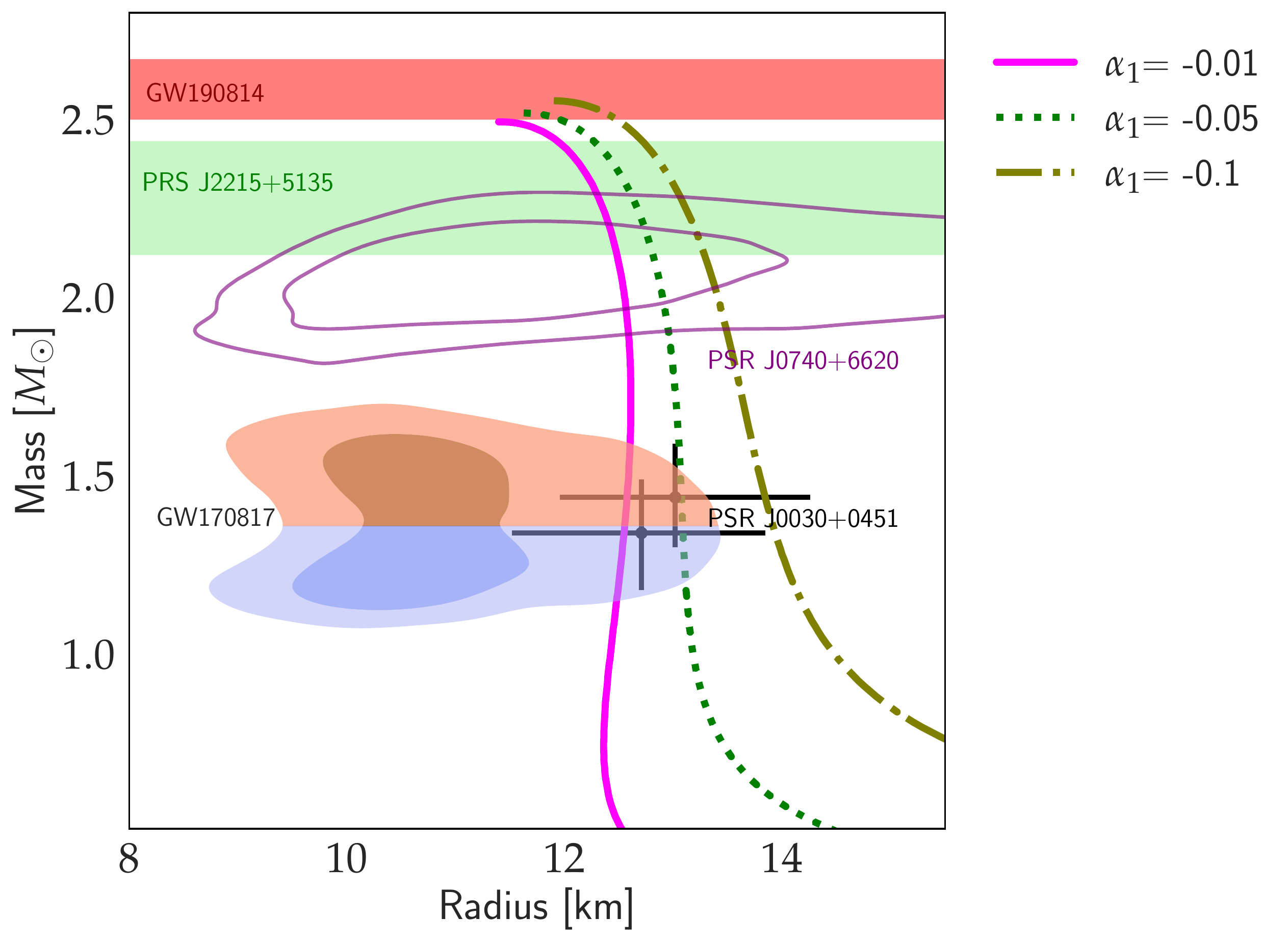}} \newline
    \subfloat[$V_2(\phi)=\frac{\alpha_2}{e^{K\phi^2}}$]{
        \includegraphics[width=0.4\textwidth]{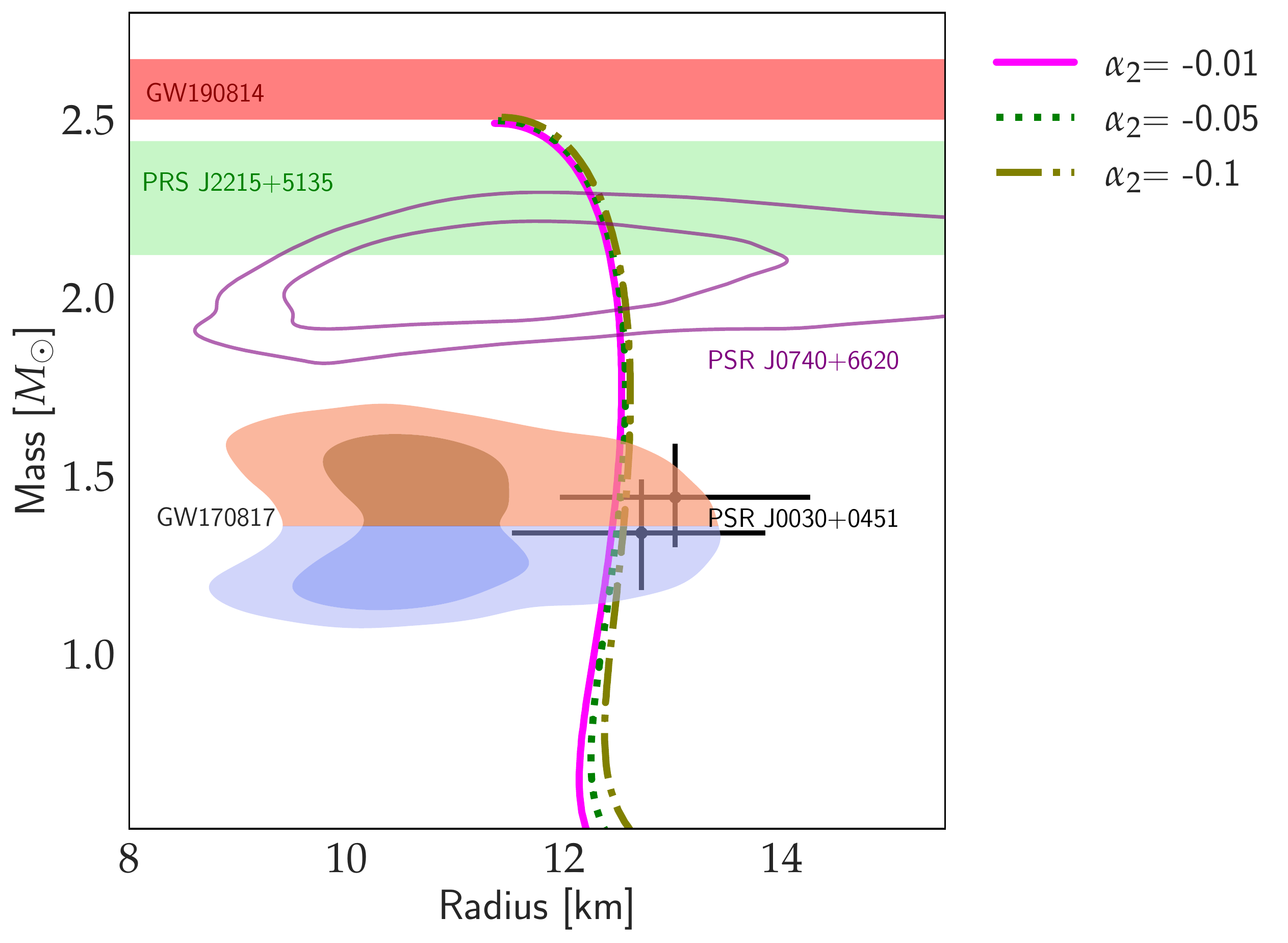} } \quad
    \subfloat[$V_2(\phi)=\frac{\alpha_2}{e^{K\phi^2}}$]{\includegraphics[width=0.4\textwidth]{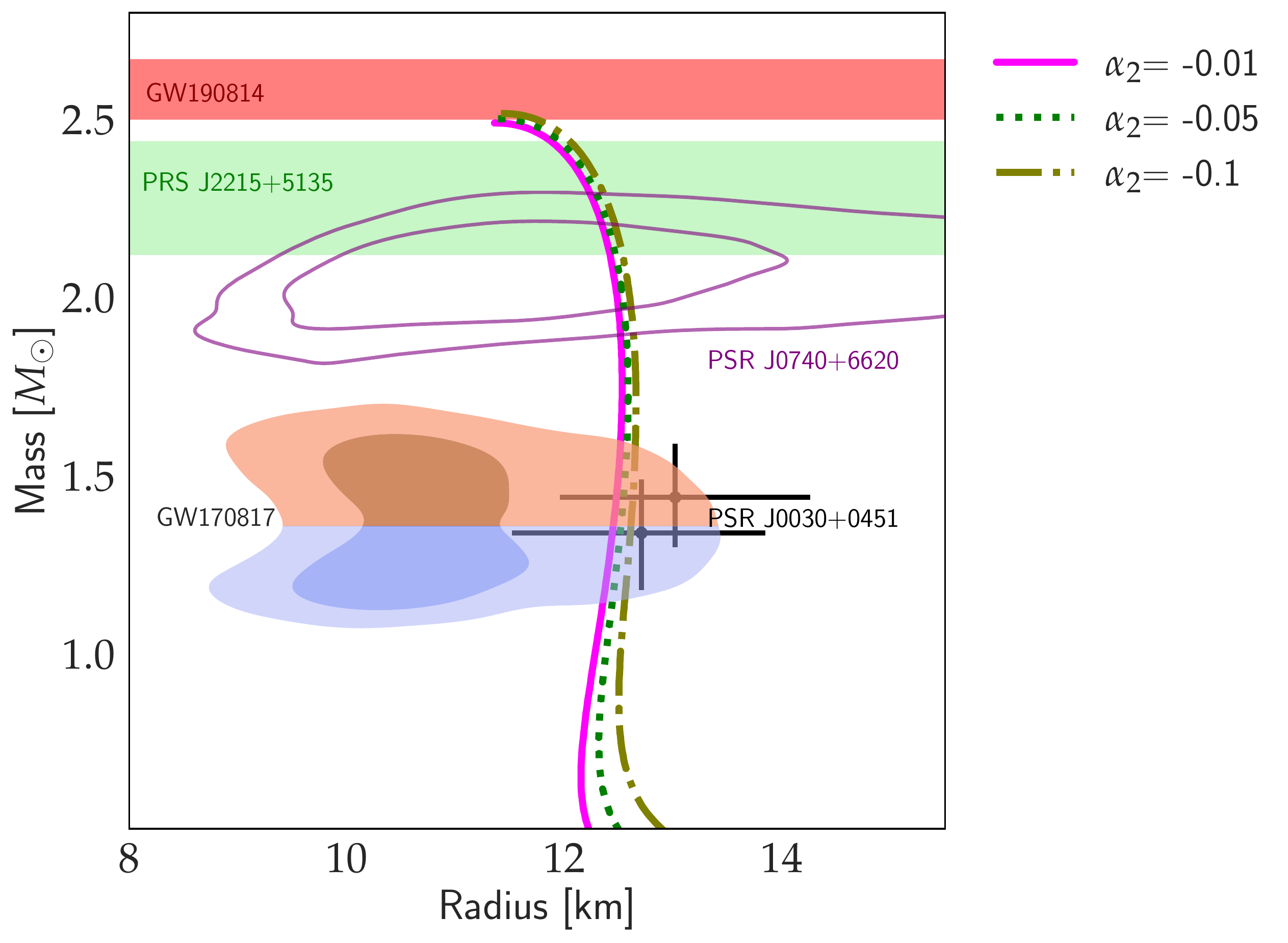} }
    \newline
    \subfloat[${V_3(\phi)=\frac{\alpha_3\phi^2}{1+e^{K\phi^2}}}$]{\includegraphics[width=0.4\textwidth]{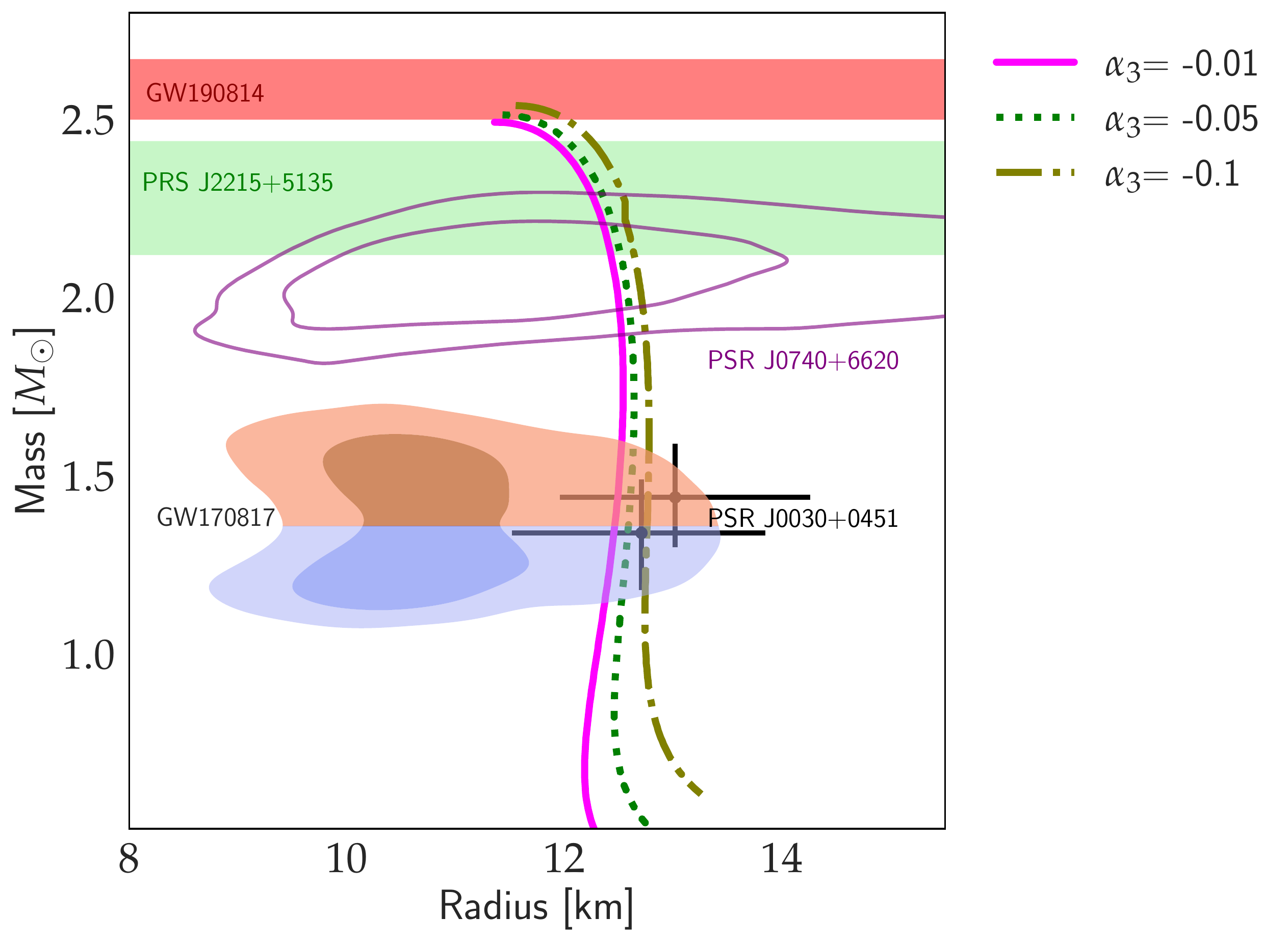}
    } \quad
    \subfloat[${V_3(\phi)=\frac{\alpha_3\phi^2}{1+e^{K\phi^2}}}$]{\includegraphics[width=0.4\textwidth]{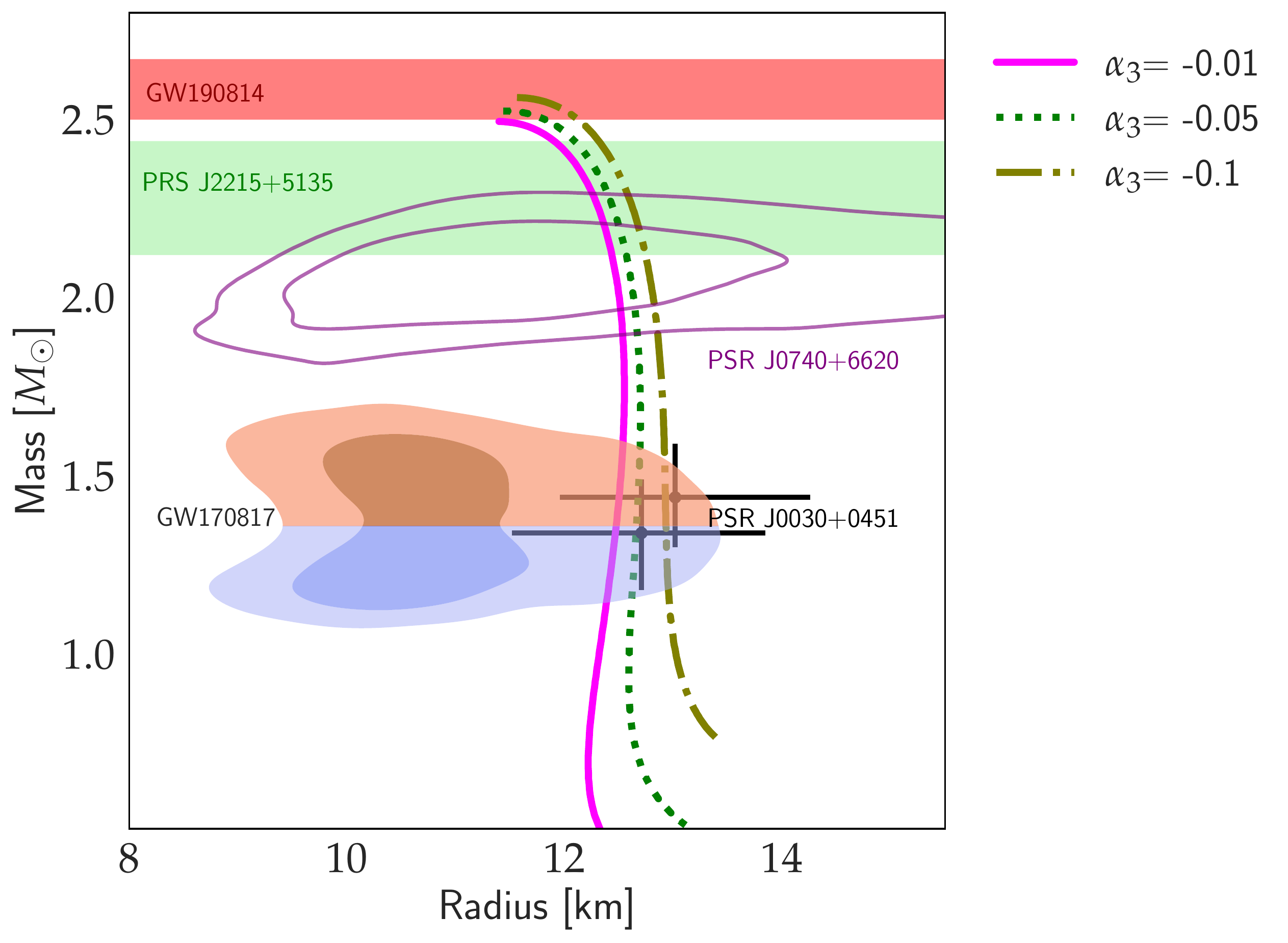}
    } \caption{The mass-radius
        relation of neutron stars in mimetic gravity for MPA1 EoS. $\protect\phi(0)=1$: left panels and $\protect%
        \phi(0)=0.8$: right panels and $K=0.5$.}
    \label{MPA1_fig}
\end{figure}

\begin{figure}
    \centering
    \subfloat[${V_1(\phi)=\frac{\alpha_1}{\phi^2}}$]{\includegraphics[width=0.4%
        \textwidth]{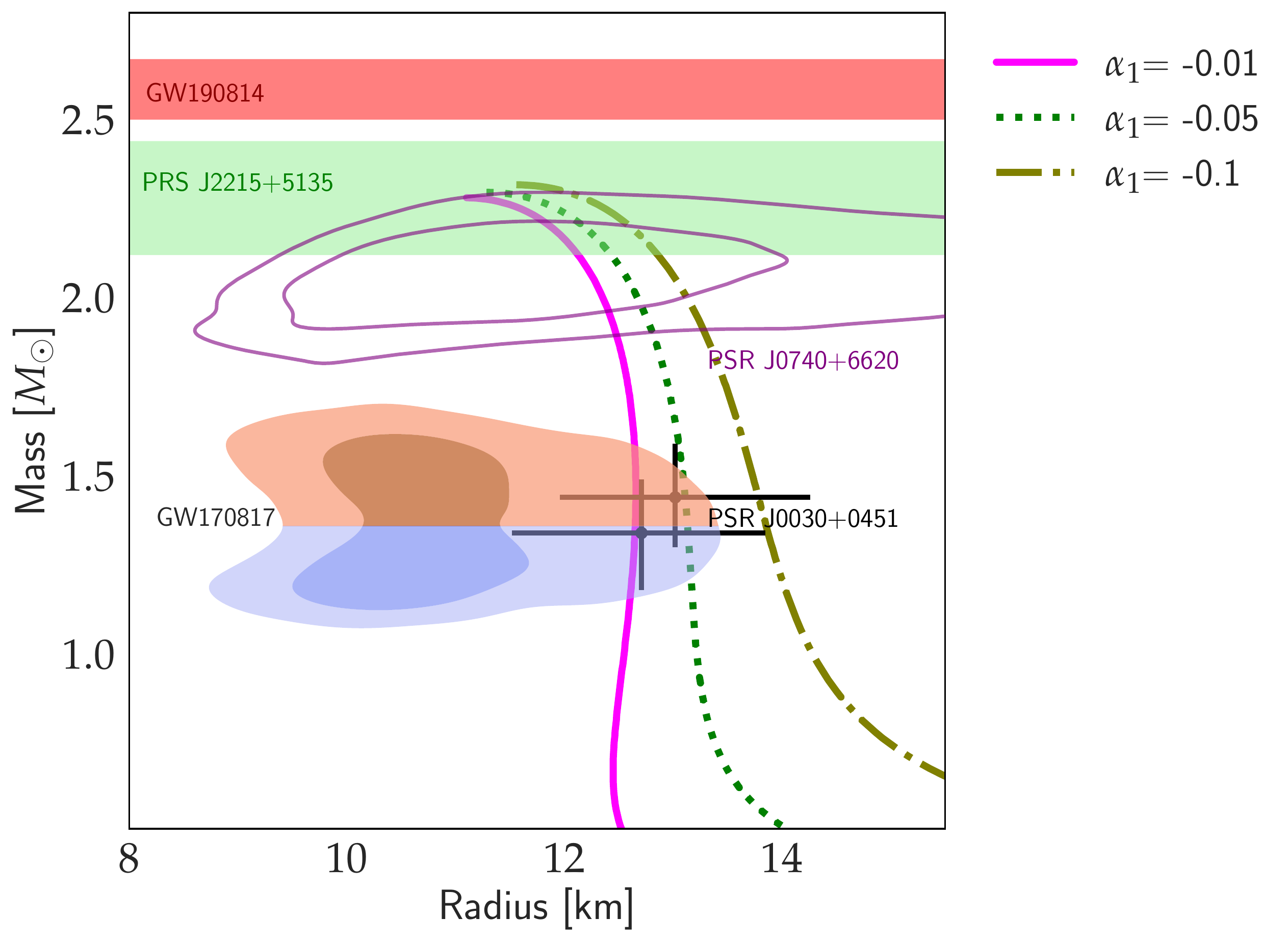}} \quad
    \subfloat[${V_1(\phi)=\frac{\alpha_1}{\phi^2}}$]{
        \includegraphics[width=0.4\textwidth]{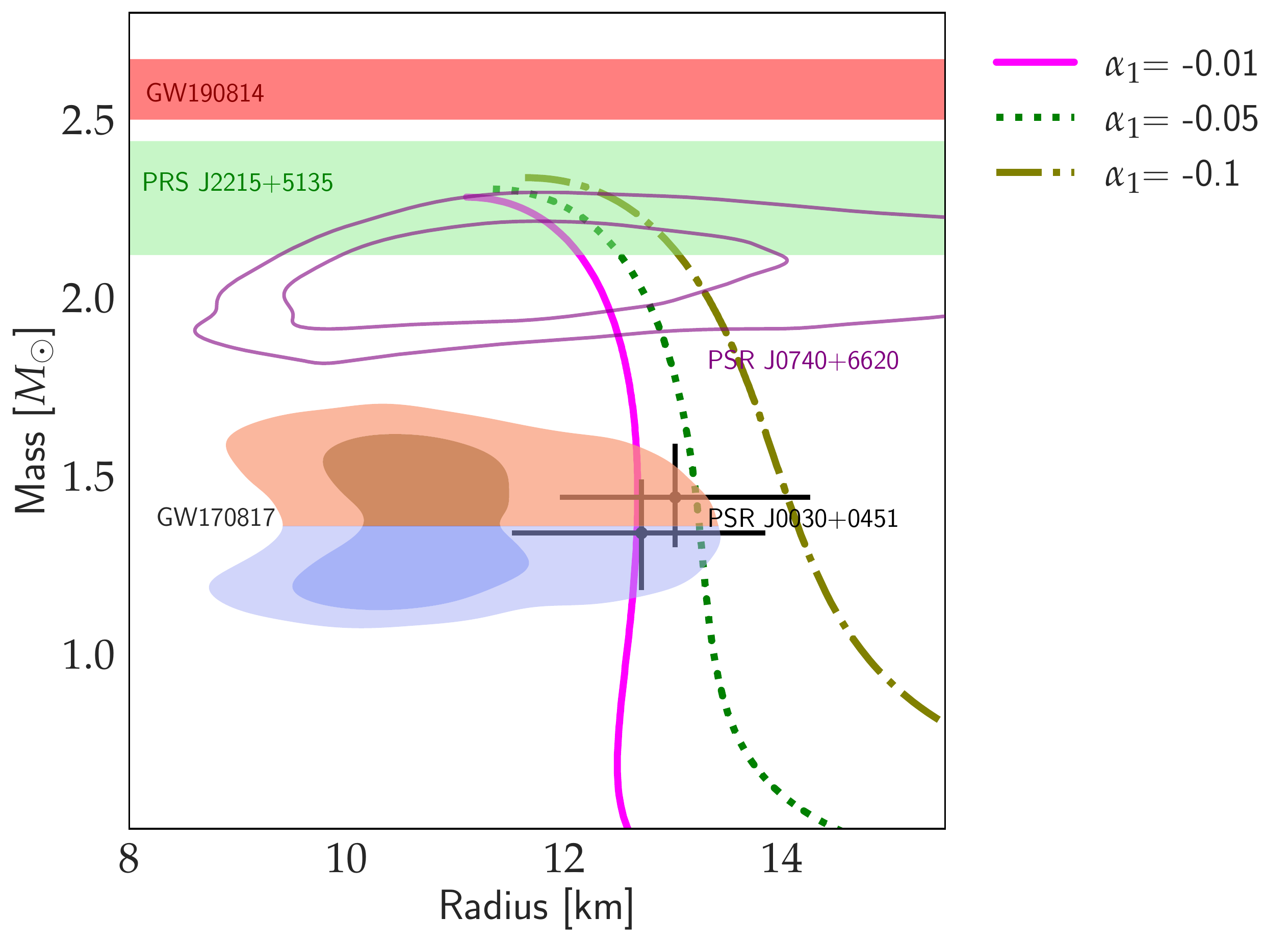}} \newline
    \subfloat[$V_2(\phi)=\frac{\alpha_2}{e^{K\phi^2}}$]{
        \includegraphics[width=0.4\textwidth]{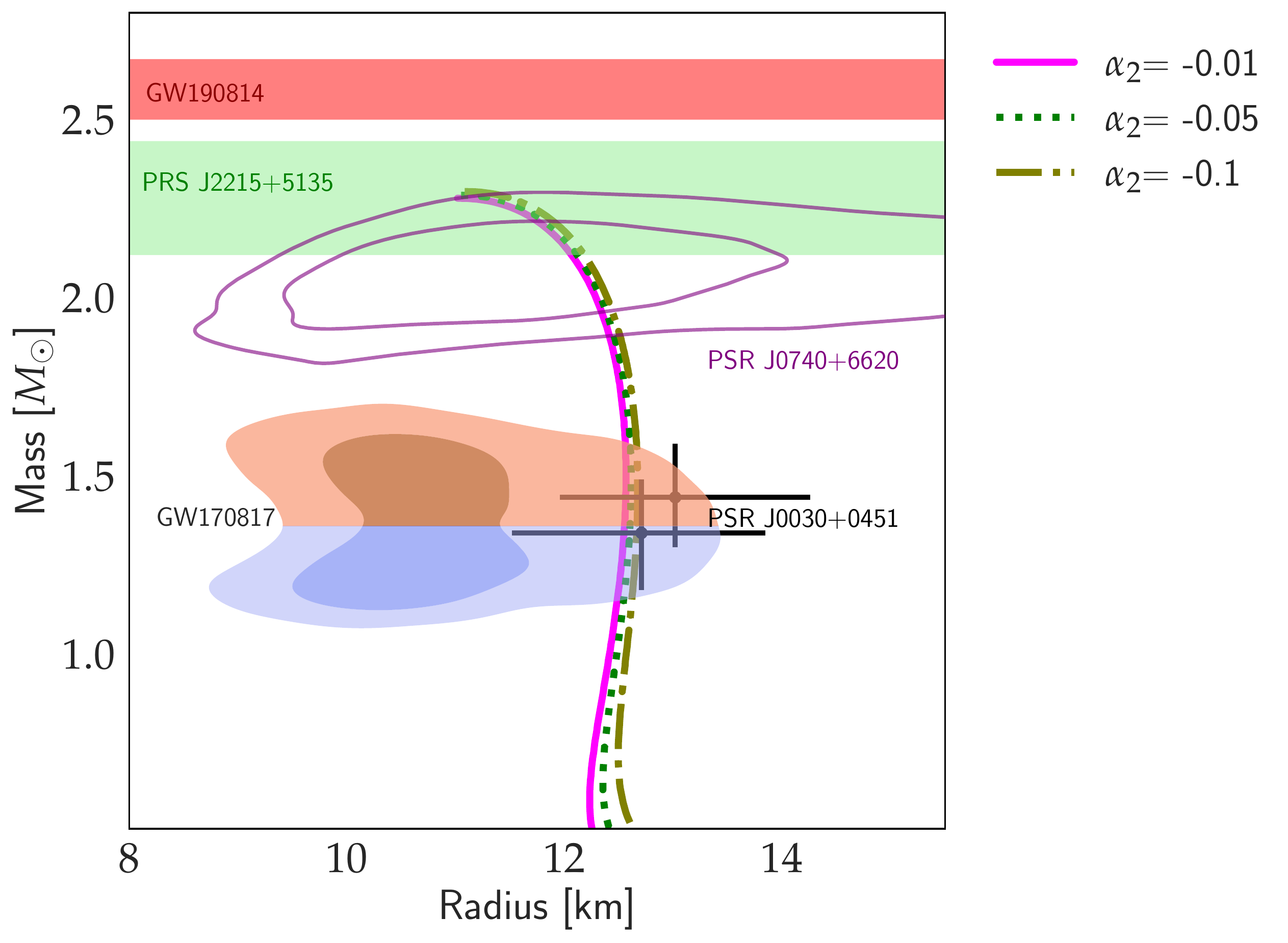} } \quad
    \subfloat[$V_2(\phi)=\frac{\alpha_2}{e^{K\phi^2}}$]{\includegraphics[width=0.4\textwidth]{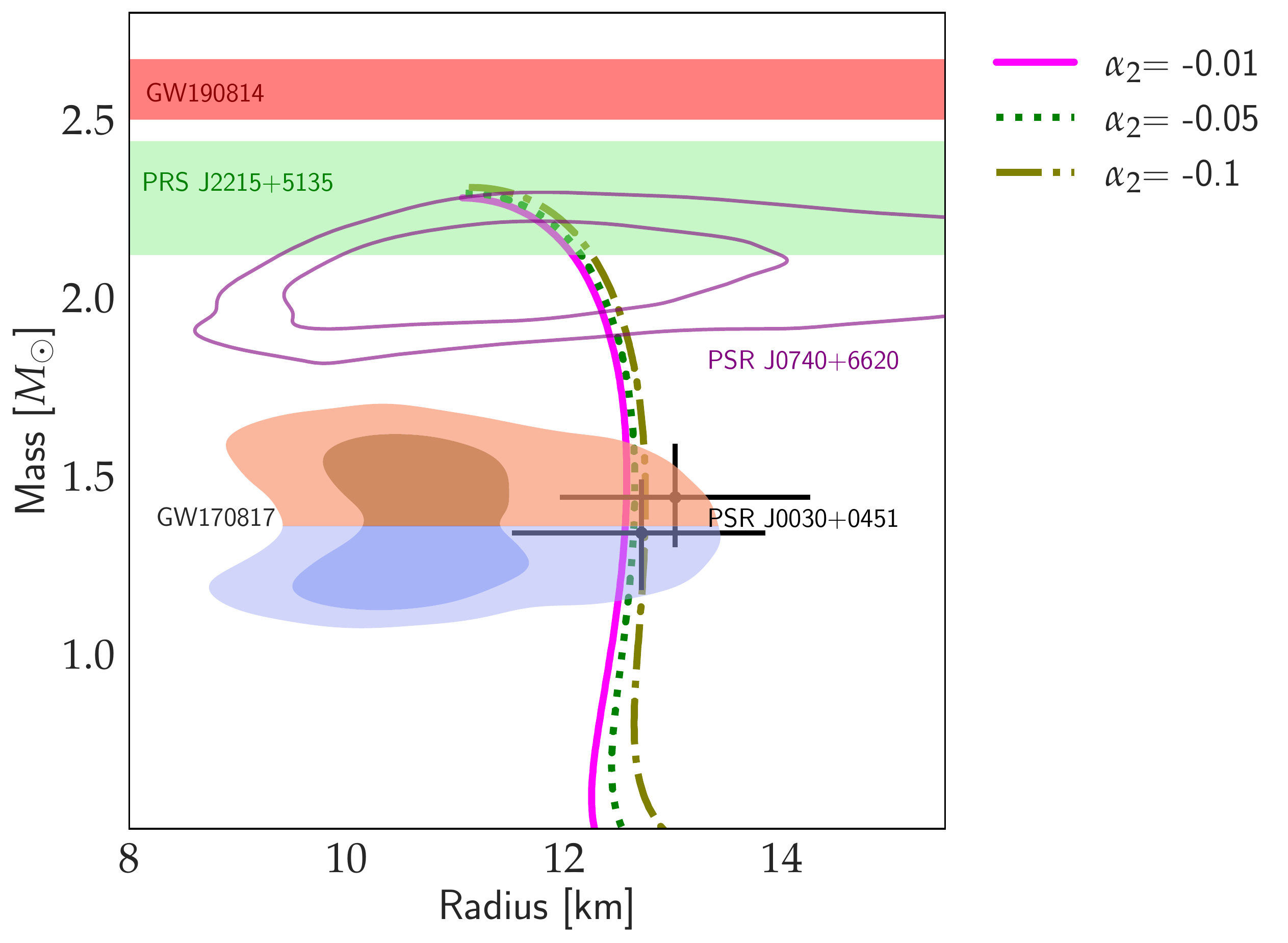} }
    \newline
    \subfloat[${V_3(\phi)=\frac{\alpha_3\phi^2}{1+e^{K\phi^2}}}$]{%
        \includegraphics[width=0.4\textwidth]{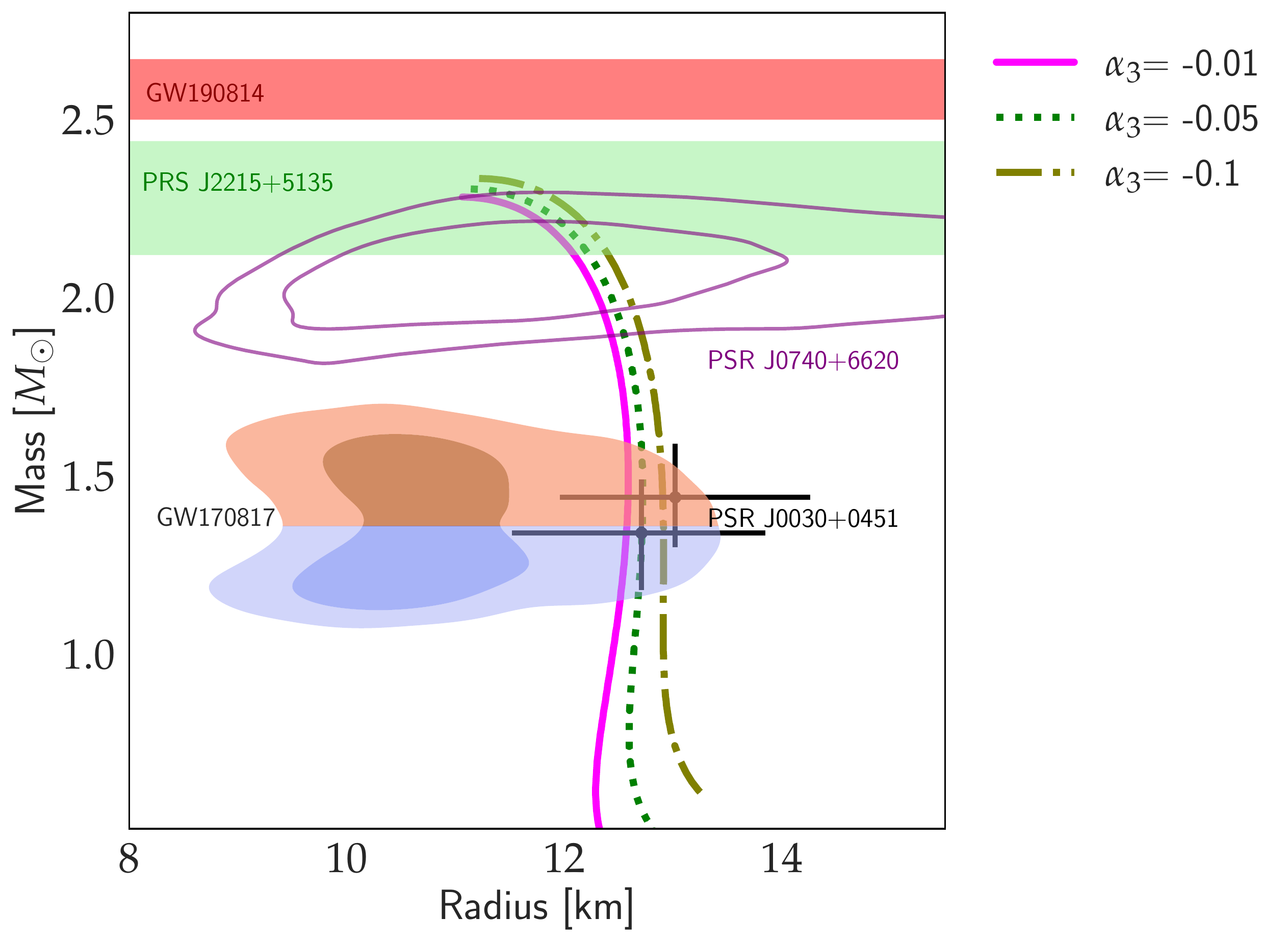} } \quad %
    \subfloat[${V_3(\phi)=\frac{\alpha_3\phi^2}{1+e^{K\phi^2}}}$]{%
        \includegraphics[width=0.4\textwidth]{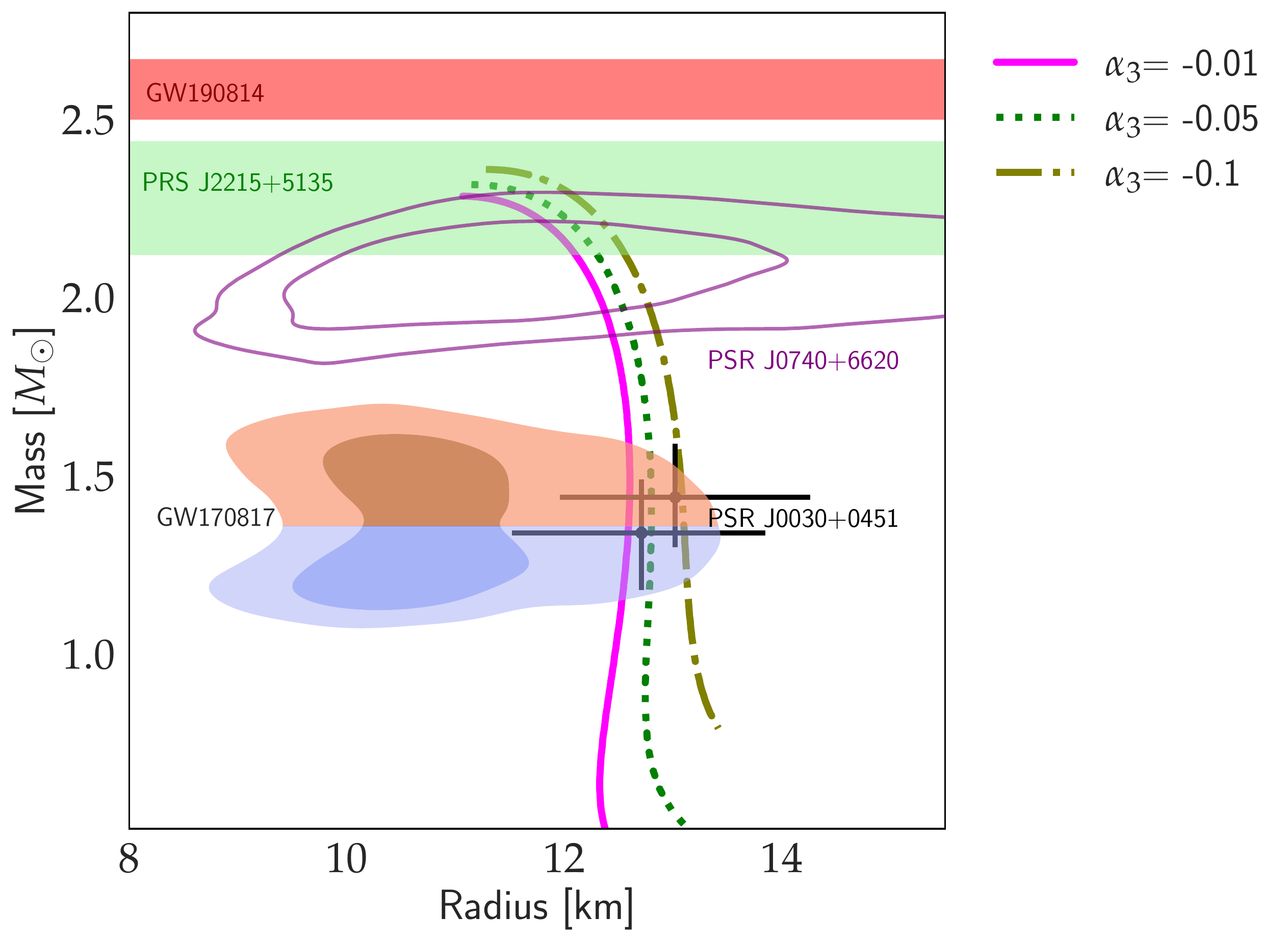} }
    \caption{The mass-radius
        relation of neutron stars in mimetic gravity for BSK21 EoS. $\protect\phi(0)=1$: left panels and $\protect%
        \phi(0)=0.8$: right panels and $K=0.5$.}
    \label{BSK21_fig}
\end{figure}

\begin{figure}[!ht]
    \centering
    \subfloat[${V_1(\phi)=\frac{\alpha_1}{\phi^2}}$]{\includegraphics[width=0.4%
        \textwidth]{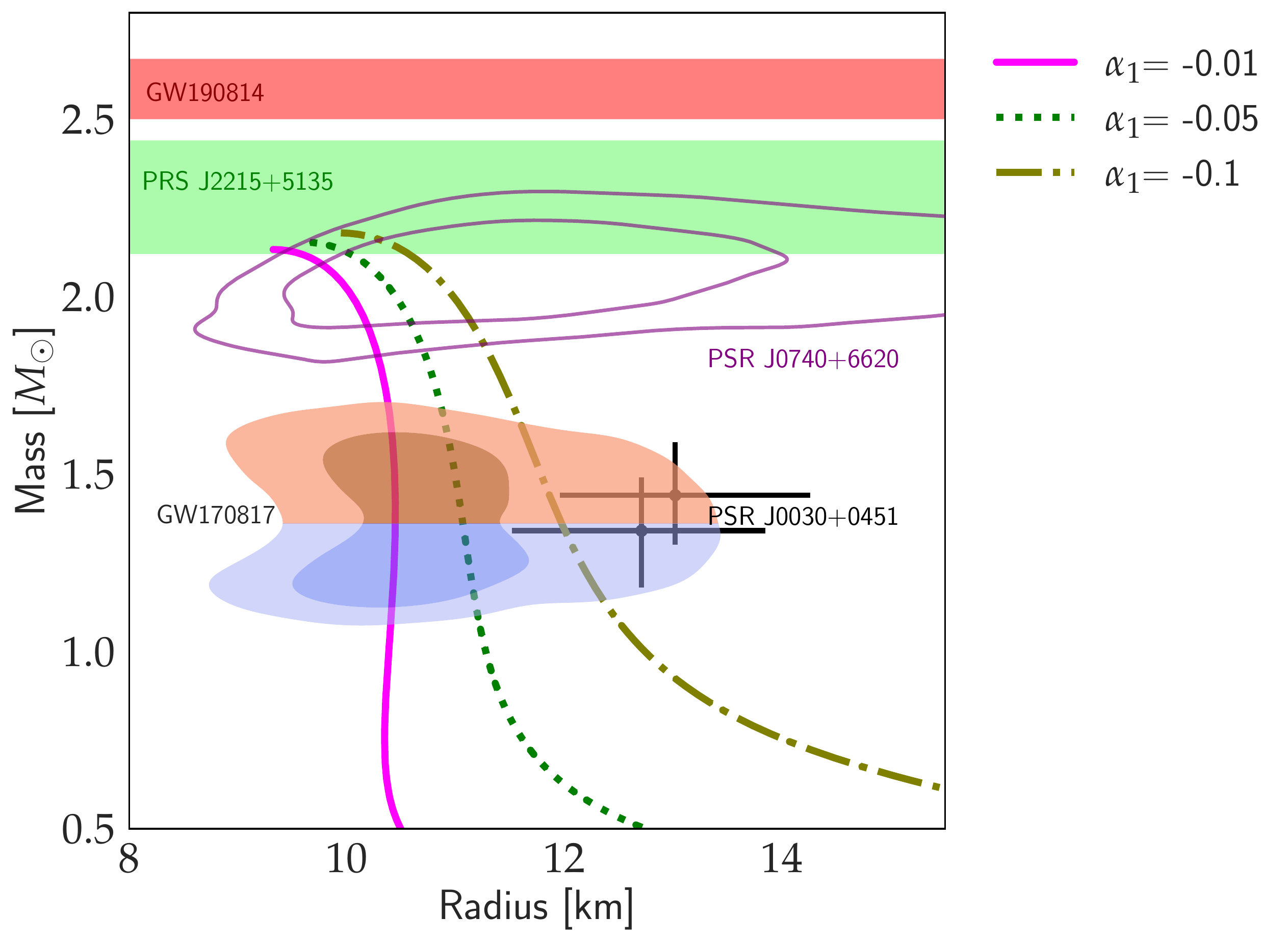}} \quad
    \subfloat[${V_1(\phi)=\frac{\alpha_1}{\phi^2}}$]{
        \includegraphics[width=0.4\textwidth]{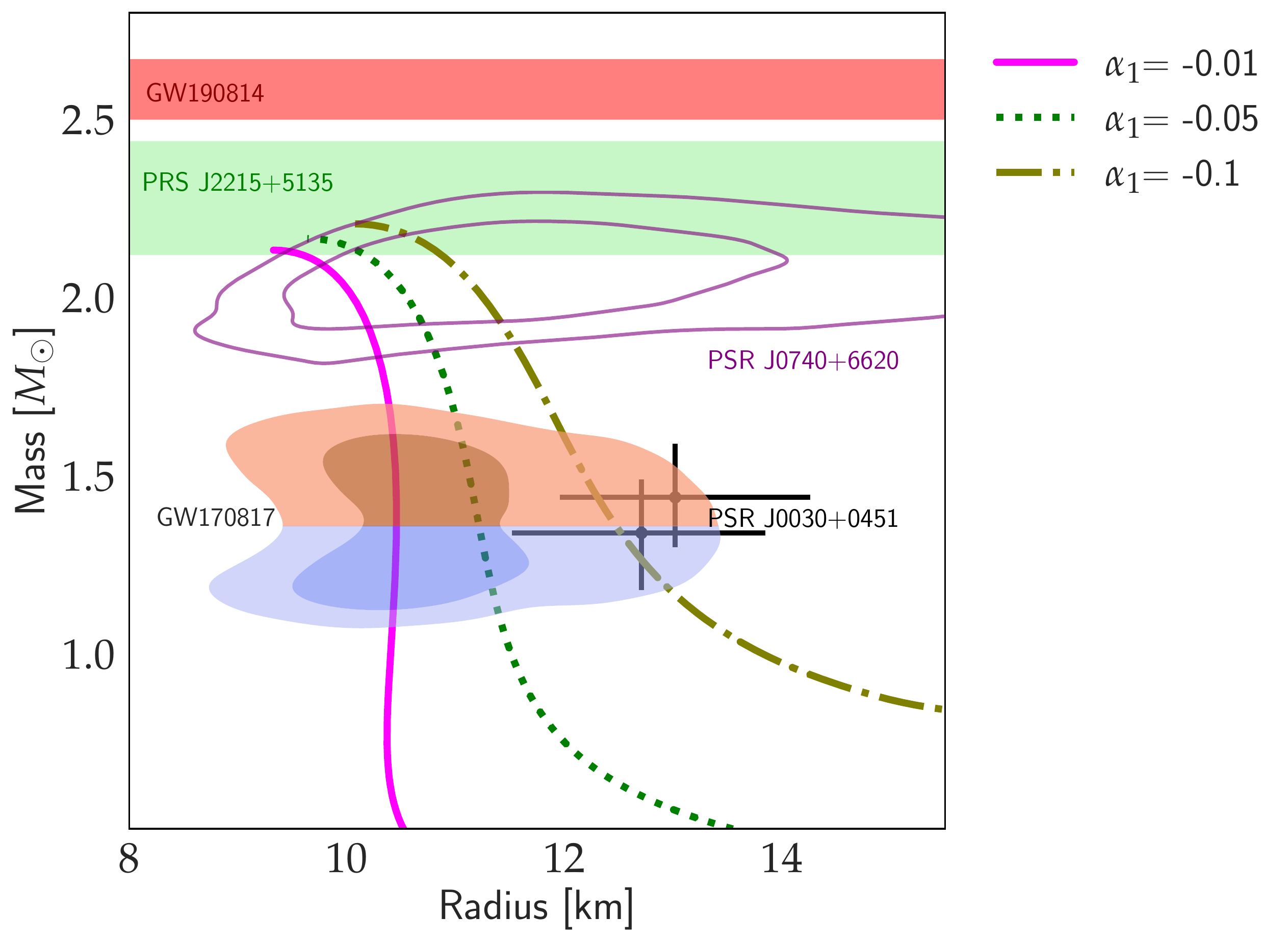}}
    \newline
    \subfloat[$V_2(\phi)=\frac{\alpha_2}{e^{K\phi^2}}$]{
        \includegraphics[width=0.4\textwidth]{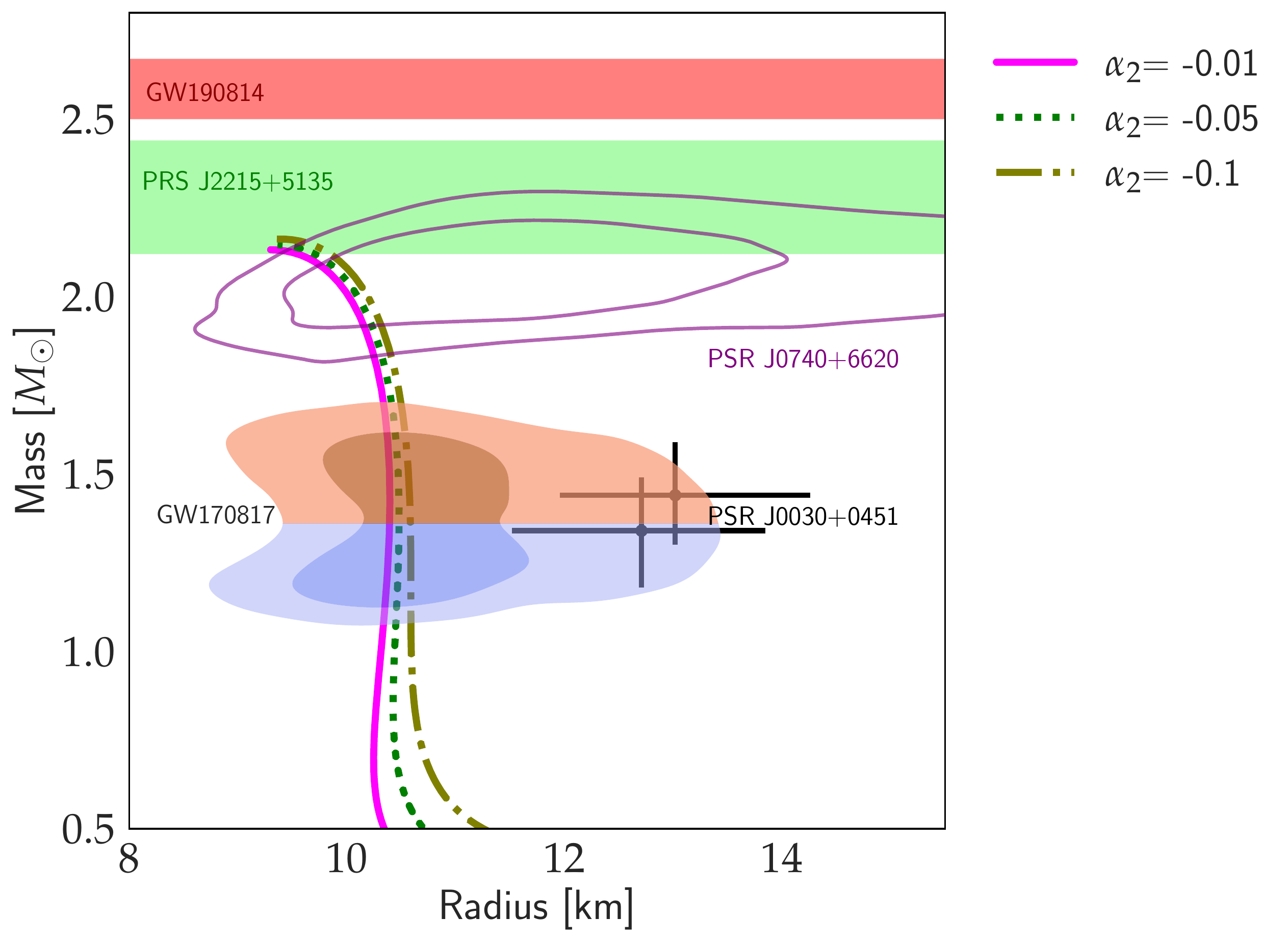}} \quad
    \subfloat[$V_2(\phi)=\frac{\alpha_2}{e^{K\phi^2}}$]{\includegraphics[width=0.4\textwidth]{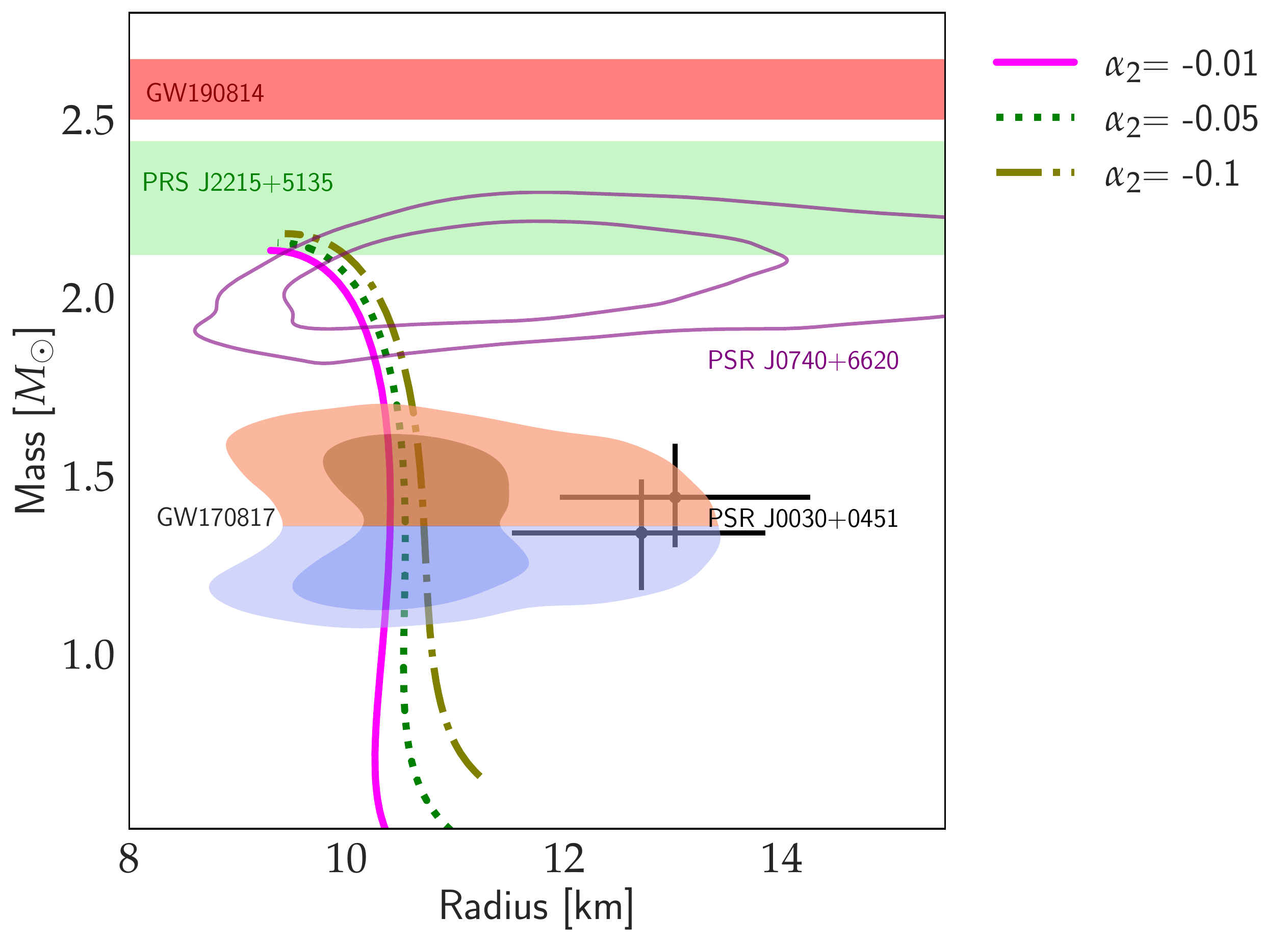}}
    \newline
    \subfloat[${V_3(\phi)=\frac{\alpha_3\phi^2}{1+e^{K\phi^2}}}$]{%
        \includegraphics[width=0.4\textwidth]{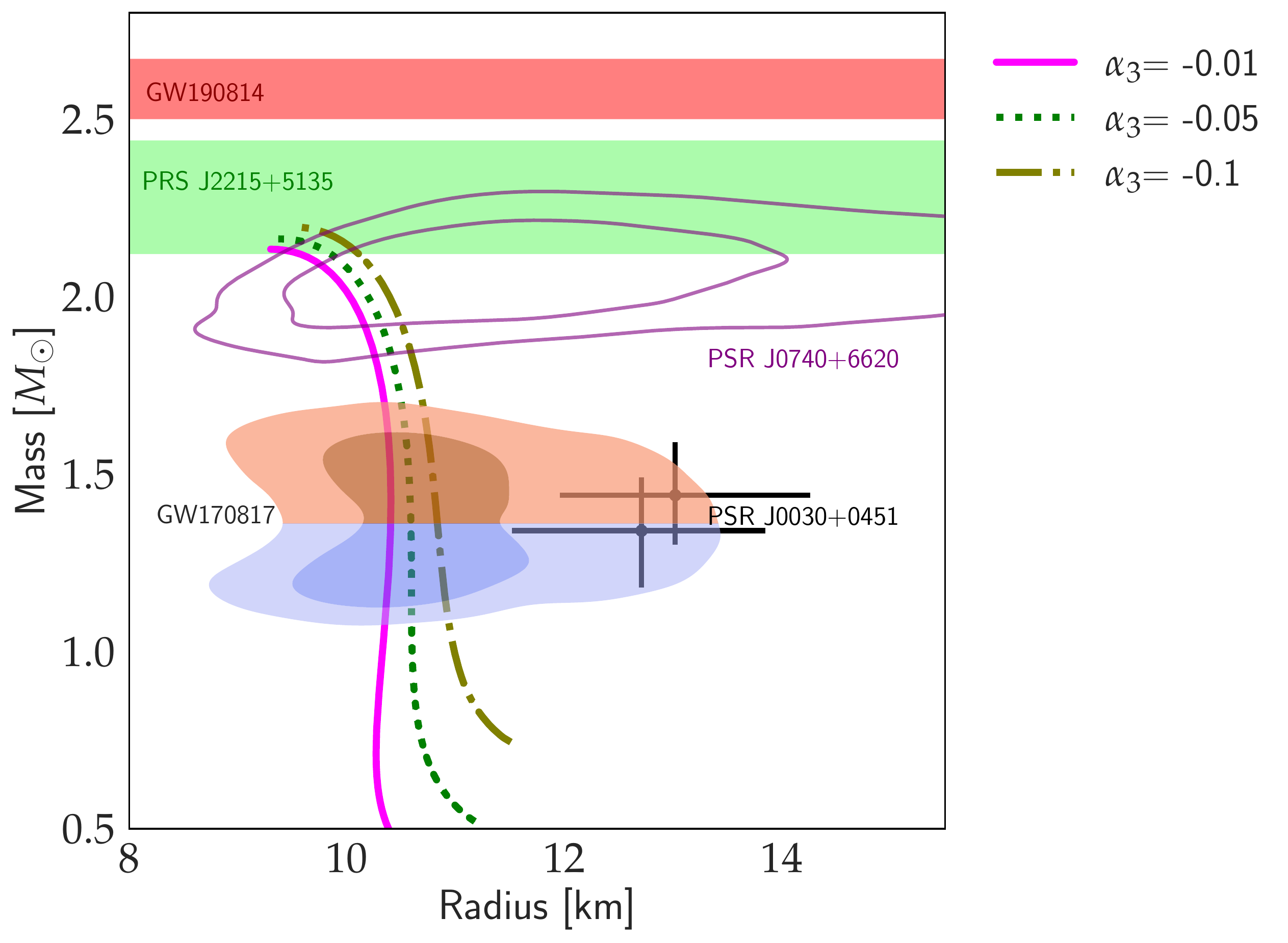} } \quad %
    \subfloat[${V_3(\phi)=\frac{\alpha_3\phi^2}{1+e^{K\phi^2}}}$]{%
        \includegraphics[width=0.4\textwidth]{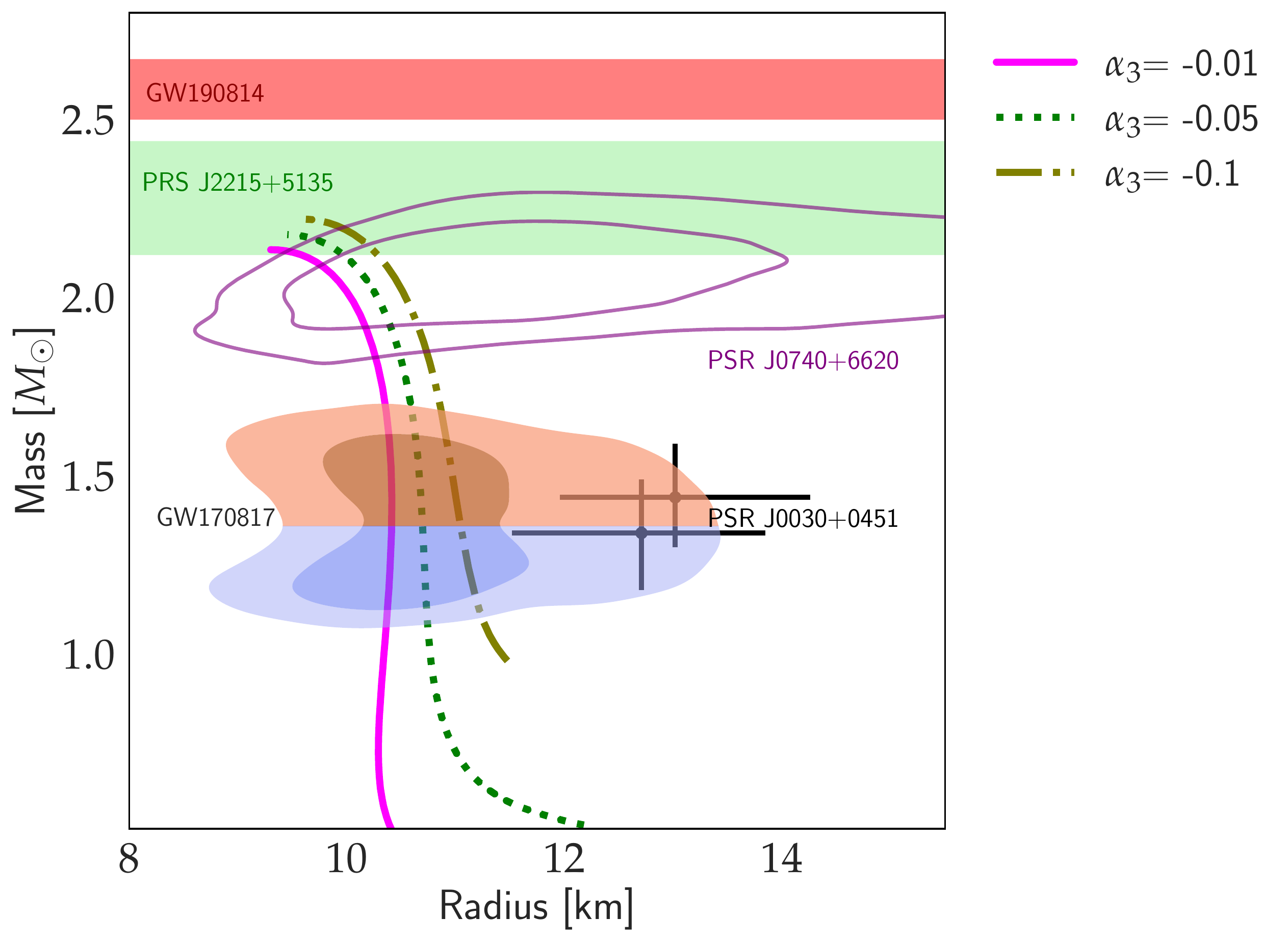}}
    \caption{The mass-radius
        relation of neutron stars in mimetic gravity for WFF1 EoS. $\protect\phi(0)=1$: left panels and $\protect%
        \phi(0)=0.8$: right panels and $K=0.5$.}
    \label{WFF1_fig}
\end{figure}

Before comparing the results of neutron star properties in GR and
mimetic gravity, we should note that in Figs.
\ref{sly4_fig0}-\ref{WFF1_fig}, we consider $1\sigma (\% 63)$ and
$2\sigma (\% 95)$ mass-radius constraints from the gravitational
wave event GW170817 by the blue and orange clouded regions
corresponding to the heavier and lighter neutron stars,
respectively \cite{abbott2018gw170817}. In these figures, the
purple region also shows the $1\sigma$ and $2\sigma$ confidential
levels of the massive pulsar PSR J0740+6620 with mass
$2.14_{-0.18}^{+.20}M_{\odot}$ \cite{cromartie2020relativistic}.
Also, we illustrate the mass range of the pulsar PSR J2215+5135
\cite{linares2018peering}, with a mass $2.27 M_{\odot}$, by light
green region in the mentioned figures. Besides, regarding these
figures, the black error bars and the red band show the NICER
(Neutron Star Interior Composition Explorer) mass-radius
measurement on PSR J0030+0451 \cite{miller2019psr,riley2019nicer}
and GW190814 event \cite{abbott2020gw190425}, respectively.

Since we have the results of different EoSs in GR, here, we use
the modified theory of gravity (mimetic gravity) to calculate the
mass-radius relation for each EoS to find matching with the
observational data. Strictly speaking, we investigate the
mass-radius relation of neutron stars in mimetic gravity for three
scalar potentials; $V_1(\phi)=\frac{\alpha_1}{\phi^2}$,
$V_2(\phi)=\frac{\alpha_2}{e^{K\phi^2}}$, and
$V_3(\phi)=\frac{\alpha_3 \phi^2}{1+e^{K\phi^2}}$. Therefore, we
have the free parameters  $\alpha_i (i=1..3)$ and $\phi(0)$ that
can affect more or less the maximum mass and radius of the neutron
stars (Note: $\phi(0)=\phi(r)|_{r=0}$ is the initial value of
mimetic scalar field at the center of the neutron star which
should be fixed for numerical analysis).

\begin{table*}[tbp]
    \caption{The maximum mass and corresponding radius for the different EoSs in GR.}\centering
%   {\footnotesize \
%       \resizebox{.28\hsize}{!}{
%           \tiny
            \begin{tabular}{|@{}c|c|c@{} |}
                \noalign{\hrule height .5pt}\hline
                GR & ${M_{max}}\ (M_{\odot})$ &$R\ (km)$ \\ \hline
                \noalign{\hrule height 1pt}
                SLy4 & $2.05$ &  $10.00$   \\ \hline
                WFF1  &   $2.12$  &  $9.48$    \\ \hline
                BSK21 & $2.28$  &  $11.02$    \\ \hline
                MPA1 &   $2.49$ &  $11.35$  \\ \hline
                DD2 &   $2.42$ &  $11.90$  \\ \hline
                $BHB\Lambda\Phi$ &   $2.10$ &  $11.58$  \\ \hline
                FSU2H & $2.37$ & $12.43$ \\ \hline
                SFHoY &   $1.99$ &  $10.52$  \\ \hline
                \noalign{\hrule height .5pt}
            \end{tabular}\label{table_GR}
\end{table*}

The first step is devoted to considering one of the well-known
EoSs, SLy4 EoS. It is easy to show that in the GR limit, this EoS
can reach $2 M_{\odot}$ and it is within the GW170817 region.
However, the mass-radius relation of this EoS could not support
one of the ranges of J0030+0451 error bars. For SLy4 EoS,
we plot the $M-R$ relation for potentials $V_1(\phi)$, $V_2(\phi)$
and $V_3(\phi)$ in Fig. \ref{sly4_fig} with $\phi(0)=1$ (left
panels) and $\phi(0)=0.8$ (right panels). We observe for
$\phi(0)=1$,  the maximum mass generally increases by increasing
the parameter $\alpha_i$. Recall that the sign of $\alpha_i$ is
negative, and hereafter, we discuss the absolute value of
$\alpha_i$ for the sake of simplicity. The increase of the mass at
$V_3(\phi)$ is greater than $V_1(\phi)$ and $V_2(\phi)$. In order
to have a easier comparison, we calculate the maximum mass and
radius for three different values for $\alpha_i$ and two values
for $\phi(0)$ in table \ref{table_sly4_1}. Considering the
reported maximum mass, one can find that the
larger one belongs to potential $V_3(\phi)$. Furthermore, according to Fig. \ref%
{sly4_fig}, we can find that by increasing of $|\alpha_i|$, there
is a small increment in the radius corresponding to potential
$V_2(\phi)$ and a bigger one corresponding to $V_1(\phi)$ and
$V_3(\phi)$. Therefore, considering the potentials $V_1(\phi)$ and
$V_3(\phi)$ for $\alpha_i = -0.05,-0.10 \; (i=1,3)$ , not only the
curves is still within the $GW170817$ region, but also it can
reach both the PSR $J0030+0451$ error bars.

We also consider the case $\phi(0)=0.8$ in right panels of
Fig. \ref{sly4_fig}. Obviously, the mass-radius yields a similar
behavior as in the previous case $\phi(0)=1$, i.e., as we increase
the value of parameter  $\alpha_i$, an enhancement is observed in
the maximum mass. However, by comparing the left and right panels
of Fig. \ref{sly4_fig}, one finds that reducing $\phi(0)$ leads to
increasing the maximum mass. Such a result can be confirmed by
comparison of maximum masses for $\phi(0)=1$ and $\phi(0)=0.8$ in
table \ref{table_sly4_1} more clearly. It is observed that for
potential $V_3(\phi)$ ($\lvert{\alpha_3}\lvert=0.1$), the maximum
mass reaches $2.14 M_{\odot}$, so in addition to the fact that it
still lies within two $GW170817$ and NICER regions, it also
reaches the PSR $J2215+5135$.

%\begin{figure*}[tbp]
%\centering
%\subfloat[${V_1(\phi)=\frac{\alpha_1}{\phi^2}}$]{\includegraphics[width=0.31%
%\textwidth]{v1phi01SLY4}} \quad
%\subfloat[$V_2(\phi)=\frac{\alpha_2}{e^{K\phi^2}}$]{
%\includegraphics[width=0.31\textwidth]{v2phi01SLY4} } \quad %
%\subfloat[${V_3(\phi)=\frac{\alpha_3\phi^2}{1+e^{K\phi^2}}}$]{%
%\includegraphics[width=0.31\textwidth]{v3phi01SLY4} }
%\caption{The mass-radius diagram for neutron stars in mimetic gravity for a
%SLy4 equation of state :$\protect\phi(0) =1$ and $K=0.5$. }
%\label{sly4_fig1}
%\end{figure*}

%%%%%%%%%%%%%%%%%%%%%%%%%%%%%%%%%%%%%%%%%%%%%%%%%%%%%%%%%%%%%%%%%%%%%%%%%%%%%%%%%%%%%%%%%%%%
\begin{table*}[tbp]
\caption{$M_{max}$ and corresponding radius for
\emph{non-relativistic EoSs} with $K=0.5$, $\protect\phi(0)=1$
\tcb{$(\protect\phi(0)=0.8)$} and various  $\alpha_i (i=1..3)$
.}\centering
%{\footnotesize \
%\resizebox{.75\hsize}{!}{
%        \tiny
        \begin{tabular}{|@{}c|c|c|c|c|c|c@{}|}
            \noalign{\hrule height.5pt} \hline
            \textbf{SLy4}& \multicolumn{2}{c|}{$V_1(\phi)=\frac{\alpha_1}{\phi^2}$}& \multicolumn{2}{c|}{$V_2(\phi)=\frac{\alpha_2}{e^{K\phi^2}}$}& \multicolumn{2}{c|}{$V_3(\phi)=\frac{\alpha_3\phi^2}{1+e^{K\phi^2}}$} \\
            \noalign{\hrule height 1pt}
            $\phi(0)=1$ \tcb{$(0.8)$}& ${M_{max}}\ (M_{\odot})$ &$R\ (km)$ & ${M_{max}}\ (M_{\odot})$  & $R\ (km)$ & ${M_{max}}\ (M_{\odot})$  &$R\ (km)$ \\ \hline
            \noalign{\hrule height 1pt}
            $\alpha_i=-0.01$  & $2.06 \tcb{(2.06)}$ & $10.02 \tcb{(10.02)}$ &  $2.05 \tcb{(2.06)}$ & $9.99 \tcb{(10.00)} $ & $2.06 \tcb{(2.06)}$ & $10.00 \tcb{(10.02)}$  \\
            $\alpha_i=-0.05$  & $2.07 \tcb{(2.07)}$ & $10.17 \tcb{(10.21)}$ &  $2.06 \tcb{(2.07)}$ & $10.01 \tcb{(10.04)}$ & $2.08 \tcb{(2.09)}$ & $10.08 \tcb{(10.12)}$   \\
            $\alpha_i=-0.10 $ & $2.08 \tcb{(2.09)}$ & $10.40 \tcb{(10.49)}$ &  $2.07 \tcb{(2.08)}$ & $10.03 \tcb{(10.10)}$ & $2.11 \tcb{(2.14)}$ & $10.18 \tcb{(10.23)}$   \\
            \noalign{\hrule height .5pt}\hline
        \end{tabular}\label{table_sly4_1}
        \begin{tabular}{|@{}c|c|c|c|c|c|c@{}|}
        \noalign{\hrule height .5pt}\hline
        \textbf{BSK21}& \multicolumn{2}{c|}{$V_1(\phi)=\frac{\alpha_1}{\phi^2}$}& \multicolumn{2}{c|}{$V_2(\phi)=\frac{\alpha_2}{e^{K\phi^2}}$}& \multicolumn{2}{c|}{$V_3(\phi)=\frac{\alpha_3\phi^2}{1+e^{K\phi^2}}$} \\
        \noalign{\hrule height 1pt} \hline
        $\phi(0)=1$ \tcb{$(0.8)$}& ${M_{max}}\ (M_{\odot})$ &$R\ (km)$ & ${M_{max}}\ (M_{\odot})$  &$R\ (km)$ & ${M_{max}}\ (M_{\odot})$  & $R\ (km)$ \\
        \noalign{\hrule height 1pt}
          $\alpha_i=-0.01$  & $2.28 \tcb{(2.28)}$ & $11.10 \tcb{(11.10)}$ &  $2.28 \tcb{(2.28)}$ & $11.02 \tcb{(11.06)}$ & $2.28 \tcb{(2.28)}$ & $11.06 \tcb{(11.07)}$  \\ \hline
          $\alpha_i=-0.05$  & $2.30 \tcb{(2.30)}$ & $11.28 \tcb{(11.34)}$ &  $2.29 \tcb{(2.29)}$ & $11.05 \tcb{(11.09)}$ & $2.31 \tcb{(2.32)}$ & $11.14 \tcb{(11.15)}$   \\ \hline
          $\alpha_i=-0.10 $ & $2.32 \tcb{(2.34)}$ & $11.56 \tcb{(11.64)}$ &  $2.30 \tcb{(2.31)}$ & $11.08 \tcb{(11.12)}$ & $2.33 \tcb{(2.36)}$ & $11.22 \tcb{(11.28)}$    \\
        \noalign{\hrule height .5pt}\hline
    \end{tabular}\label{table_BSK21}
    \begin{tabular}{|@{}c|c|c|c|c|c|c@{}|}
        \noalign{\hrule height .5pt}\hline
        \textbf{WFF1}& \multicolumn{2}{c|}{$V_1(\phi)=\frac{\alpha_1}{\phi^2}$}& \multicolumn{2}{c|}{$V_2(\phi)=\frac{\alpha_2}{e^{K\phi^2}}$}& \multicolumn{2}{c|}{$V_3(\phi)=\frac{\alpha_3\phi^2}{1+e^{K\phi^2}}$} \\
        \noalign{\hrule height 1pt} \hline
        $\phi(0)=1$ \tcb{$(0.8)$}& ${M_{max}}\ (M_{\odot})$ &$R\ (km)$ & ${M_{max}}\ (M_{\odot})$  &$R\ (km)$ & ${M_{max}}\ (M_{\odot})$  & $R\ (km)$ \\
        \noalign{\hrule height 1pt}
        $\alpha_i=-0.01$  & $2.13 \tcb{(2.15)}$ & $9.32 \tcb{(9.33)} $ &  $ 2.13 \tcb{(2.13)}$ & $ 9.30 \tcb{(9.30)}$ & $2.13 \tcb{(2.13)}$ & $ 9.30 \tcb{(9.30)}$  \\ \hline
        $\alpha_i=-0.05$  & $2.15 \tcb{(2.17)}$ & $9.61 \tcb{(9.64)} $ &  $ 2.15 \tcb{(2.15)}$ & $ 9.36 \tcb{(9.36)}$ & $2.16 \tcb{(2.18)}$ & $ 9.37 \tcb{(9.45)}$   \\ \hline
        $\alpha_i=-0.10 $ & $2.18 \tcb{(2.21)}$ & $9.95 \tcb{(10.08)}$ &  $ 2.16 \tcb{(2.18)}$ & $ 9.36 \tcb{(9.43)}$ & $2.20 \tcb{(2.22)}$ & $ 9.51 \tcb{(9.60)}$    \\
        \noalign{\hrule height .5pt}\hline
    \end{tabular}\label{table_WFF1}
\end{table*}

In order to investigate the effect of MPA1 EoS on the neutron star
structure, we plot the corresponding mass-radius relation in Fig. \ref%
{MPA1_fig}. By increasing parameter $\alpha_i$ or reducing
$\phi(0)$, the mass-radius relation yields a similar behavior such
as previous case (the SLy4 case). Also, similar to the SLy4 case,
we observe a greater maximum mass for potential $V_3(\phi)$
comparing to the potentials $V_2(\phi)$ and $V_1(\phi)$ (see table
\ref{table_MPA1_1} for more details). It is notable that the
curves for all three potentials can support all the observation
regions. However, some of the curves for parts (a) and (b) of Fig.
\ref{MPA1_fig} is discredited, since it lies out of the LIGO-VIRGO
region. As a final result, it is worth mentioning that to support
all observational evidence within this EoS, the best value of
parameter $\alpha_1$ is about $-0.05$ for potential $V_1(\phi)$.
%We report the values of maximum mass and radius for this EOS in Tables~ \ref{table_MPA1_1}- \ref{table_MPA1_8} for  $\phi(0)=1$ and $\phi(0)=0.8$ respectively. We can see that the maximum mass even reach  $2.56 M_{\odot}$ for $A=0.1$ and $\phi(0)=0.8$.

%%%%%%%%%%%%%%%%%%%%%%%%%%%%%%%%%%%%%%%%%%%%%%%%%%%%%%%%%%%%%%%%%%%%%%%%
\begin{table*}[tbp]
\caption{$M_{max}$ and corresponding radius for \emph{relativistic
EoSs} with $K=0.5$, $\protect\phi(0)=1$
 \tcb{$(\protect\phi(0)=0.8)$} and various  $\alpha_i (i=1..3)$
.}\centering
%{\footnotesize \
%\resizebox{.75\hsize}{!}{
%        \tiny
        \begin{tabular}{|@{}c|c|c|c|c|c|c@{}|}
            \noalign{\hrule height .5pt}\hline
            \textbf{MPA1}& \multicolumn{2}{c|}{$V_1(\phi)=\frac{\alpha_1}{\phi^2}$}& \multicolumn{2}{c|}{$V_2(\phi)=\frac{\alpha_2}{e^{K\phi^2}}$}& \multicolumn{2}{c|}{$V_3(\phi)=\frac{\alpha_3\phi^2}{1+e^{K\phi^2}}$} \\
            \noalign{\hrule height 1pt} \hline
            $\phi(0)=1$ \tcb{$(0.8)$}& ${M_{max}}\ (M_{\odot})$ &$R\ (km)$ & ${M_{max}}\ (M_{\odot})$  &$R\ (km)$ & ${M_{max}}\ (M_{\odot})$  & $R\ (km)$ \\
            \noalign{\hrule height 1pt}
            $\alpha_i=-0.01$  & $2.50 \tcb{(2.49)}$ & $11.39 \tcb{(11.40)}$ &  $2.49 \tcb{(2.49)}$ & $ 11.36 \tcb{(11.36)}$ & $2.49 \tcb{(2.50)}$ & $11.36 \tcb{(11.40)}$  \\
            $\alpha_i=-0.05$  & $2.51 \tcb{(2.52)}$ & $11.57 \tcb{(11.63)}$ &  $2.50 \tcb{(2.50)}$ & $ 11.39 \tcb{(11.39)}$ & $2.51 \tcb{(2.52)}$ & $11.43 \tcb{(11.44)}$  \\
            $\alpha_i=-0.10 $ & $2.53 \tcb{(2.55)}$ & $11.83 \tcb{(11.90)}$ &  $2.51 \tcb{(2.52)}$ & $ 11.42 \tcb{(11.42)}$ & $2.54 \tcb{(2.56)}$ & $11.51 \tcb{(11.56)}$   \\
            \noalign{\hrule height .5pt}\hline
        \end{tabular}\label{table_MPA1_1}
        \begin{tabular}{|@{}c|c|c|c|c|c|c@{}|}
        \noalign{\hrule height.5pt} \hline
        \textbf{DD2}& \multicolumn{2}{c|}{$V_1(\phi)=\frac{\alpha_1}{\phi^2}$}& \multicolumn{2}{c|}{$V_2(\phi)=\frac{\alpha_2}{e^{K\phi^2}}$}& \multicolumn{2}{c|}{$V_3(\phi)=\frac{\alpha_3\phi^2}{1+e^{K\phi^2}}$} \\
        \noalign{\hrule height 1pt}
        $\phi(0)=1$ \tcb{$(0.8)$}& ${M_{max}}\ (M_{\odot})$ &$R\ (km)$ & ${M_{max}}\ (M_{\odot})$  & $R\ (km)$ & ${M_{max}}\ (M_{\odot})$  &$R\ (km)$ \\ \hline
        \noalign{\hrule height 1pt}
        $\alpha_i=-0.01$  & $2.42 \tcb{(2.43)}$ & $11.93 \tcb{(11.93)}$ &  $2.42 \tcb{(2.42)}$ & $11.89 \tcb{(11.89)}$ & $2.43 \tcb{(2.44)}$ & $11.89 \tcb{(11.89)}$  \\
        $\alpha_i=-0.05$  & $2.44 \tcb{(2.46)}$ & $12.18 \tcb{(12.20)}$ &  $2.43 \tcb{(2.44)}$ & $11.88 \tcb{(11.87)}$ & $2.45 \tcb{(2.47)}$ & $11.97 \tcb{(11.97)}$   \\
        $\alpha_i=-0.10$  & $2.47 \tcb{(2.51)}$ & $12.38 \tcb{(12.52)}$ &  $2.44 \tcb{(2.45)}$ & $11.95 \tcb{(11.95)}$ & $2.48 \tcb{(2.52)}$ & $11.94 \tcb{(12.05)}$    \\
        \noalign{\hrule height .5pt}\hline
    \end{tabular}\label{table_DD2}
    \begin{tabular}{|@{}c|c|c|c|c|c|c@{}|}
        \noalign{\hrule height.5pt} \hline
        \textbf{FSU2H}& \multicolumn{2}{c|}{$V_1(\phi)=\frac{\alpha_1}{\phi^2}$}& \multicolumn{2}{c|}{$V_2(\phi)=\frac{\alpha_2}{e^{K\phi^2}}$}& \multicolumn{2}{c|}{$V_3(\phi)=\frac{\alpha_3\phi^2}{1+e^{K\phi^2}}$} \\
        \noalign{\hrule height 1pt}
        $\phi(0)=1$ \tcb{$(0.8)$}& ${M_{max}}\ (M_{\odot})$ &$R\ (km)$ & ${M_{max}}\ (M_{\odot})$  & $R\ (km)$ & ${M_{max}}\ (M_{\odot})$  &$R\ (km)$ \\ \hline
        \noalign{\hrule height 1pt}
        $\alpha_i=-0.01$  & $2.40 \tcb{(2.40)}$ & $12.57 \tcb{(12.58)}$ &  $2.39 \tcb{(2.39)}$ & $12.53 \tcb{(12.52)}$ & $2.40 \tcb{(2.41)}$ & $12.56 \tcb{(12.56)}$  \\
        $\alpha_i=-0.05$  & $2.42 \tcb{(2.43)}$ & $12.81 \tcb{(12.87)}$ &  $2.40 \tcb{(2.40)}$ & $12.51 \tcb{(12.55)}$ & $2.42 \tcb{(2.43)}$ & $12.59 \tcb{(12.59)}$  \\
        $\alpha_i=-0.10$  & $2.45 \tcb{(2.49)}$ & $13.15 \tcb{(13.30)}$ &  $2.41 \tcb{(2.43)}$ & $12.59 \tcb{(12.57)}$ & $2.48 \tcb{(2.51)}$ & $12.66 \tcb{(12.71)}$  \\
        \noalign{\hrule height .5pt}\hline
    \end{tabular}\label{table_fsu2h}
\end{table*}
%%%%%%%%%%%%%%%%%%%%%%%%%%%%%%%%%%%%%%%%%%%%%%%%%%%%%%%%%%%%%%%%%%%%%%%%%%%%%

%\begin{table*}[tbp]
%\caption{The maximum mass and corresponding radius for various values of $\alpha_i (i=1..3)$, $\protect\phi(0)$=0.8 and $K=0.5$ (for MPA1 EoS)}\centering
%%{\footnotesize \
%%\resizebox{.75\hsize}{!}{
%%        \tiny
%        \begin{tabular}{@{}c|c|c|c|c|c|c@{}}
%            \noalign{\hrule height .5pt}\hline\hline
%            MPA1& \multicolumn{2}{c|}{$V_1(\phi)=\frac{\alpha_1}{\phi^2}$}& \multicolumn{2}{c|}{$V_2(\phi)=\frac{\alpha_2}{e^{K\phi^2}}$}& \multicolumn{2}{c}{$V_3(\phi)=\frac{\alpha_3\phi^2}{1+e^{K\phi^2}}$} \\
%            \noalign{\hrule height 1pt}
%            Mimetic gravity& ${M_{max}}\ (M_{\odot})$ &$R\ (km)$ & ${M_{max}}\ (M_{\odot})$  &$R\ (km)$ & ${M_{max}}\ (M_{\odot})$  & $R\ (km)$ \\
%            \noalign{\hrule height 1pt}
%            $\alpha_i=-0.01$  &  $  2.49 $ &  $11.40 $&  $ 2.49$ & $ 11.36$ & $  2.50$ &   $  11.40$  \\
%            $\alpha_i=-0.05$  & $   2.52$ & $  11.63$ &  $  2.50$ & $  11.39$  & $  2.52$ &   $  11.44$   \\
%            $\alpha_i=-0.10 $ & $2.55$ & $11.90$ &  $ 2.52$ & $  11.42$&$  2.56$&$  11.56$    \\
%            \noalign{\hrule height .5pt}\hline\hline
%        \end{tabular}\label{table_MPA1_8}
%\end{table*}

Following the previous calculations, we investigate the
mass-radius behaviour of neutron stars for the BSK21 EoS in Fig.
\ref{BSK21_fig} and table \ref{table_MPA1_1}. It is worth
mentioning that similar to the case of MPA1 EoS, there are some
curves for parts (a) and (b) of Fig. \ref{BSK21_fig} that are out
of LIGO-VIRGO data, and therefore, parameter $\alpha_i$ can be
considered around $-0.05$ which is similar to MPA1 EoS.

We desire to investigate the mass-radius relation for WFF1 EoS
that is reflected in Fig. \ref{WFF1_fig}. The general behavior of
the mass-radius relation is similar to those of the previous EoSs,
i.e, increasing $|\alpha_i|$ leads to increasing maximum mass (see
table \ref{table_sly4_1}). It is also observed that for GR limit
this EoS could not support the pulsar J2215+5135 and its radius is
small (see Fig. \ref{sly4_fig0}). So it could not reach NICER
observations. Taking Fig. \ref{WFF1_fig} into account, we find
that in mimetic gravity for all three potentials, for large
$\alpha_i$, the maximum mass is within the pulsar J2215+5135
region. Besides, for potential $V_1(\phi)$ and by considering
$\phi(0)=1$, the mass-radius plot supports the first NICER error
bar, while by changing $\phi(0)$ to $0.8$, we find that both of
NICER error bars are supported by mass-radius diagram.

\begin{figure}[!ht]
    \centering
    \subfloat[${V_1(\phi)=\frac{\alpha_1}{\phi^2}}$]{\includegraphics[width=0.4%
        \textwidth]{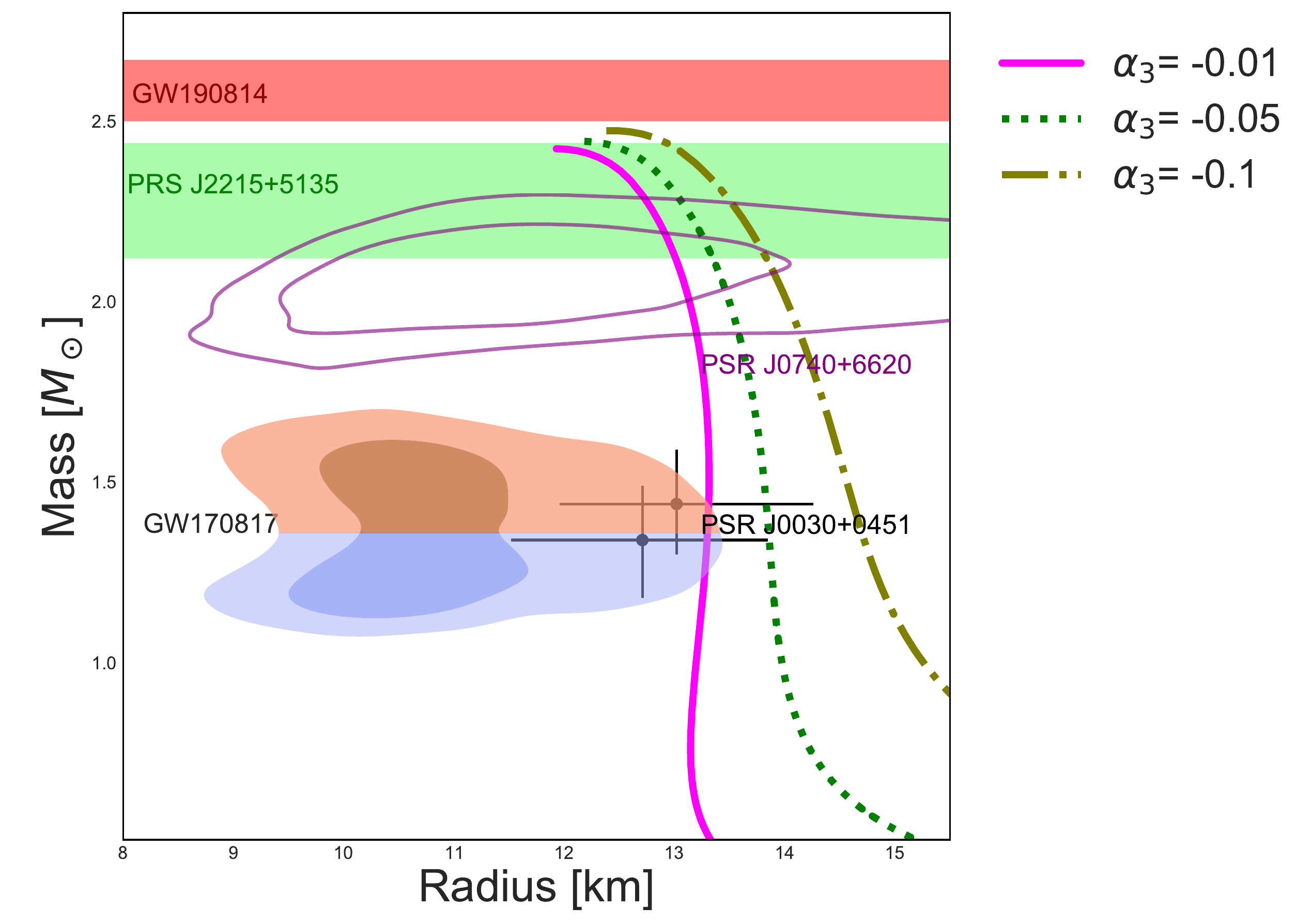}} \quad
    \subfloat[${V_1(\phi)=\frac{\alpha_1}{\phi^2}}$]{
        \includegraphics[width=0.4\textwidth]{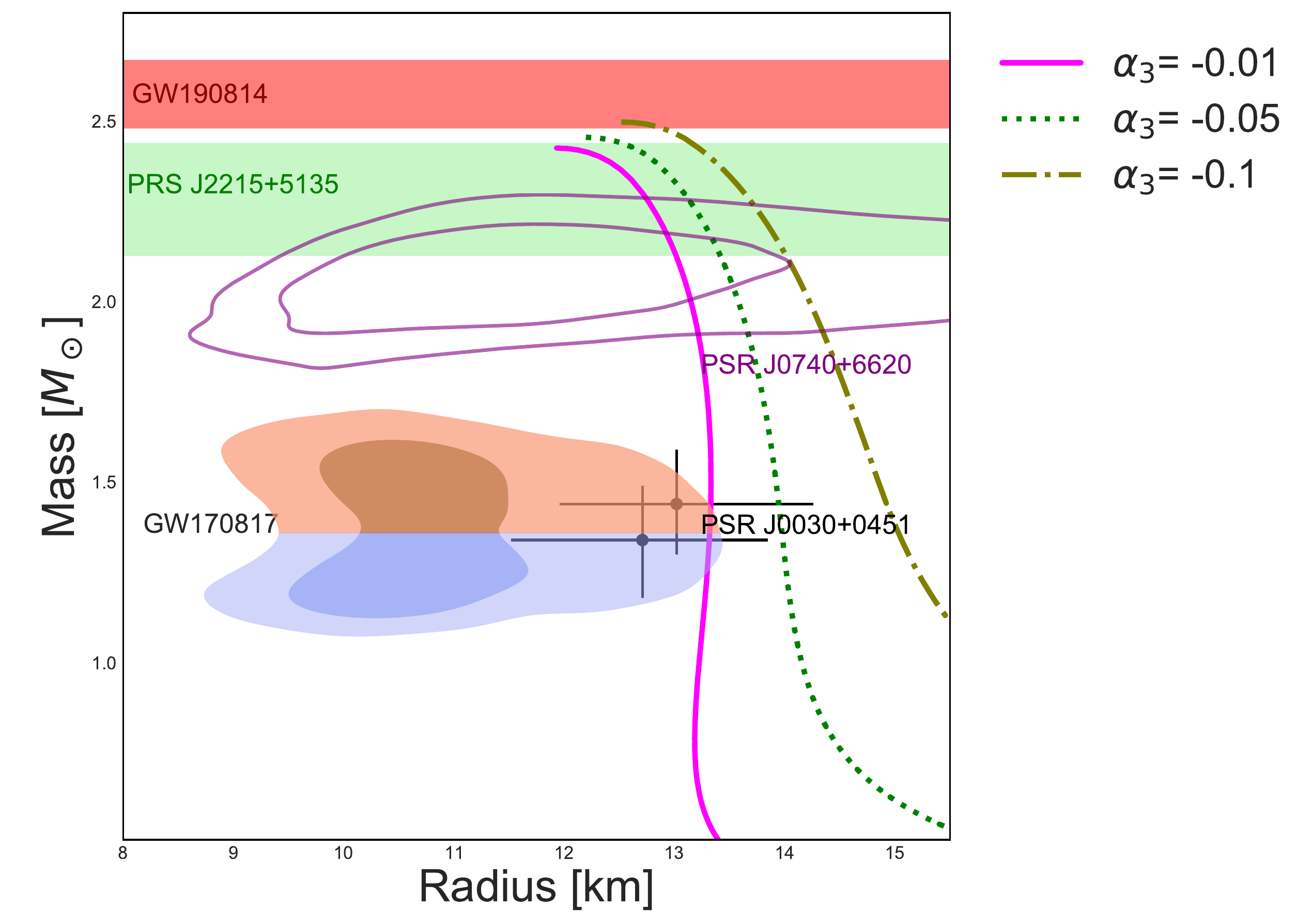}}
    \newline
    \subfloat[$V_2(\phi)=\frac{\alpha_2}{e^{K\phi^2}}$]{
        \includegraphics[width=0.4\textwidth]{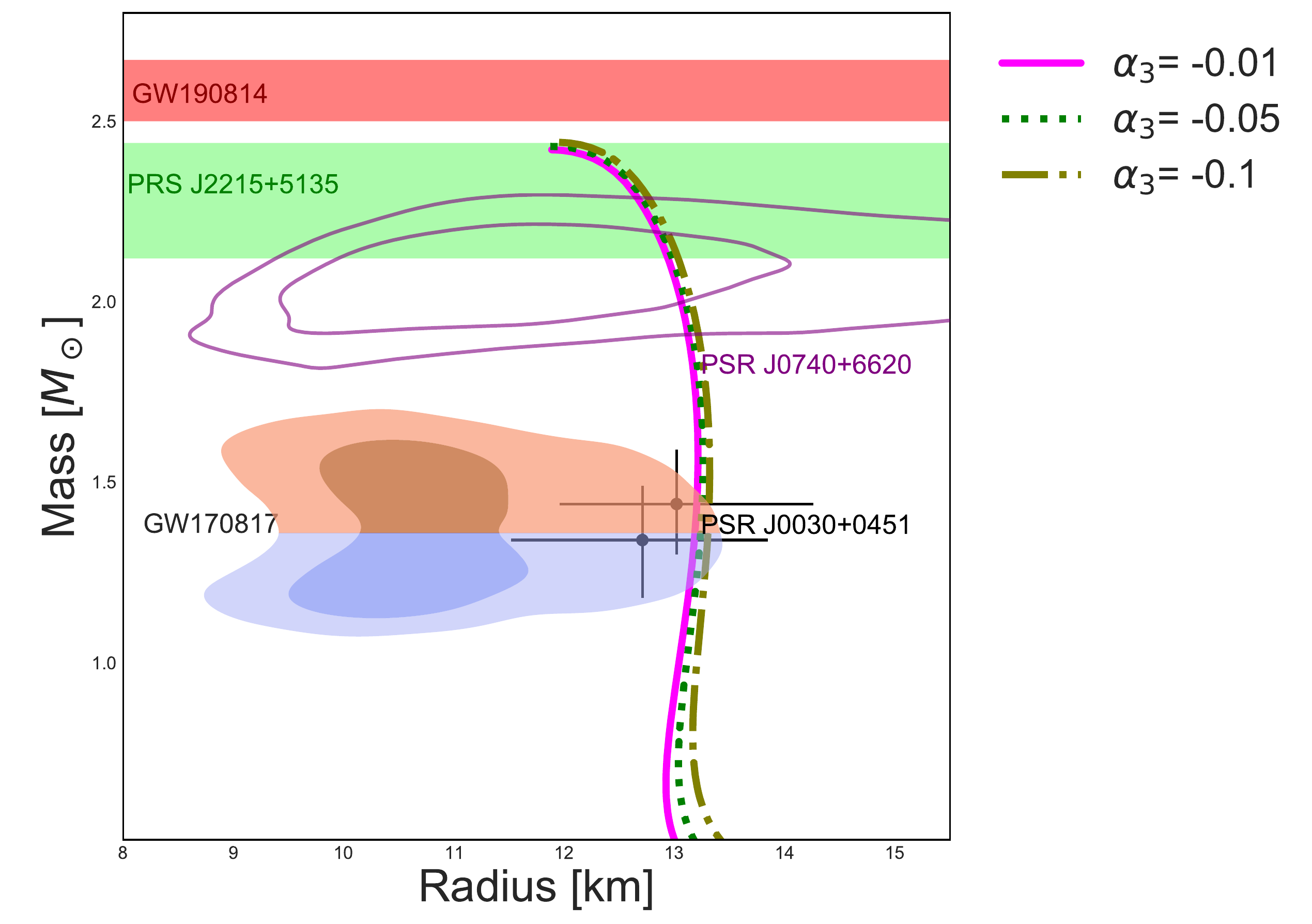}} \quad
    \subfloat[$V_2(\phi)=\frac{\alpha_2}{e^{K\phi^2}}$]{\includegraphics[width=0.4\textwidth]{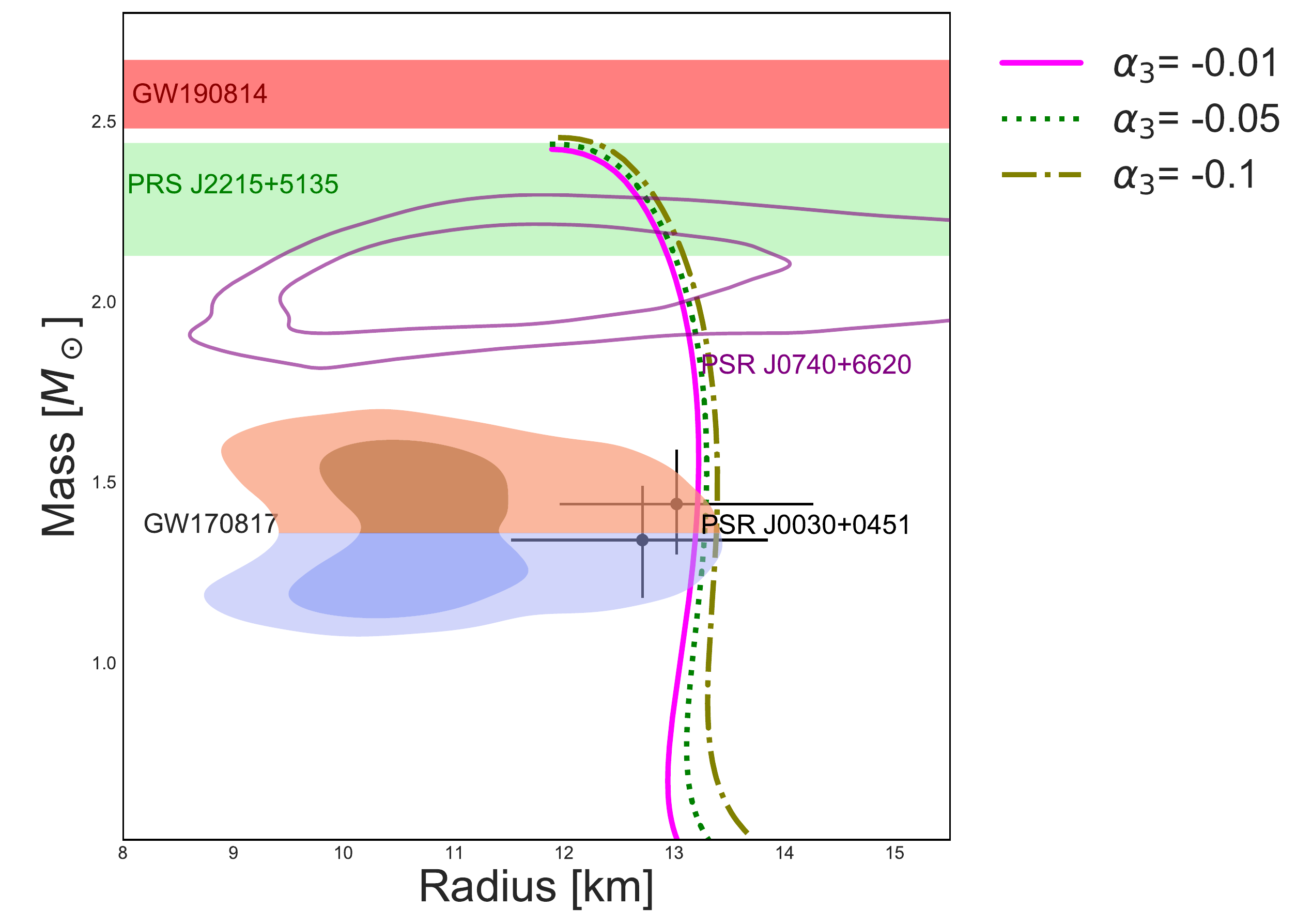}}
    \newline
    \subfloat[${V_3(\phi)=\frac{\alpha_3\phi^2}{1+e^{K\phi^2}}}$]{%
        \includegraphics[width=0.4\textwidth]{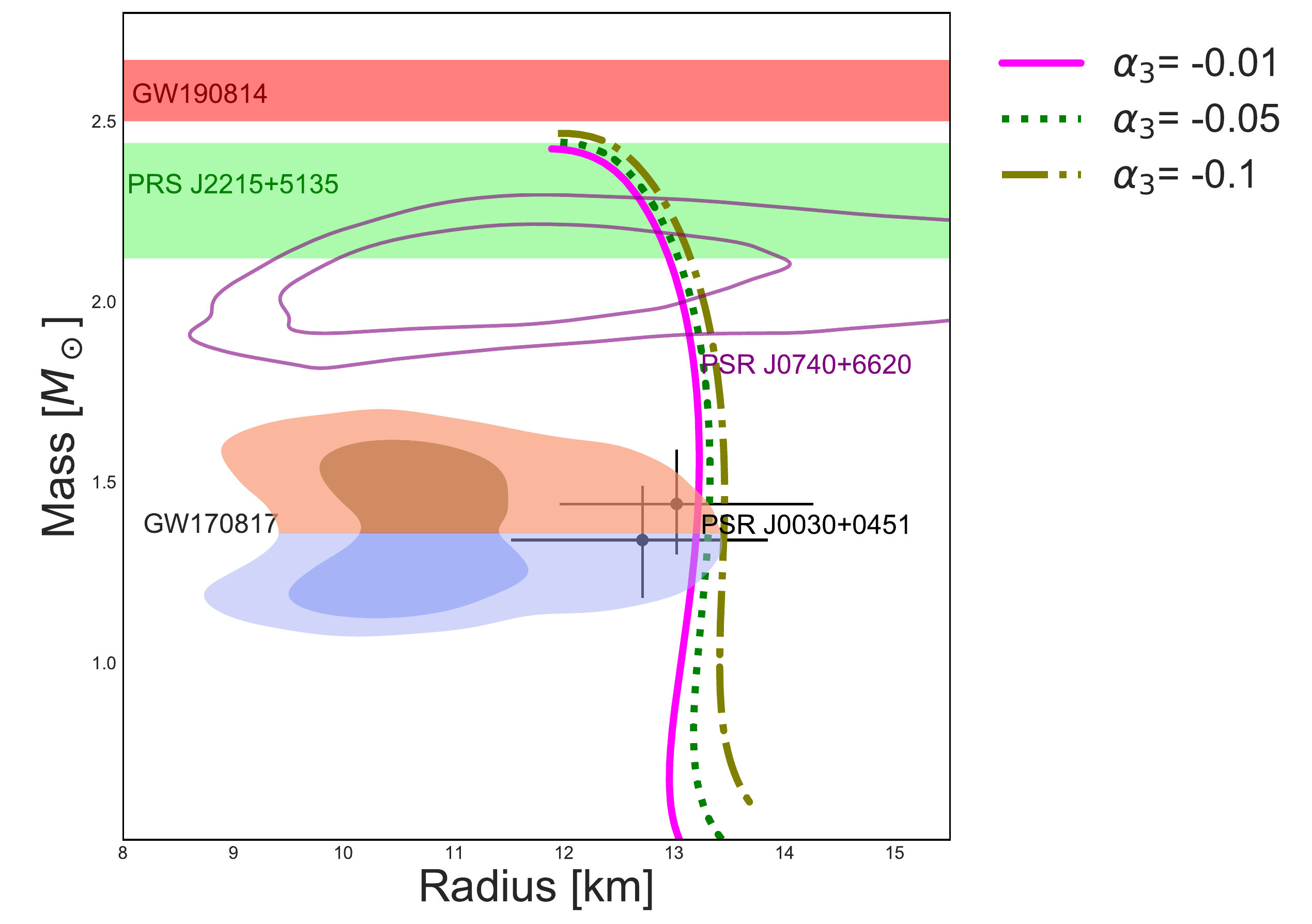} } \quad %
    \subfloat[${V_3(\phi)=\frac{\alpha_3\phi^2}{1+e^{K\phi^2}}}$]{%
        \includegraphics[width=0.4\textwidth]{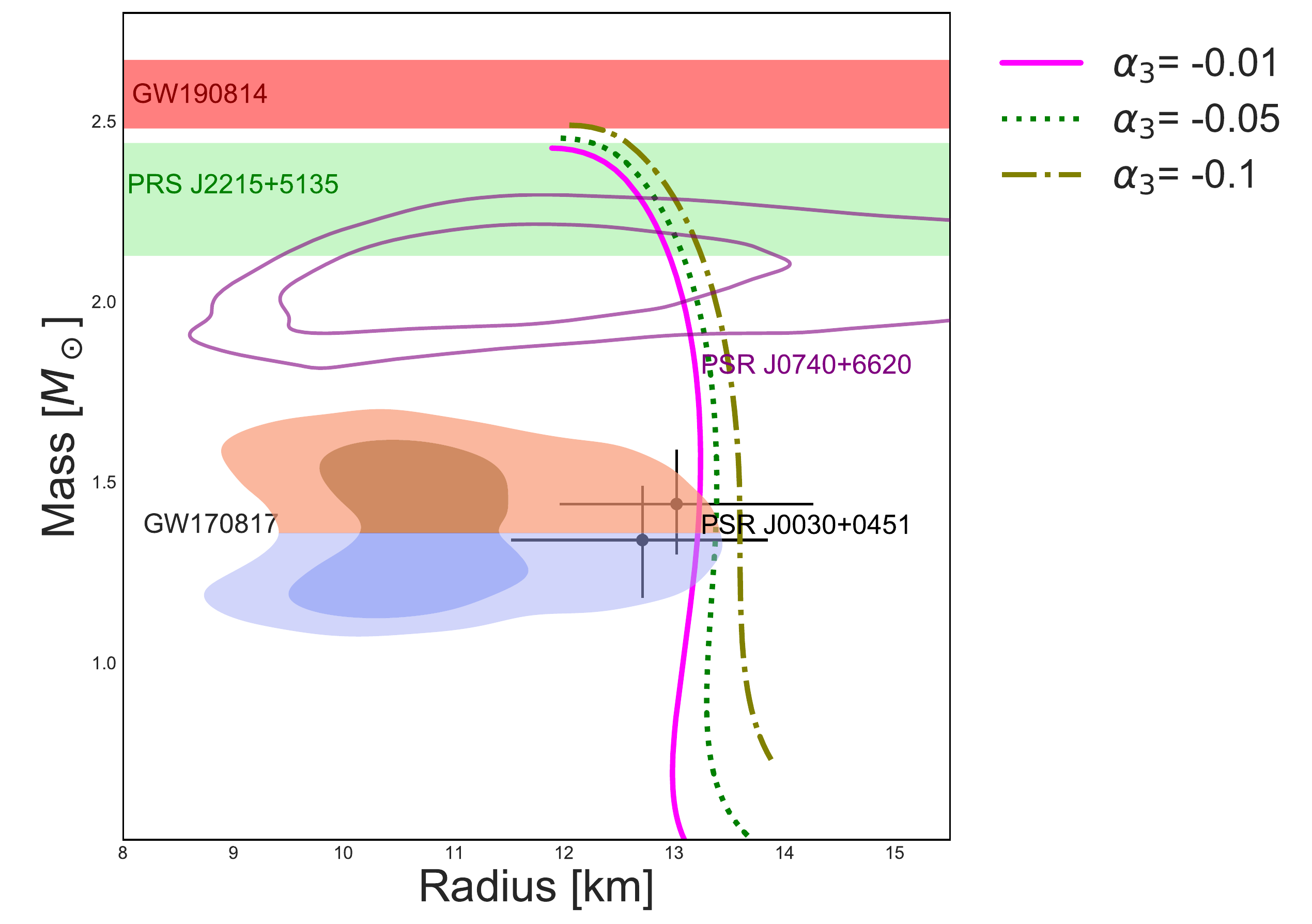}}
    \caption{The mass-radius
        relation of neutron stars in mimetic gravity for DD2 EoS. $\protect\phi(0)=1$: left panels and $\protect%
        \phi(0)=0.8$: right panels and $K=0.5$.}
    \label{DD2_fig2}
\end{figure}

\begin{figure}[!ht]
    \centering
    \subfloat[${V_1(\phi)=\frac{\alpha_1}{\phi^2}}$]{\includegraphics[width=0.4%
        \textwidth]{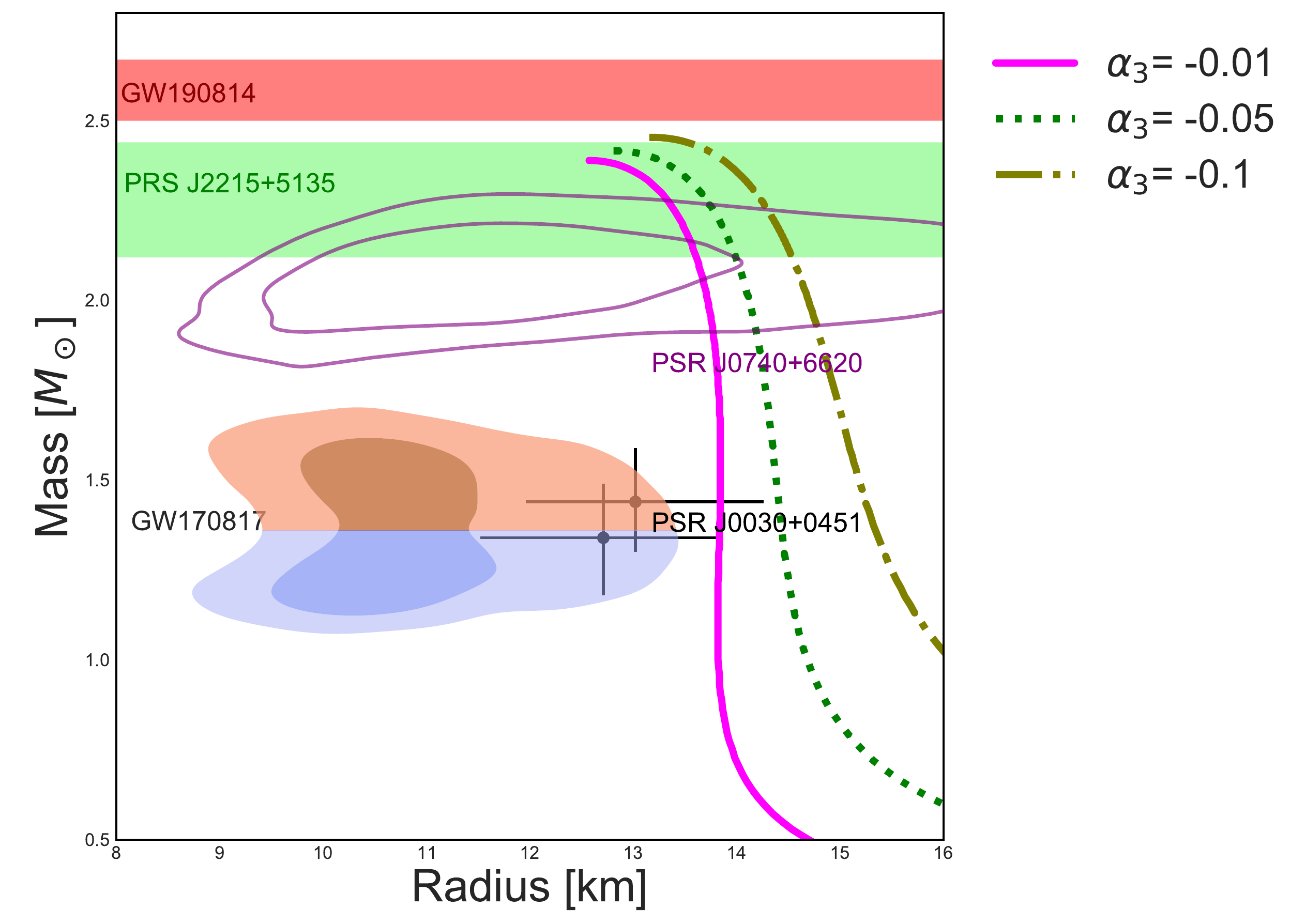}} \quad
    \subfloat[${V_1(\phi)=\frac{\alpha_1}{\phi^2}}$]{
        \includegraphics[width=0.4\textwidth]{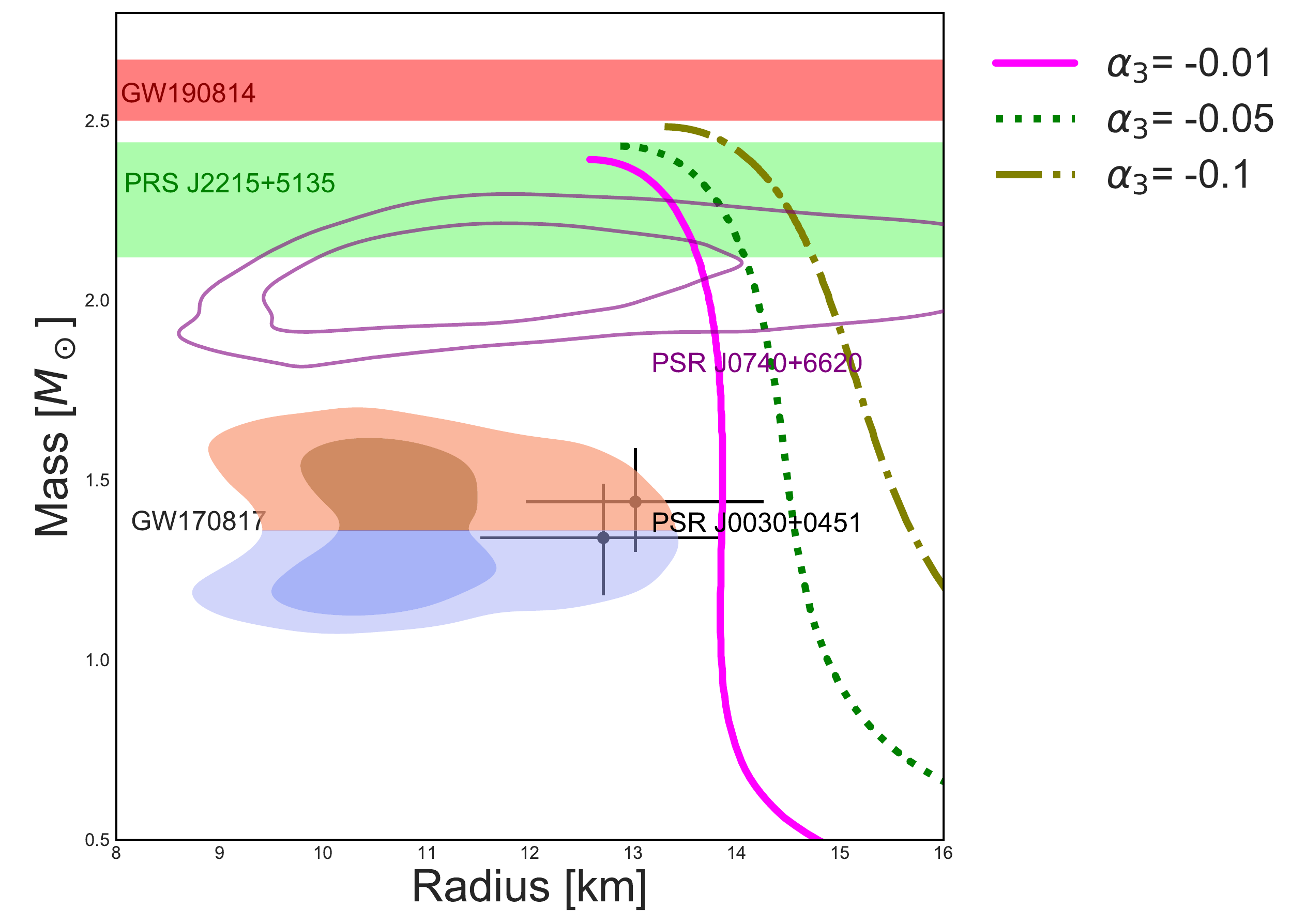}}
    \newline
    \subfloat[$V_2(\phi)=\frac{\alpha_2}{e^{K\phi^2}}$]{
        \includegraphics[width=0.4\textwidth]{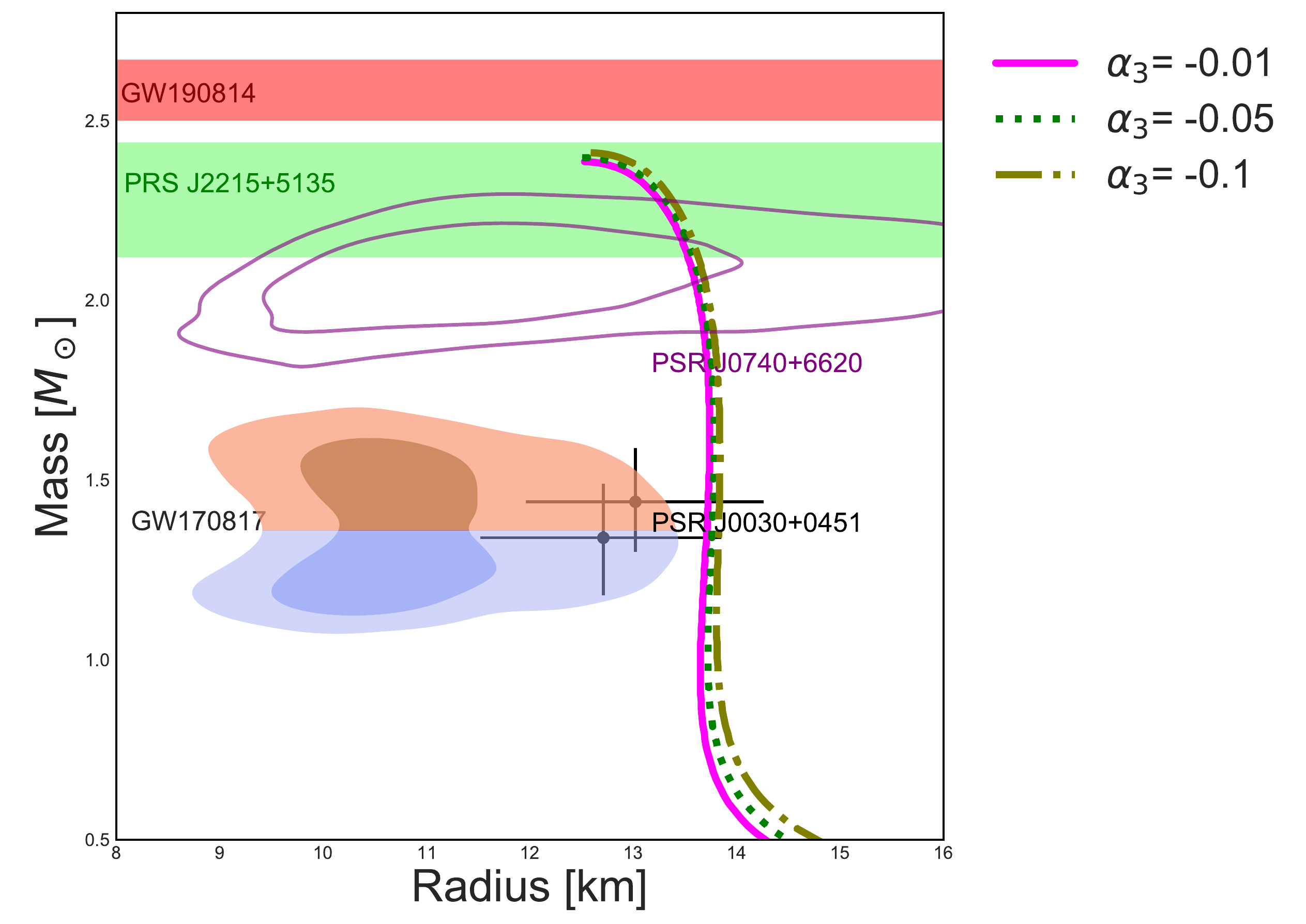}} \quad
    \subfloat[$V_2(\phi)=\frac{\alpha_2}{e^{K\phi^2}}$]{\includegraphics[width=0.4\textwidth]{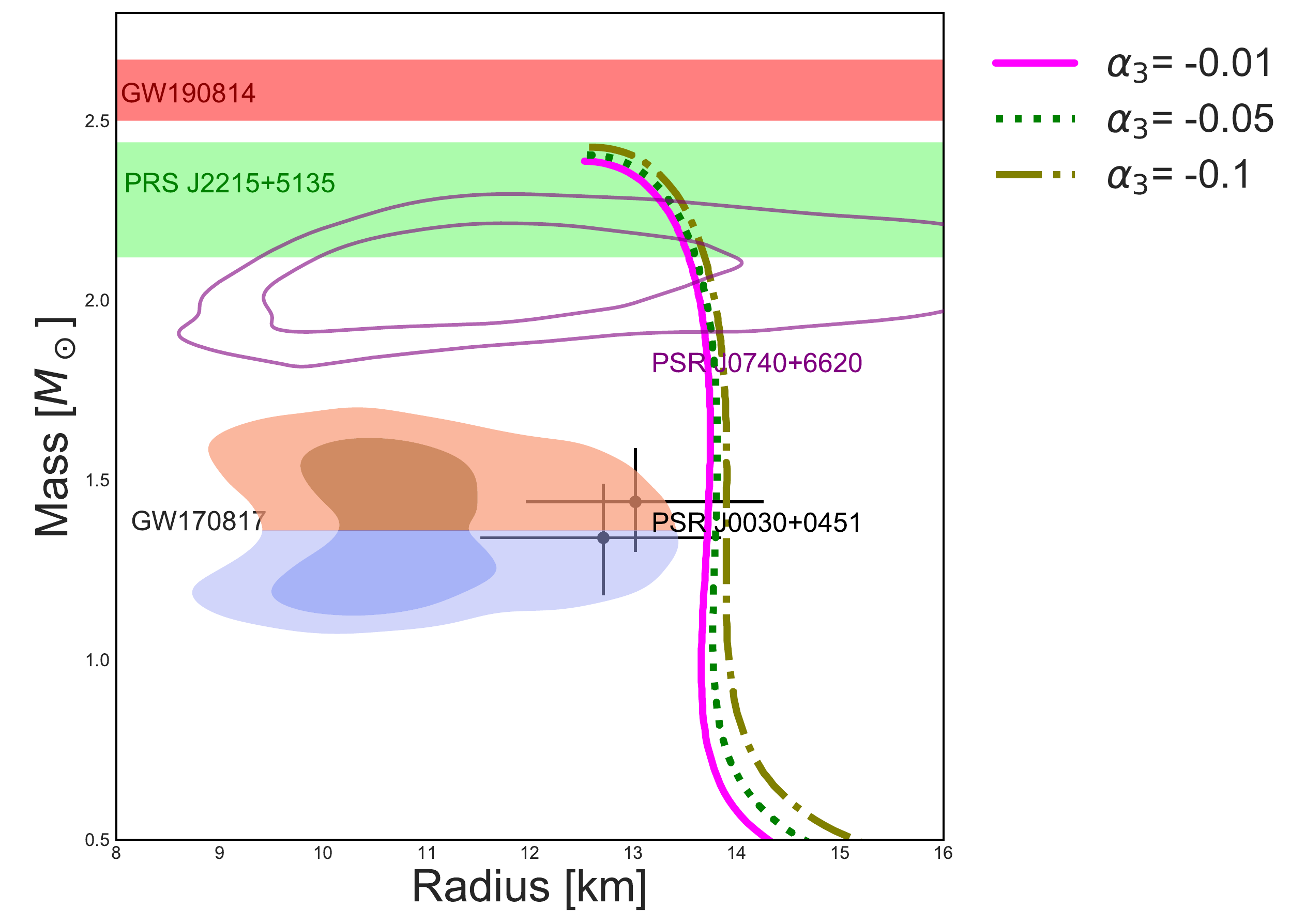}}
    \newline
    \subfloat[${V_3(\phi)=\frac{\alpha_3\phi^2}{1+e^{K\phi^2}}}$]{%
        \includegraphics[width=0.4\textwidth]{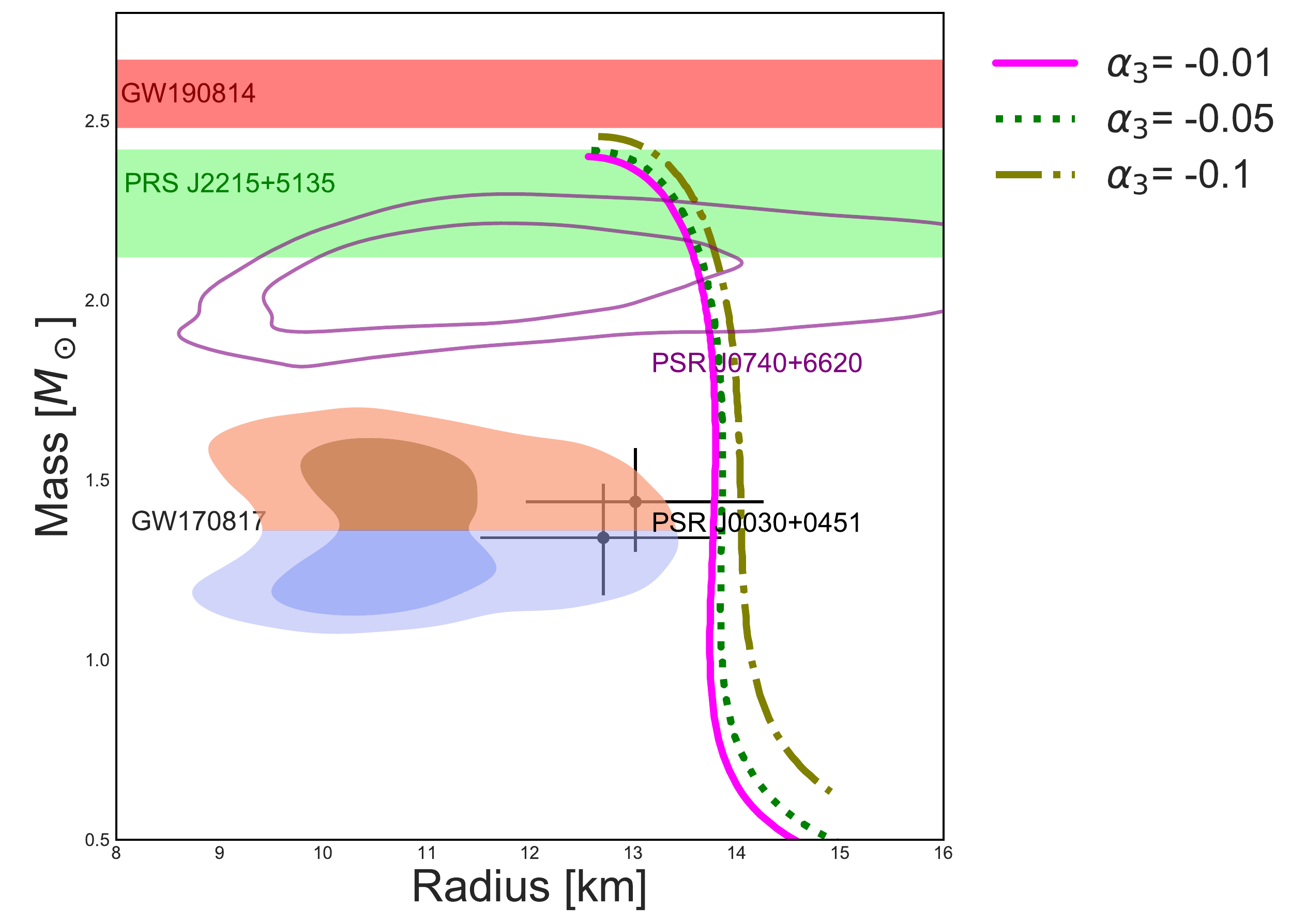} } \quad %
    \subfloat[${V_3(\phi)=\frac{\alpha_3\phi^2}{1+e^{K\phi^2}}}$]{%
        \includegraphics[width=0.4\textwidth]{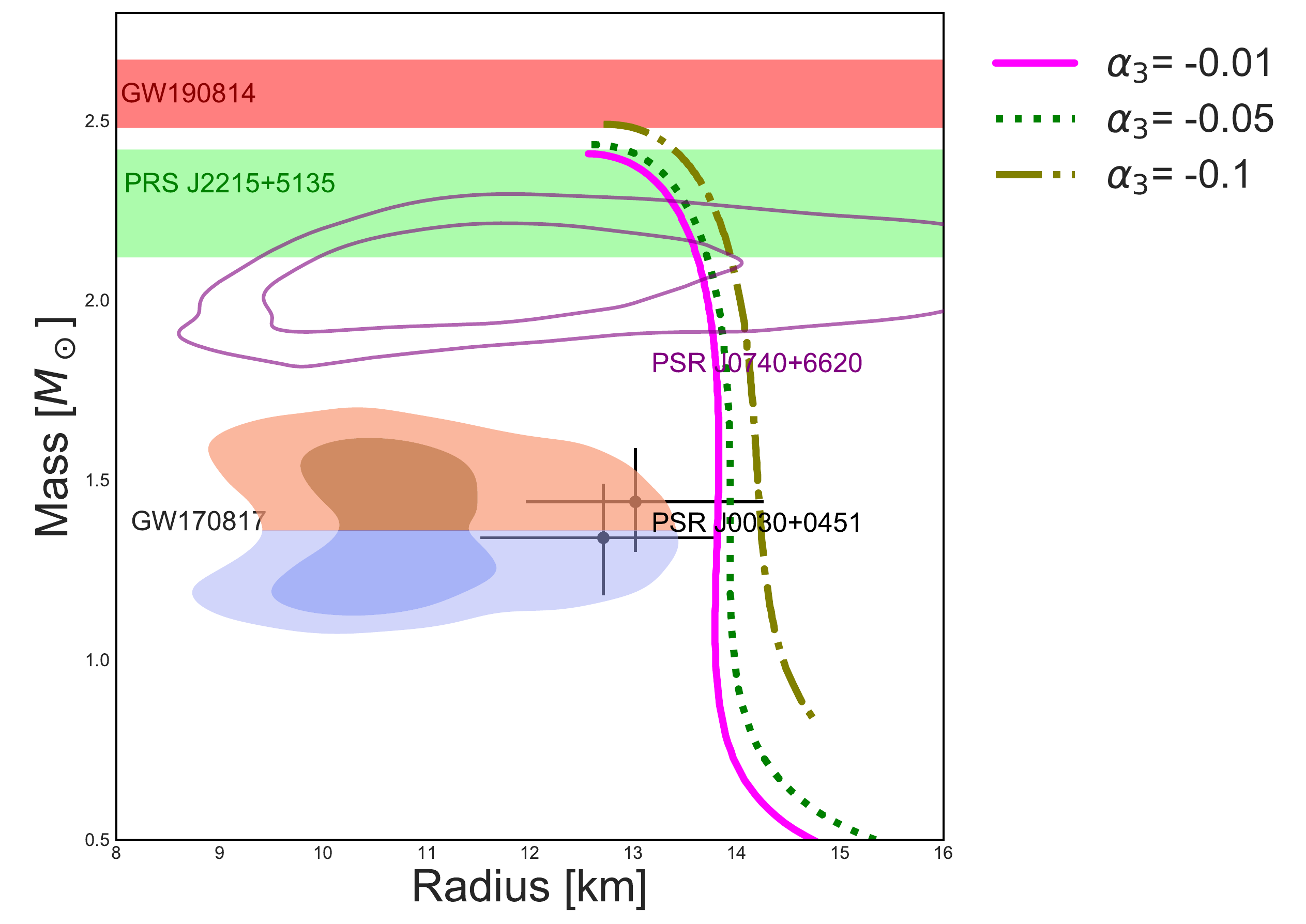}}
    \caption{The mass-radius
        relation of neutron stars in mimetic gravity for FSU2H EoS. $\protect\phi(0)=1$: left panels and $\protect%
        \phi(0)=0.8$: right panels and $K=0.5$.}
    \label{fsu2h_fig2}
\end{figure}

Also, we investigate the DD2 EoS in Fig. \ref{DD2_fig2} in
mimetic gravity. It is seen that this EoS can reach the GW190814
event for all potentials. However, for $V_1(\phi)$ model with
$\alpha_1=-0.1$, we observe that radii are large and the
mass-radius diagram cannot support the $GW170817$ region. As for
the final nucleonic EoS, we consider the FSU2H EoS in Fig.
\ref{fsu2h_fig2}. It can be seen that for $V_3(\phi)$, by
increasing $\alpha$, the maximum mass lies in the GW190814
observations while it could not reach this region in GR limit.
Furthermore, for all of potentials, the radii of this hadronic EoS
increase and they are not in the GW170817 region anymore.

\begin{figure}[!ht]
    \centering
    \subfloat[${V_1(\phi)=\frac{\alpha_1}{\phi^2}}$]{\includegraphics[width=0.4%
        \textwidth]{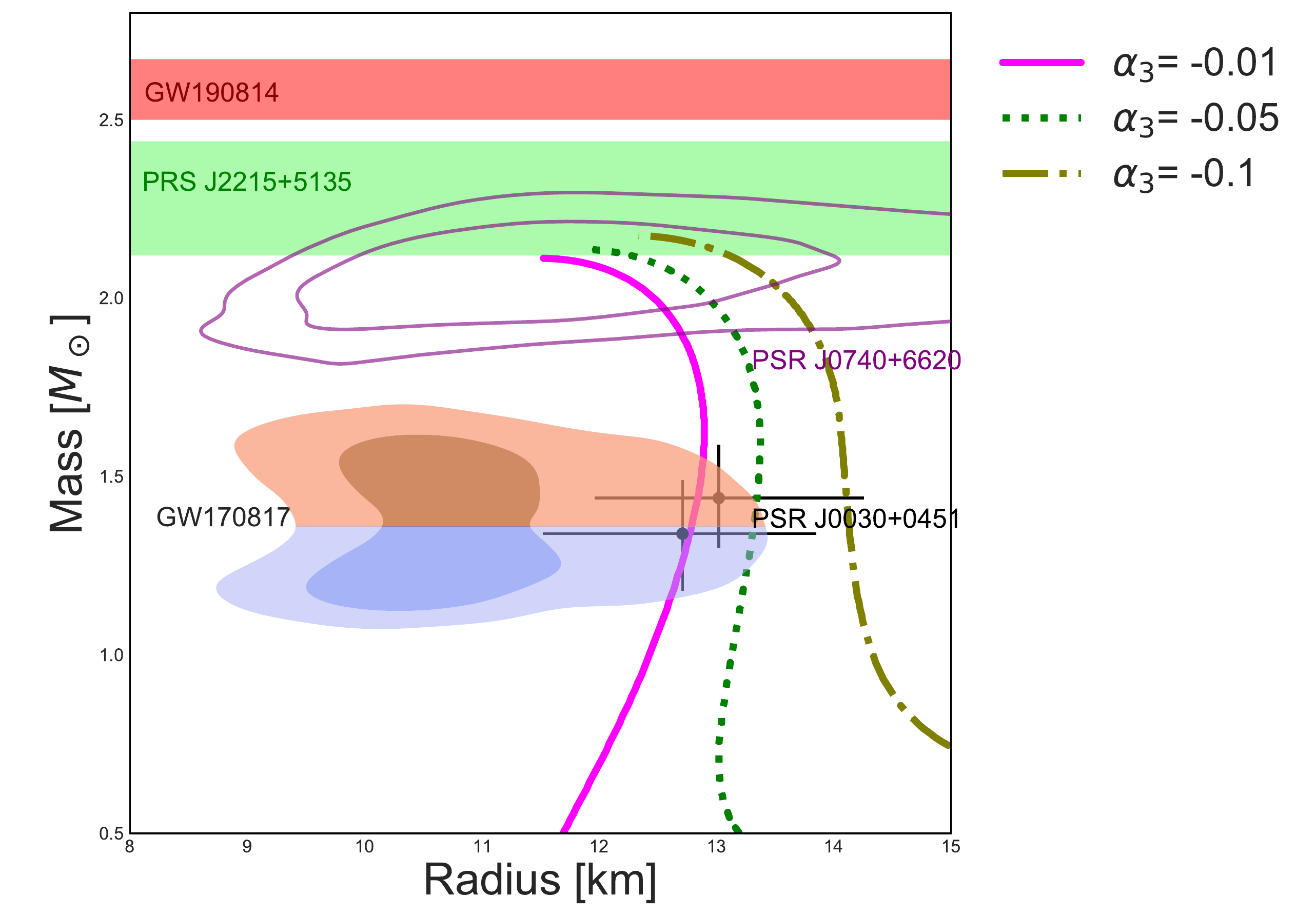}} \quad
    \subfloat[${V_1(\phi)=\frac{\alpha_1}{\phi^2}}$]{
        \includegraphics[width=0.4\textwidth]{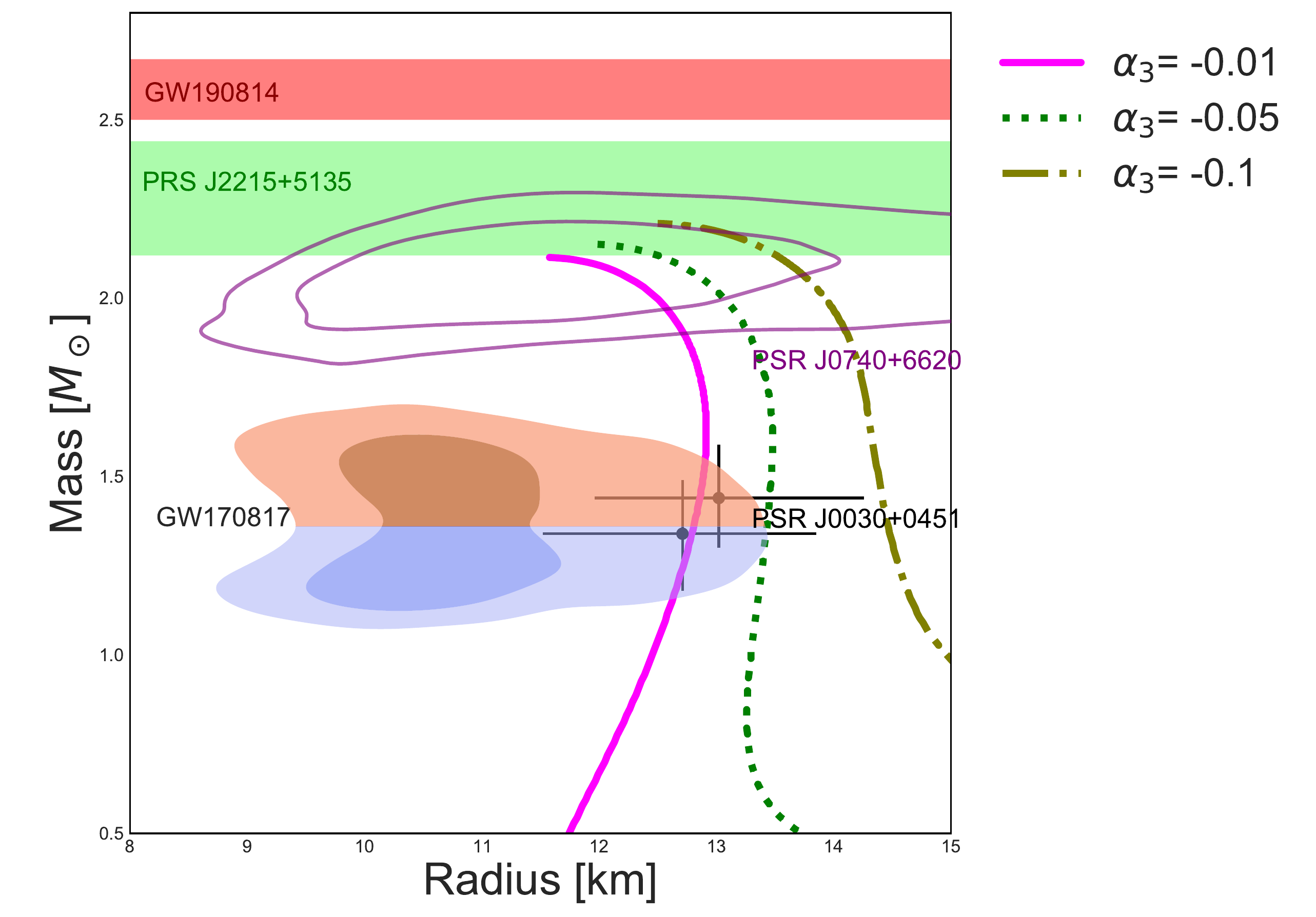}}
    \newline
    \subfloat[$V_2(\phi)=\frac{\alpha_2}{e^{K\phi^2}}$]{
        \includegraphics[width=0.4\textwidth]{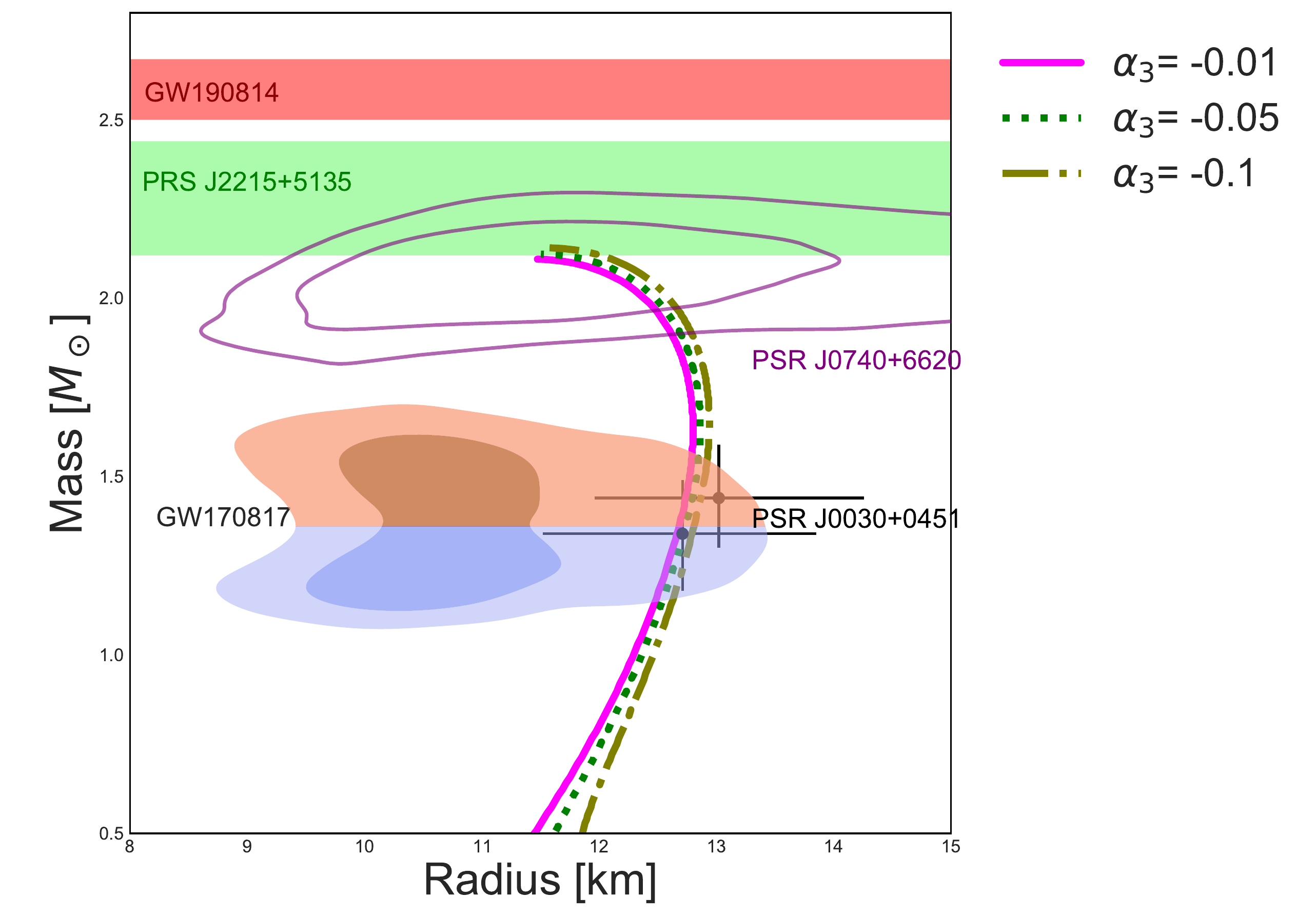}} \quad
    \subfloat[$V_2(\phi)=\frac{\alpha_2}{e^{K\phi^2}}$]{\includegraphics[width=0.4\textwidth]{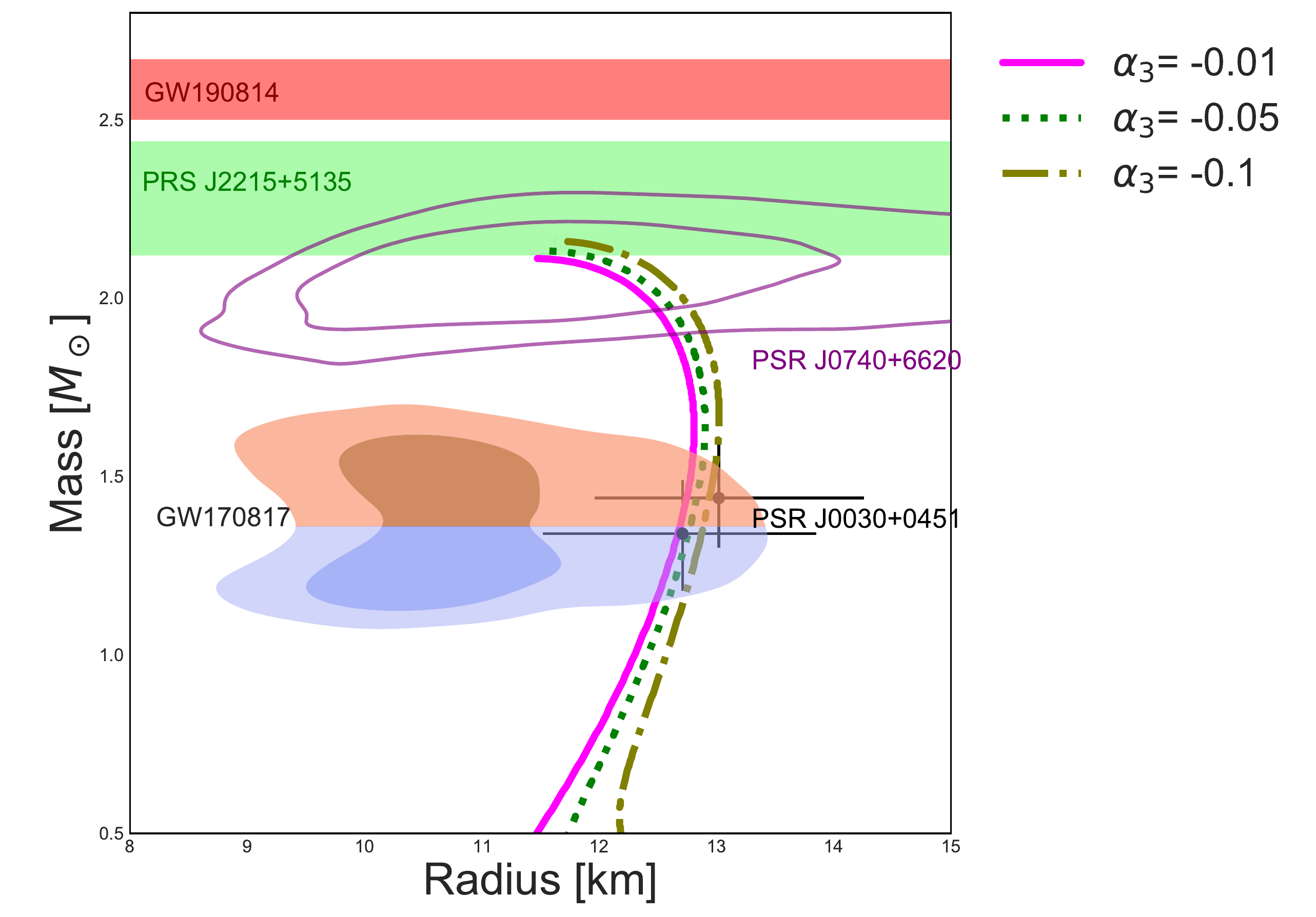}}
    \newline
    \subfloat[${V_3(\phi)=\frac{\alpha_3\phi^2}{1+e^{K\phi^2}}}$]{%
        \includegraphics[width=0.4\textwidth]{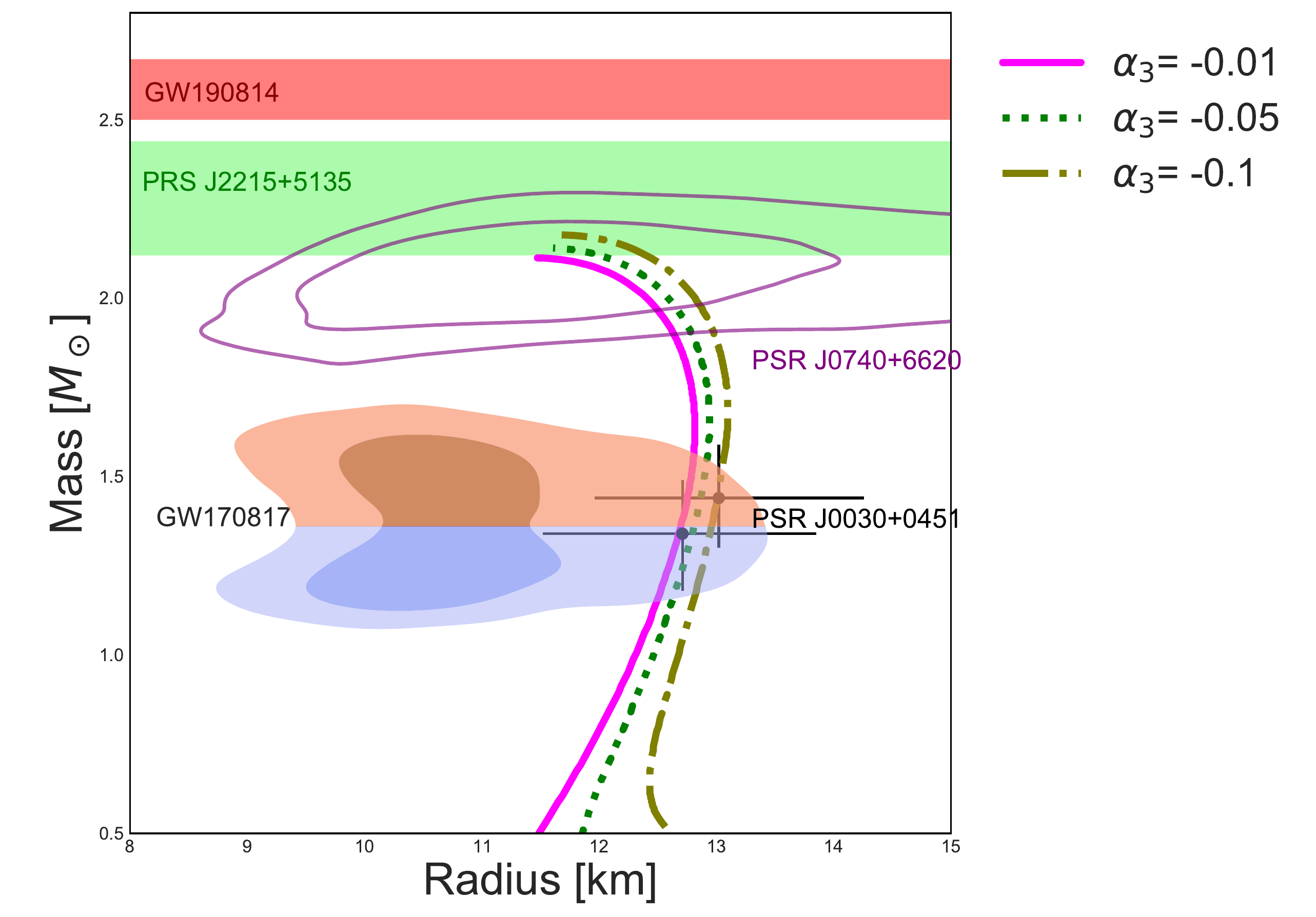} } \quad %
    \subfloat[${V_3(\phi)=\frac{\alpha_3\phi^2}{1+e^{K\phi^2}}}$]{%
        \includegraphics[width=0.4\textwidth]{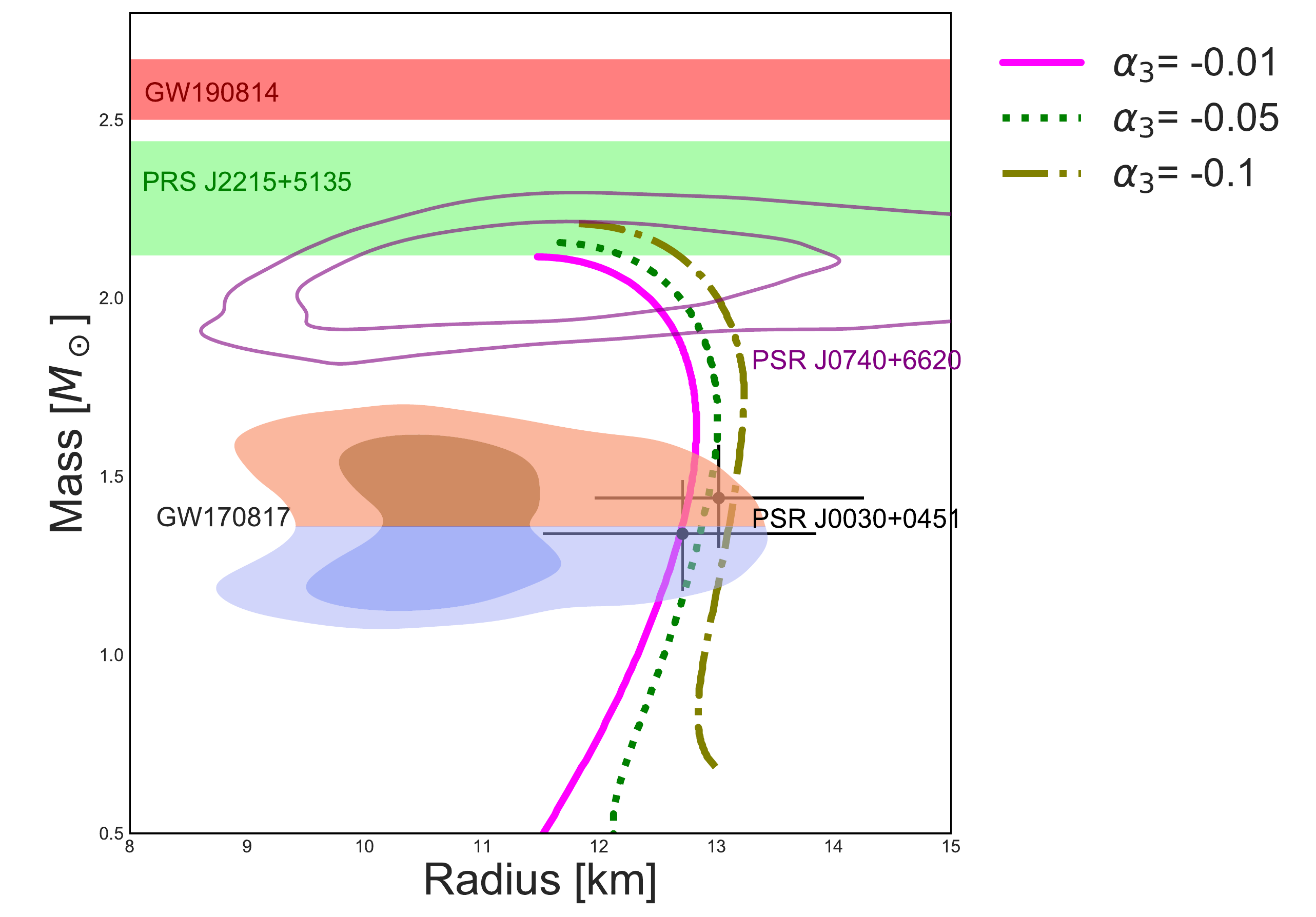}}
    \caption{The mass-radius
        relation of neutron stars in mimetic gravity for $BHB\Lambda\phi$ EoS. $\protect\phi(0)=1$: left panels and $\protect%
        \phi(0)=0.8$: right panels and $K=0.5$.}
    \label{BHB_fig2}
\end{figure}

\begin{figure}[!ht]
    \centering
    \subfloat[${V_1(\phi)=\frac{\alpha_1}{\phi^2}}$]{\includegraphics[width=0.4%
        \textwidth]{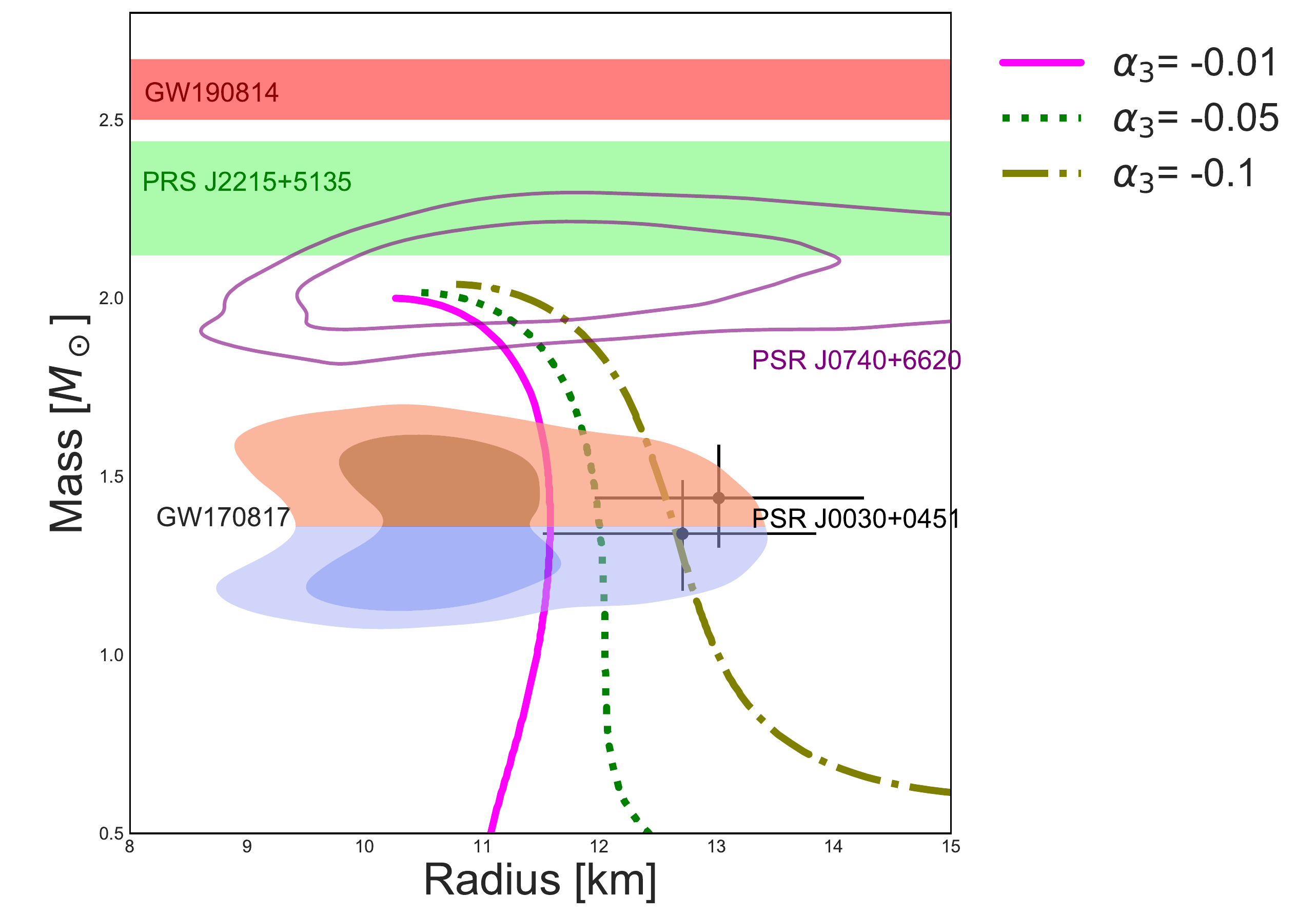}} \quad
    \subfloat[${V_1(\phi)=\frac{\alpha_1}{\phi^2}}$]{
        \includegraphics[width=0.4\textwidth]{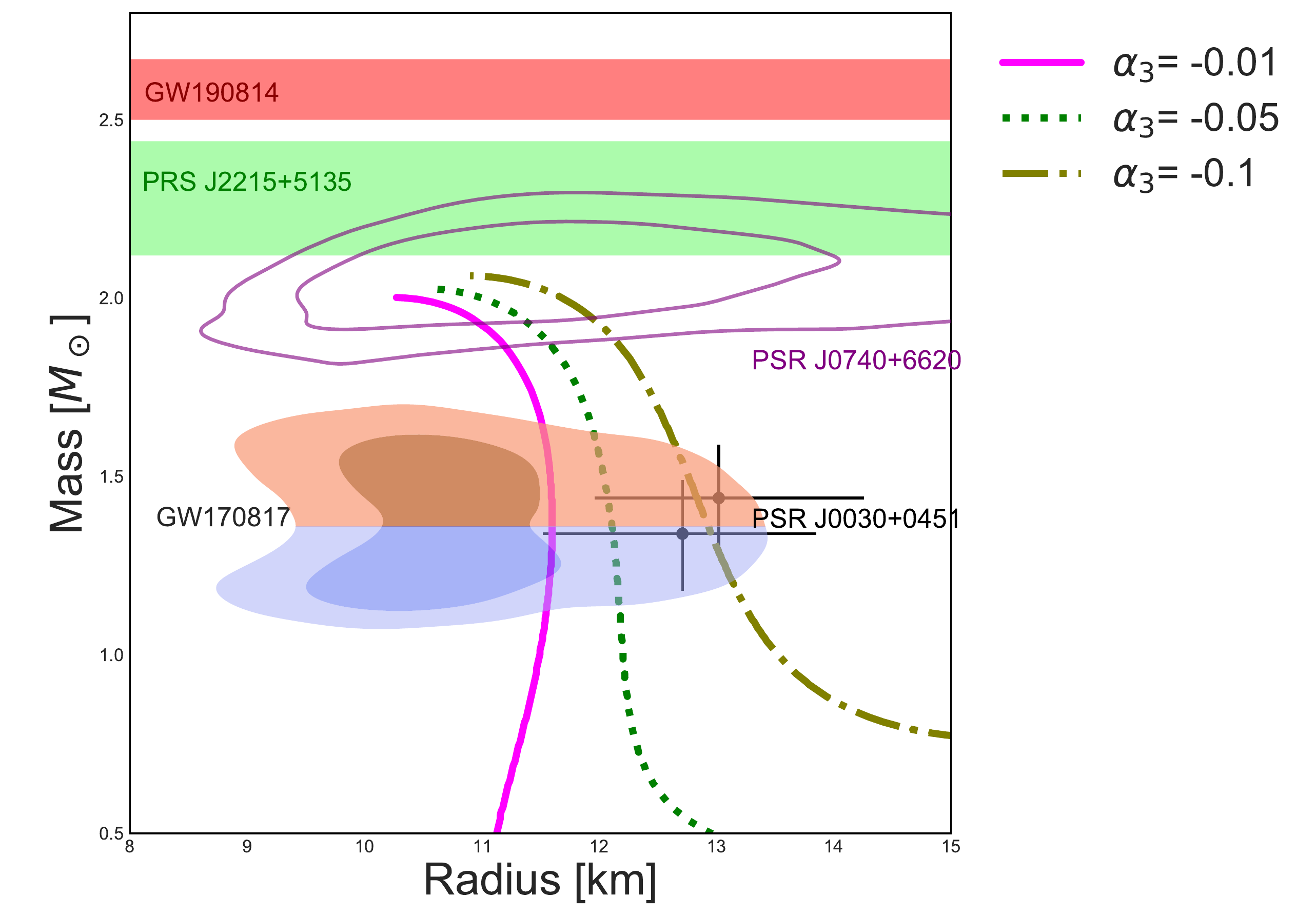}}
    \newline
    \subfloat[$V_2(\phi)=\frac{\alpha_2}{e^{K\phi^2}}$]{
        \includegraphics[width=0.4\textwidth]{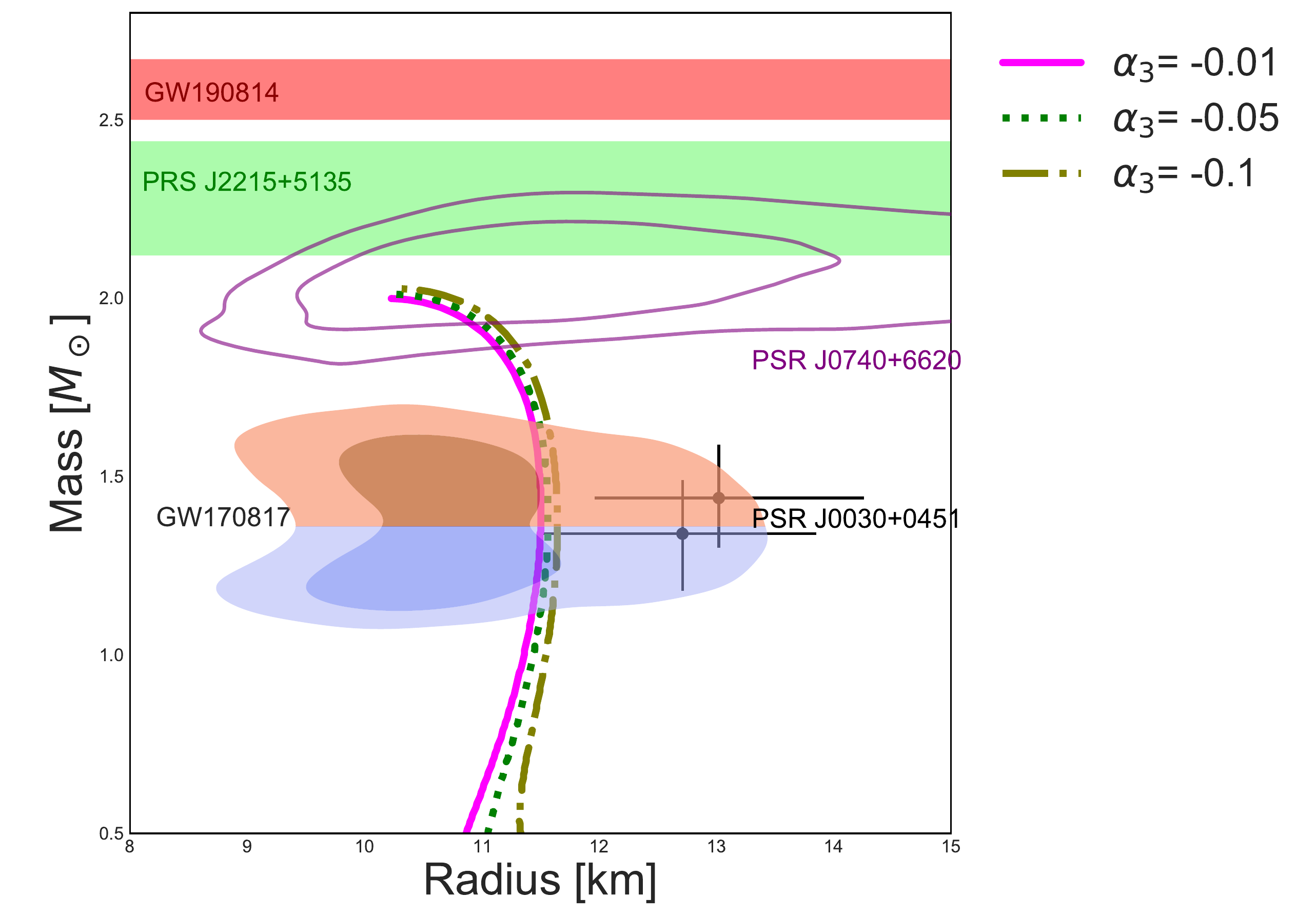}} \quad
    \subfloat[$V_2(\phi)=\frac{\alpha_2}{e^{K\phi^2}}$]{\includegraphics[width=0.4\textwidth]{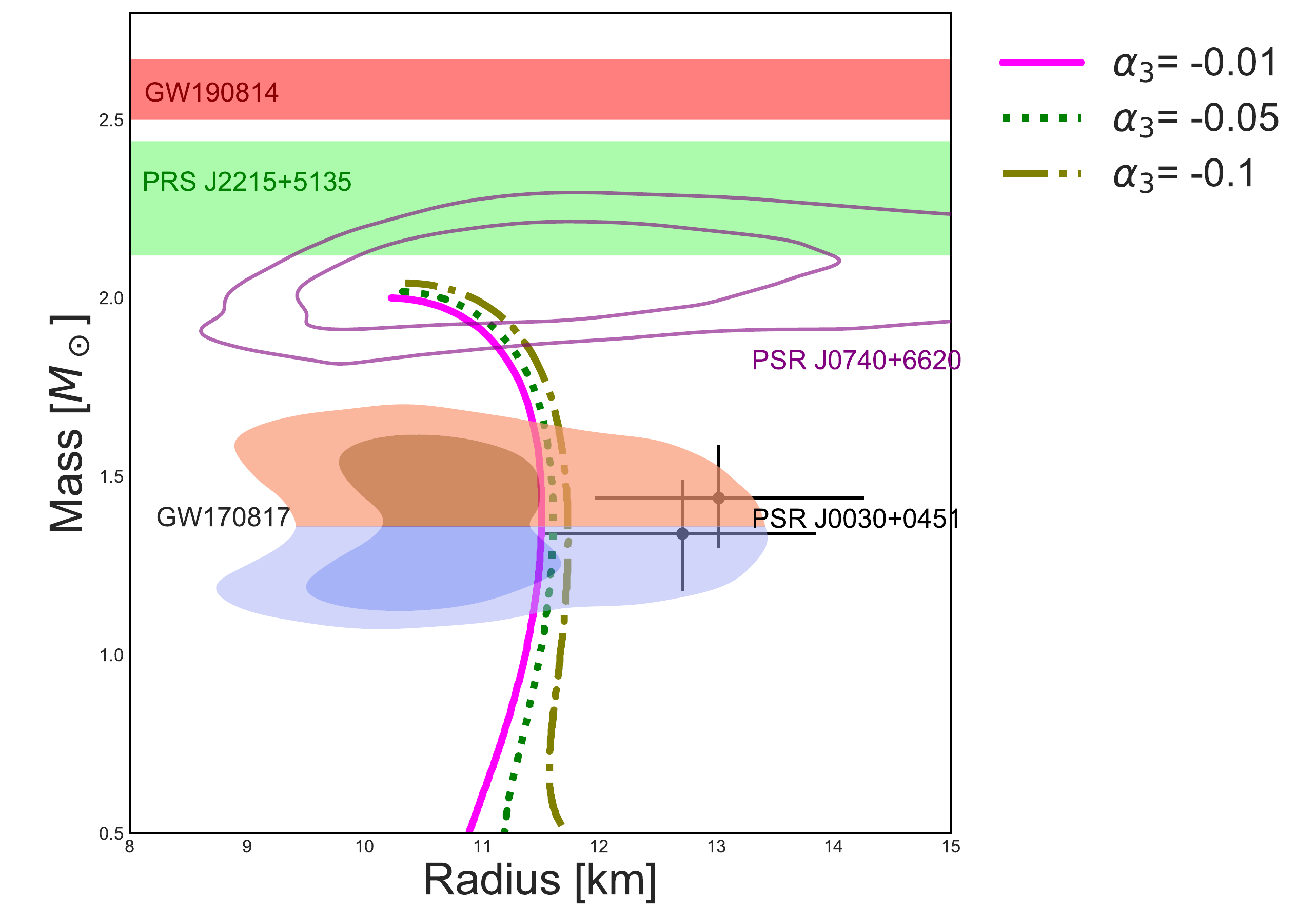}}
    \newline
    \subfloat[${V_3(\phi)=\frac{\alpha_3\phi^2}{1+e^{K\phi^2}}}$]{%
        \includegraphics[width=0.4\textwidth]{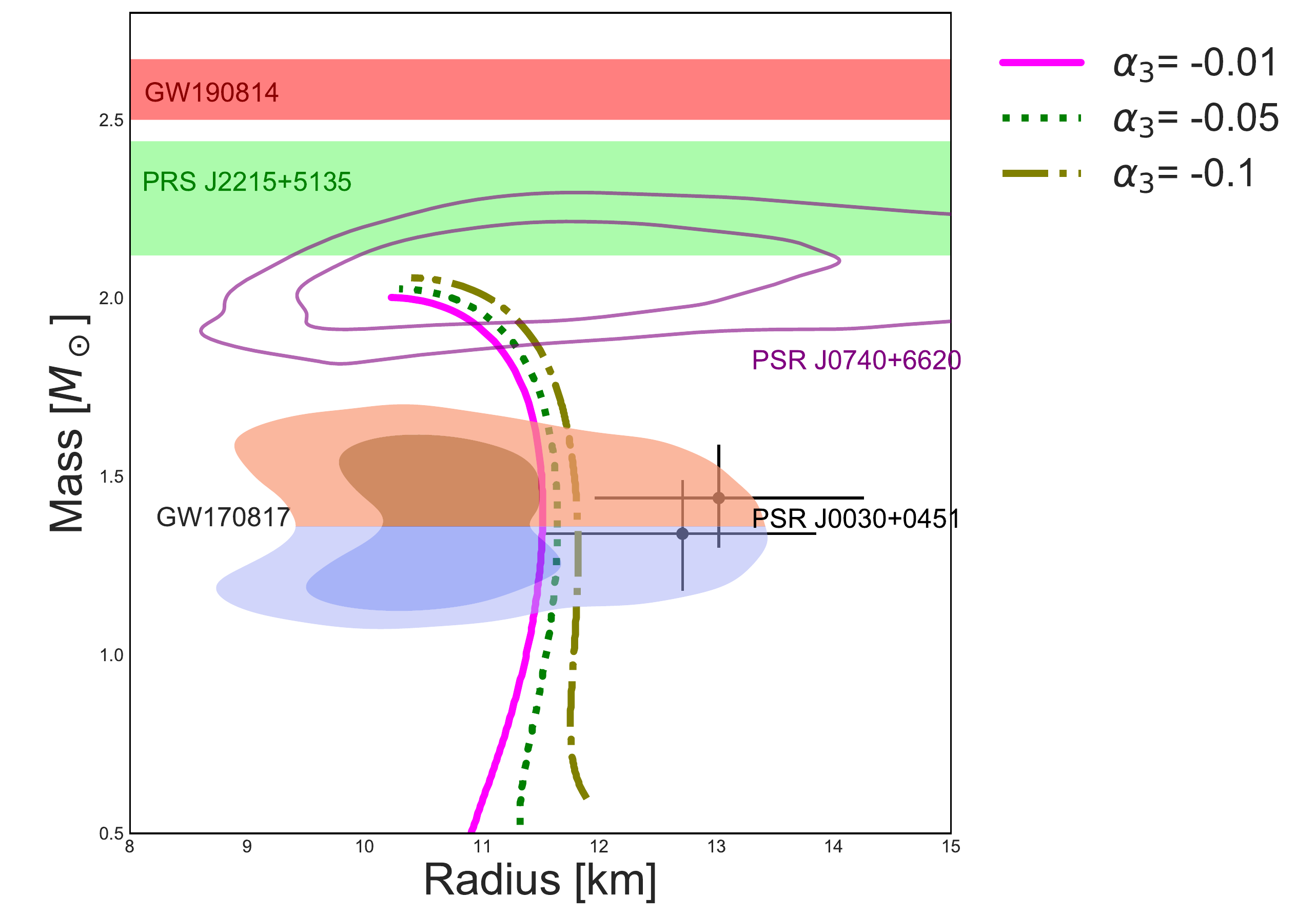} } \quad %
    \subfloat[${V_3(\phi)=\frac{\alpha_3\phi^2}{1+e^{K\phi^2}}}$]{%
        \includegraphics[width=0.4\textwidth]{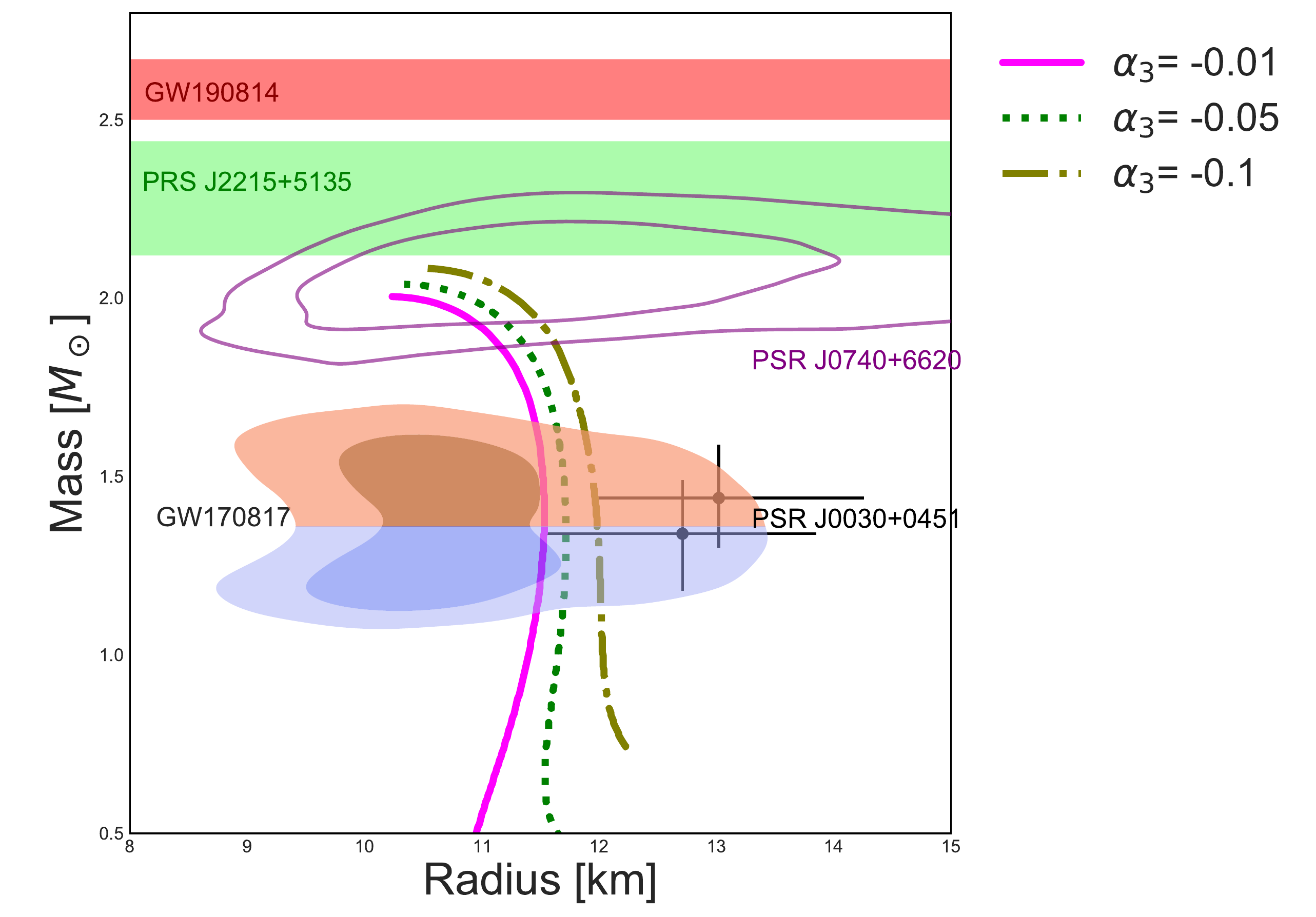}}
    \caption{The mass-radius
        relation of neutron stars in mimetic gravity for SFHoY EoS. $\protect\phi(0)=1$: left panels and $\protect%
        \phi(0)=0.8$: right panels and $K=0.5$.}
    \label{fshoy_fig2}
\end{figure}

In Fig. \ref{BHB_fig2} and \ref{fshoy_fig2}, the
mass-radius curves are exhibited for hadronic EoSs
$BHB\Lambda\Phi$ and SFHoY. We calculate maximum mass and radius
for different values of $\alpha$ and potentials in table
\ref{table_BHB}. According to Fig. \ref{sly4_fig0}, in the GR
limit, the maximum mass of star for $BHB\lambda\Phi$ could not
reach the PSR J2215+5135. However, according to Fig.
\ref{BHB_fig2}  we observe that the maximum mass is reached in
this region for all  of $V_1(\phi)$, $V_2(\phi)$ and $V_3(\phi)$
potentials.  But the radius for $V_1(\phi)$ is expanded and it is
out of GW170817 event. Nevertheless, for the $V_2(\phi)$ and
$V_3(\phi)$ potentials, the curves still lie in the GW170817
region.

The second hadronic EoS is SFHoY EoS. In Fig
\ref{fshoy_fig2}, it can be observed this EoS supports the first
or even second NICER error bars in mimetic gravity by increasing
$|\alpha|$. But in GR limit, it does not lie in this region. In
addition, for all of choice of $\alpha$ parameter and all
potentials $V_1(\phi)$, $V_2(\phi)$ and $V_3(\phi)$, the curves
support the GW170817 event.

%%%%%%%%%%%%%%%%%%%%%%%%%%%%%%%%%%%%%%%%%%%%%%%%%%%%%%%%%%%%%%%%%%%%%%%%%%%%%%
\begin{table*}[tbp]
     \caption{$M_{max}$ and corresponding radius for \emph{hadronic EoSs} with $K=0.5$, $\protect\phi(0)=1$ \tcb{$(\protect\phi(0)=0.8)$} and various  $\alpha_i (i=1..3)$ .}\centering
    %{\footnotesize \
    %\resizebox{.75\hsize}{!}{
    %        \tiny
    \begin{tabular}{|@{}c|c|c|c|c|c|c@{}|}
        \noalign{\hrule height.5pt} \hline
        \textbf{$BHB\Lambda\Phi$}& \multicolumn{2}{c|}{$V_1(\phi)=\frac{\alpha_1}{\phi^2}$}& \multicolumn{2}{c|}{$V_2(\phi)=\frac{\alpha_2}{e^{K\phi^2}}$}& \multicolumn{2}{c|}{$V_3(\phi)=\frac{\alpha_3\phi^2}{1+e^{K\phi^2}}$} \\
        \noalign{\hrule height 1pt}
        $\phi(0)=1$ \tcb{$(0.8)$}& ${M_{max}}\ (M_{\odot})$ &$R\ (km)$ & ${M_{max}}\ (M_{\odot})$  & $R\ (km)$ & ${M_{max}}\ (M_{\odot})$  &$R\ (km)$ \\ \hline
        \noalign{\hrule height 1pt}
        $\alpha_i=-0.01$ & $2.11 \tcb{(2.12)}$ & $11.52 \tcb{(11.58)}$ &  $2.11 \tcb{(2.11)}$ & $11.47 \tcb{(11.47)}$ & $2.11 \tcb{(2.12)}$ & $11.47 \tcb{(11.47)}$  \\
        $\alpha_i=-0.05$ & $2.13 \tcb{(2.15)}$ & $11.86 \tcb{(11.93)}$ &  $2.12 \tcb{(2.13)}$ & $11.50 \tcb{(11.55)}$ & $2.14 \tcb{(2.16)}$ & $11.61 \tcb{(11.61)}$   \\
        $\alpha_i=-0.10$ & $2.17 \tcb{(2.21)}$ & $12.34 \tcb{(12.05)}$ &  $2.14 \tcb{(2.16)}$ & $11.58 \tcb{(11.62)}$ & $2.18 \tcb{(2.22)}$ & $11.68 \tcb{(11.81)}$    \\
        \noalign{\hrule height .5pt}\hline
    \end{tabular}\label{table_BHB}
    \begin{tabular}{|@{}c|c|c|c|c|c|c@{}|}
        \noalign{\hrule height.5pt} \hline
        \textbf{SFHoY}& \multicolumn{2}{c|}{$V_1(\phi)=\frac{\alpha_1}{\phi^2}$}& \multicolumn{2}{c|}{$V_2(\phi)=\frac{\alpha_2}{e^{K\phi^2}}$}& \multicolumn{2}{c|}{$V_3(\phi)=\frac{\alpha_3\phi^2}{1+e^{K\phi^2}}$} \\
        \noalign{\hrule height 1pt}
        $\phi(0)=1$ \tcb{$(0.8)$}& ${M_{max}}\ (M_{\odot})$ &$R\ (km)$ & ${M_{max}}\ (M_{\odot})$  & $R\ (km)$ & ${M_{max}}\ (M_{\odot})$  &$R\ (km)$ \\ \hline
        \noalign{\hrule height 1pt}
        $\alpha_i=-0.01$ & $2.00 \tcb{(2.00)}$ & $10.27 \tcb{(10.27)}$ &  $2.00 \tcb{(2.00)}$ & $10.23 \tcb{(10.23)}$ & $ 2.00 \tcb{(2.00)}$ & $ 10.23 \tcb{(10.23)}$  \\
        $\alpha_i=-0.05$ & $2.02 \tcb{(2.03)}$ & $10.48 \tcb{(10.53)}$ &  $2.01 \tcb{(2.02)}$ & $10.26 \tcb{(10.29)}$ & $ 2.03 \tcb{(2.04)}$ & $ 10.29 \tcb{(10.34)}$   \\
        $\alpha_i=-0.10$ & $2.04 \tcb{(2.06)}$ & $10.78 \tcb{(10.90)}$ &  $2.03 \tcb{(2.04)}$ & $10.32 \tcb{(10.34)}$ & $ 2.06 \tcb{(2.08)}$ & $ 10.40 \tcb{(10.49)}$    \\
        \noalign{\hrule height .5pt}\hline
    \end{tabular}\label{table_fshoy}
\end{table*}

%%%%%%%%%%%%%%%%%%%%%%%%%%%%%%%%%%%%%%%%%%%%%%%%%%%%%%%%%%%%%%%%%%%%%%%%%%%%%%%%%

In order to compare the effectiveness of different EoSs, we
collect the results of GR and mimetic gravity in table \ref{tab1}.
In general, we find that mimetic gravity can improve the results
and it can support more observational evidence than GR case.
Besides, it is evident that MPA1, DD2 and FSU2H EoSs in the
context of mimetic gravity can support the mentioned observational
data as indicated in table \ref{tab1}. According to this
interesting result, we focus on these three EoSs to drive other
physical properties of neutron stars such as Schwarzschild radius,
redshift and compactness.
\begin{table}[!ht]
    \caption{The different observation data for the different EoSs in GR and mimetic gravity.}
    \centering
    {\footnotesize \
        \resizebox{.95\hsize}{!}{
            \begin{tabular}{|@{}c|c|c|c|c|c|c@{}|}
                \hline \noalign{\hrule height .5pt} \noalign{\hrule height 1pt}
                GR & GW170817 & GW190814& PSR J0740+6620 & PSR J2215+5135 & PSR J0030+04511 & PSR J0030+04511 \\
                && & &&(the first error bar)&(the second error bar)\\ \hline
                SLy4  &    \checkmark   & $\times$ &  \checkmark & $\times$ & \checkmark &$\times$ \\ \hline
                MPA1  & \checkmark & $\times$ &  \checkmark & \checkmark & \checkmark  & \checkmark  \\ \hline
                BSK21 &\checkmark& $\times$ &  \checkmark &\checkmark&\checkmark & \checkmark   \\ \hline
                WFF1 & \checkmark & $\times$  &  \checkmark  &$\times$ &$\times$  &$\times$   \\ \hline
                DD2  & \checkmark & $\times$  &  \checkmark  & \checkmark &\checkmark &\checkmark   \\ \hline
                $BHB\Lambda\Phi$  & \checkmark & $\times$  & \checkmark   & $\times$ &\checkmark &\checkmark   \\ \hline
                FSU2H  & \checkmark & $\times$  &  \checkmark  & \checkmark &\checkmark &\checkmark  \\ \hline
                SFHoY  & \checkmark & $\times$  &  \checkmark  &  $\times$ &\checkmark & $\times$  \\\hline\hline  \noalign{\hrule height .5pt}
                Mimetic gravity& GW170817 & GW190814 & PSR J0740+6620 & PSR J2215+5135 & PSR J0030+0451 & PSR J0030+0451 \\
                && & &&(the first error bar)&(the second error bar)\\  \hline
                SLy4  &    \checkmark   & $\times$ &  $  \checkmark$ & \checkmark &  \checkmark& \checkmark \\ \hline
                MPA1  & \checkmark & \checkmark &  \checkmark & \checkmark & \checkmark& \checkmark  \\ \hline
                BSK21  &\checkmark& $\times$ &  \checkmark &\checkmark&\checkmark &\checkmark  \\ \hline
                WFF1  & \checkmark & $\times$  &  \checkmark  & \checkmark &\checkmark &\checkmark   \\ \hline
                DD2  & \checkmark & \checkmark  &  \checkmark  & \checkmark &\checkmark &\checkmark   \\ \hline
                $BHB\Lambda\Phi$  & \checkmark & $\times$  & \checkmark   & $\times$ &\checkmark &\checkmark   \\ \hline
                FSU2H  & \checkmark  & \checkmark  &  \checkmark  & \checkmark &\checkmark  &\checkmark   \\ \hline
                SFHoY  & \checkmark & $\times$  &  \checkmark  &  $\times$ &\checkmark & \checkmark  \\\hline
                \noalign{\hrule height .5pt}
            \end{tabular}\label{tab1}} }
\end{table}

\textbf{Schwarzschild radius}: considering the obtained function
$g(r)$ in Eq. (\ref{g(r)}), and using the horizon radius
constraint ($\left. g\left(
r\right) \right\vert _{r=R}=0$), we can extract the Schwarzschild radius $%
\left( R_{Sch}\right) $. It is clear that by applying the mimetic scalar
field (Eq. (\ref{Phi})), and using $\left. g\left( r\right) \right\vert
_{r=R}=0$, the mentioned scalar potentials become zero on surface of neutron
star, i.e., $\left. V(\phi )\right\vert _{r=R}=0$. Therefore, we obtain the
Schwarzschild radius for the mimetic gravity (which is the same GR gravity)
as
\begin{equation}
R_{Sch}=\frac{2GM}{c^{2}}.  \label{Rsch}
\end{equation}%

%%%%%%%%%%%%%%%%%%%%%%%%%%%%%%%%%%%%%%%%%%%%%%%%%%%%%%%%%%%%%%%%%%%%%%%
%%%%%%%%%%%%%%%%%%%%%%%%%%%%%%%%%%%%%%%%%%%%%%%%%%%%%%%%%
\begin{table*}[!ht]
\caption{Structure properties of neutron star in mimetic gravity
for MPA1 EoS with $\protect\phi(0)=1$
\tcb{($\protect\phi(0)=0.8$)} and $K=0.5$.} \label{tabPhysP1}
\begin{center}
\begin{tabular}{|c|c|c|c|c|c|}
\hline
$V_{1}(\phi)=\frac{\alpha_1}{\phi^2}$ & ${M_{max}}\ (M_{\odot})$ & $R\ (km)$ & $%
R_{Sch}\ (km)$ & $\sigma (10^{-1})$ & $z(10^{-1})$ \\ \hline
$\alpha_1=-0.01$ & $2.50 \tcb{(2.49)}$ & $11.39 \tcb{(11.40)}$ &$7.37 \tcb{(7.34)}$ & $3.24 \tcb{(3.22)}$ & $6.83 \tcb{(6.76)}$ \\
$\alpha_1=-0.05$ & $2.51 \tcb{(2.52)}$ & $11.57 \tcb{(11.63)}$ & $7.40 \tcb{(7.43)}$ & $3.20 \tcb{(3.19)}$ & $6.66 \tcb{(6.64)}$ \\
$\alpha_1=-0.10$ & $2.53 \tcb{(2.55)}$ & $11.83 \tcb{(11.90)}$ & $7.46 \tcb{(7.52)}$ & $3.16 \tcb{(3.16)}$ & $6.45 \tcb{(6.48)}$ \\
 \hline\hline$V_{2}(\phi)=\frac{\alpha_2}{e^{K\phi^2}}$ & ${M_{max}}\ (M_{\odot})$ & $R\ (km)$ & $R_{Sch}\
(km)$ & $\sigma (10^{-1})$ & $z(10^{-1})$ \\ \hline
$\alpha_2=-0.01$ & $2.49 \tcb{(2.49)}$ & $11.36 \tcb{(11.36)}$ & $7.34 \tcb{(7.34)}$ & $3.23 \tcb{(3.23)}$ & $6.81 \tcb{(6.81)}$ \\
$\alpha_2=-0.05$ & $2.50 \tcb{(2.50)}$ & $11.39 \tcb{(11.39)}$ & $7.37 \tcb{(7.37)}$ & $3.24 \tcb{(3.24)}$ & $6.83 \tcb{(6.83)}$ \\
$\alpha_2=-0.10$ & $2.51 \tcb{(2.52)}$ & $11.42 \tcb{(11.42)}$ &$7.40 \tcb{(7.43)}$ & $3.24 \tcb{(3.25)}$ & $6.85 \tcb{(6.92)}$ \\
 \hline\hline$V_{3}(\phi)=\frac{\alpha_3\phi^2}{1+e^{K\phi^2}}$ & ${M_{max}}\ (M_{\odot})$ & $R\
(km) $ & $R_{Sch}\ (km)$ & $\sigma (10^{-1})$ & $z(10^{-1})$ \\
\hline
$\alpha_3=-0.01$ & $2.49 \tcb{(2.50)}$ & $11.36 \tcb{(11.40)}$ &$7.34 \tcb{(7.37)}$ & $3.23 \tcb{(3.23)}$ & $6.81 \tcb{(6.82)}$ \\
$\alpha_3=-0.05$ & $2.51 \tcb{(2.52)}$ & $11.43 \tcb{(11.44)}$ & $7.40 \tcb{(7.43)}$ & $3.24 \tcb{(3.25)}$ & $6.84 \tcb{(6.89)}$ \\
$\alpha_3=-0.10$ & $2.54 \tcb{(2.56)}$ & $11.51 \tcb{(11.56)}$ &
$7.49 \tcb{(7.55)}$ & $3.26 \tcb{(3.26)}$ & $6.92 \tcb{(6.98)}$ \\
\hline
\end{tabular}
\end{center}
\end{table*}

\begin{table*}[!ht]
\caption{Structure properties of neutron star in mimetic gravity
for DD2 EoS with $\protect\phi(0)=1$ \tcb{($\protect\phi(0)=0.8$)}
and $K=0.5$.} \label{tabPhysP3}
\begin{center}
\begin{tabular}{|c|c|c|c|c|c|}
\hline
$V_{1}(\phi)=\frac{\alpha_1}{\phi^2}$ & ${M_{max}}\ (M_{\odot})$ & $R\ (km)$ & $%
R_{Sch}\ (km)$ & $\sigma (10^{-1})$ & $z(10^{-1})$ \\ \hline
$\alpha_1=-0.01$ & $2.42 \tcb{(2.43)} $ & $11.93 \tcb{(11.93)}$ &$7.13 \tcb{(7.16)}$ & $2.99 \tcb{(3.00)}$ & $5.78 \tcb{(5.82)}$ \\
$\alpha_1=-0.05$ & $2.44 \tcb{(2.46)}$ & $12.18 \tcb{(12.20)}$ & $7.19 \tcb{(7.25)}$ & $2.95 \tcb{(2.97)}$ & $5.63 \tcb{(5.70)}$ \\
$\alpha_1=-0.10$ & $2.47 \tcb{(2.51)}$ & $12.38 \tcb{(12.52)}$ &
$7.28 \tcb{(7.40)}$ & $2.94 \tcb{(2.96)}$ & $5.58 \tcb{(5.64)}$ \\
\hline\hline $V_{2}(\phi)=\frac{\alpha_1}{e^{K\phi^2}}$ &
${M_{max}}\ (M_{\odot})$ & $R\ (km)$ & $R_{Sch}\ (km)$ & $\sigma
(10^{-1})$ & $z(10^{-1})$ \\ \hline
$\alpha_2=-0.01$ & $2.42 \tcb{(2.42)}$ & $11.89 \tcb{(11.89)}$ &  $7.13 \tcb{(7.13)}$ & $3.00 \tcb{(3.00)}$ & $5.81 \tcb{(5.81)}$ \\
$\alpha_2=-0.05$ & $2.43 \tcb{(2.44)}$ & $11.88 \tcb{(11.87)}$ &$7.16 \tcb{(7.19)}$ & $3.01 \tcb{(3.03)}$ & $5.87 \tcb{(5.93)}$ \\
$\alpha_2=-0.10$ & $2.44 \tcb{(2.45)}$ & $11.95 \tcb{(11.95)}$
&$7.19 \tcb{(7.22)}$ & $3.01 \tcb{(3.02)}$ & $5.85 \tcb{(5.90)}$
\\ \hline\hline $V_{3}(\phi)=\frac{\alpha_3\phi^2}{1+e^{K\phi^2}}$
& ${M_{max}}\ (M_{\odot})$ & $R\ (km) $ & $R_{Sch}\ (km)$ &
$\sigma (10^{-1})$ & $z(10^{-1})$ \\ \hline
$\alpha_3=-0.01$ & $2.43 \tcb{(2.44)}$ & $11.89 \tcb{(11.89)}$ & $7.16 \tcb{(7.19)}$ & $3.01 \tcb{(3.03)}$ & $5.86 \tcb{(5.91)}$ \\
$\alpha_3=-0.05$ & $2.45 \tcb{(2.47)}$ & $11.97 \tcb{(11.97)}$ & $7.22 \tcb{(7.28)}$ & $3.02 \tcb{(3.04)}$ & $5.88 \tcb{(5.98)}$ \\
$\alpha_3=-0.10$ & $2.48 \tcb{(2.52)}$ & $11.94 \tcb{(12.05)}$ &
$7.31 \tcb{(7.43)}$ & $3.06 \tcb{(3.08)}$ & $6.06 \tcb{(6.15)}$ \\
\hline
\end{tabular}
\end{center}
\end{table*}
\begin{table*}[!ht]
\caption{Structure properties of neutron star in mimetic gravity
for FSU2H EoS with $\protect\phi(0)=1$
\tcb{($\protect\phi(0)=0.8$)} and $K=0.5$.} \label{tabPhysP5}
\begin{center}
\begin{tabular}{|c|c|c|c|c|c|}
\hline
$V_{1}(\phi)=\frac{\alpha_1}{\phi^2}$ & ${M_{max}}\ (M_{\odot})$ & $R\ (km)$ & $%
R_{Sch}\ (km)$ & $\sigma (10^{-1})$ & $z(10^{-1})$ \\ \hline
$\alpha_1=-0.01$ & $2.40 \tcb{(2.39)}$ & $12.57 \tcb{(12.58)}$ &  $7.08 \tcb{(7.05)}$ & $2.81 \tcb{(2.80)}$ & $5.13 \tcb{(5.08)}$ \\
$\alpha_1=-0.05$ & $2.42 \tcb{(2.43)}$ & $12.81 \tcb{(12.87)}$ & $7.13 \tcb{(7.16)}$ & $2.78 \tcb{(2.79)}$ & $5.02 \tcb{(5.02)}$ \\
$\alpha_1=-0.10$ & $2.45 \tcb{(2.49)}$ & $13.15 \tcb{(13.30)}$ &
$7.22 \tcb{(7.34)}$ & $2.75 \tcb{(2.76)}$ & $4.90 \tcb{(4.94)}$ \\
\hline\hline $V_{2}(\phi)=\frac{\alpha_1}{e^{K\phi^2}}$ &
${M_{max}}\ (M_{\odot})$ & $R\ (km)$ & $R_{Sch}\ (km)$ & $\sigma
(10^{-1})$ & $z(10^{-1})$ \\ \hline
$\alpha_2=-0.01$ & $2.39 \tcb{(2.39)}$ & $12.53 \tcb{(12.52)}$ & $7.05 \tcb{(7.05)}$ & $2.81 \tcb{(2.81)}$ & $5.12 \tcb{(5.12)}$ \\
$\alpha_2=-0.05$ & $2.40 \tcb{(2.40)}$ & $12.51 \tcb{(12.55)}$ &$7.08 \tcb{(7.08)}$ & $2.83 \tcb{(2.82)}$ & $5.17 \tcb{(5.14)}$ \\
$\alpha_2=-0.10$ & $2.41 \tcb{(2.43)}$ & $12.59 \tcb{(12.57)}$ &
$7.11 \tcb{(7.16)}$ & $2.82 \tcb{(2.85)}$ & $5.15 \tcb{(5.25)}$ \\
\hline\hline $V_{3}(\phi)=\frac{\alpha_3\phi^2}{1+e^{K\phi^2}}$ &
${M_{max}}\ (M_{\odot})$ & $R\ (km) $ & $R_{Sch}\ (km)$ & $\sigma
(10^{-1})$ & $z(10^{-1})$ \\ \hline
$\alpha_3=-0.01$ & $2.40 \tcb{(2.41)}$ & $12.56 \tcb{(12.56)}$ &  $7.08 \tcb{(7.11)}$ & $2.82 \tcb{(2.83)}$ & $5.13 \tcb{(5.17)}$ \\
$\alpha_3=-0.05$ & $2.42 \tcb{(2.43)}$ & $12.59 \tcb{(12.59)}$ & $7.13 \tcb{(7.16)}$ & $2.83 \tcb{(2.85)}$ & $5.19 \tcb{(5.23)}$ \\
$\alpha_3=-0.10$ & $2.48 \tcb{(2.51)}$ & $12.66 \tcb{(12.71)}$ &
$7.31 \tcb{(7.40)}$ & $2.89 \tcb{(2.91)}$ & $5.38 \tcb{(5.47)}$ \\
\hline
\end{tabular}
\end{center}
\end{table*}

As one can see in the equation (\ref{Rsch}) and tables
\ref{tabPhysP1}-\ref{tabPhysP5}, by increasing maximum mass of
neutron star, the Schwarzschild radius increases.

\textbf{Gravitational redshift and compactness:} other important quantities
in which we intend to extract them, are related to the gravitational
redshift and compactness of neutron stars in the mimetic gravity. These
quantities may give us information about the strength of gravity of neutron
star.

We use from definition of the gravitational redshift and this fact that the
mentioned scalar potentials are zero on surface of neutron star, we extract
the gravitational redshift as
\begin{equation}
z=\frac{1}{\sqrt{1-\frac{2GM}{c^{2}R}}}-1,  \label{z}
\end{equation}%
where $R$ is radius of neutron star.

The compactness of a spherical compact object may be defined by the ratio of
the mass to radius of that compact object
\begin{equation}
\sigma =\frac{GM}{c^2R}.  \label{sigma}
\end{equation}

Now, we are in a position to investigate the gravitational
redshift and compactness of the obtained neutron stars of MPA1 EoS
in the mimetic gravity. Considering Eqs. (\ref{z}), (\ref{sigma}),
and the obtained results in table \ref{table_MPA1_1}, we find an
interesting property of neutron stars. The gravitational redshift
and compactness of neutron star in the mimetic gravity are
dependent on the scalar potential models. In other words, there
are two different manners for these quantities. They decrease by
increasing the maximum mass of neutron star in $V_{1}(\phi)$
model. However, they increase by increasing the maximum mass for
$V_{2}(\phi)$ and $V_{3}(\phi)$ models (see table
\ref{tabPhysP1}, for more details). In addition, there are the
same behavior for DD2 and FSU2H EoSs (see tables
\ref{tabPhysP3}-\ref{tabPhysP5}, for more details). Therefore the
gravitational redshift and compactness of neutron stars in the
mimetic gravity are completely depending on the scalar potential
models.

\section{Conclusion}

The study of neutron stars, one of the possible endpoints of
stellar evolution, is an interesting hot topic in physics
communities since they provide a natural laboratory for the
investigation of high density and pressure medium. The neutron
star structure can be considered from theoretical and
observational perspectives. In recent years, different properties
of diverse neutron stars are accessible from the observational
evidence, such as GW170817, PSR J0740+6620, PSR J2215+5135, NICER
data on PSR J0030+0451 and GW190814. However, from theoretical
point of view, different EoSs in the context of GR and other
modified theories of gravitation could not support all
observational data, simultaneously. Therefore, to describe the
neutron stars, research on suitable EoS and appropriate theory of
gravitation, consistent with astrophysical observations, is
ongoing with great motivation.

This paper is devoted to study the structure of neutron stars
through the use of different state equations in mimetic gravity.
Various models of EoS are examined in the context of GR and
mimetic gravity. Differentiating between the models of EoS
indicated by the measurements of the radius and mass of neutron
stars in a theory of gravitation. It was shown that considering
a few suitable EoSs, the maximum mass and related radius
in mimetic gravity can support the mentioned observational data.

In this paper, we examined eight
relativistic/non-relativistic EoSs that seem more realistic and
have already received more attention in literature. For the
non-relativistic category of EoSs, we considered SLy4, BSK21 and
WFF1 models. While for the relativistic set of EoSs, we regarded
MPA1, HS (DD2) and FSU2H cases. We also used two hadronic EoSs
known as $BHB\Lambda\Phi$ and SFHoY as other realistic models.
According to the obtained results reflected in different figures
and tables, we have found that, in general, the results of mimetic
gravity are more consistent with observational data than the GR.
Besides, we have shown that the results of mimetic gravity are
more or less depending on the functional form of potential and its
free parameters. Nonetheless, we have found that the results of
full relativistic mean-field-based models of EoSs are in better
agreement with observational data than non-relativistic models.

For the three non-relativistic models of EoSs, SLy4, WFF1
and BSK21, we have shown that the mentioned state equations can be
improved when we generalize the GR to mimetic gravity. Strictly
speaking, regardless of GW190814, other observational evidence can
be supported by the results of theoretical counterparts in mimetic
gravity for SLy4 and WFF1 models. We should note that the same
result can be obtained, equivalently, for BSK21 in GR and mimetic
gravity. After that, a relativistic model of EoS can be regarded
to examine the possible support of GW190814 data. We have taken
the relativistic EoSs MPA1, DD2 and FSU2H into account in GR and
mimetic gravity, and found that although such models could not
solve our problem in GR, they could remove the inconsistency
between the GW190814 data and the theoretical results of previous
models in mimetic gravity. In other words, considering these three
EoSs are excellent candidates for investigation of neutron star
structures in mimetic gravity.

According to the presented tables, one can find that for numerical
analysis, we have considered the effects of mimetic gravity
generalization as a small changes with respect to GR. In other
words, we have investigated small values of potential since we did
not have the allowed region of free parameters. So, it is
interesting to find permitted region of free parameters of mimetic
gravity (its potential) based on observational cosmology or
fundamental conceptions of high energy physics, and study the
effects of mimetic gravity when it is far from the GR. Besides,
taking the modification of the mass-radius relation in mimetic
gravity into account, it will be interesting to study the thermal
evolution of neutron stars. In other words, it is attractive to
investigate the cooling process of isolated neutron stars to
compare mimetic gravity and GR. So, one can study the possibility
of the direct Urca mechanism for rapid cooling or modern theory of
cooling based on the nucleon superfluidity depending on the chosen
EoS. Keeping the results of this paper and the mentioned
suggestions in mind, we can analyze the behavior of hybrid neutron
stars or quark stars as well as other compact objects in mimetic
gravity. All these points are under considerations.

As a final point, Astashenok et al. in Ref.
\cite{AstashenokO2016} have studied the structure of neutron stars
in mimetic gravity. They have obtained some interesting results
about the structure of massive neutron stars. But, they had not
compared their results with observational data. Nonetheless,
observational data such as GW170817, PSR J0740+6620, PSR
J2215+5135, NICER data on PSR J0030+0451, and GW190814 give us
essential information on massive compact objects e.g., neutron
stars. In this work, we extended the work of Astashenok et al. by
comparing observational data with mimetic theory of gravity. In
other words, to obtain a good agreement between theory and
observational data, we evaluated the structure of neutron stars in
GR and mimetic gravity.

\begin{acknowledgements}
We are grateful to the anonymous referee for the insightful
comments. SHH and HN thank Shiraz University Research Council. The
work of BEP has been supported by University of Mazandaran under
title "Evolution of the Masses of Celestial Compact Objects in
Various Gravity".
\end{acknowledgements}

%\bibliographystyle{unsrt}
%\bibliography{refs}

\end{document}